\documentstyle[epsf]{ioplppt}

\def\openone{\leavevmode\hbox{\small1\kern-3.3pt\normalsize1}}
\def\losim{\mathrel{\hbox{\lower1ex\hbox{\rlap{$\sim$}\raise1ex\hbox{$<$}}}}}
\def\grsim{\mathrel{\hbox{\lower1ex\hbox{\rlap{$\sim$}\raise1ex\hbox{$>$}}}}}

\begin{document}
\jl{3}
\review{Muon spin rotation and relaxation in magnetic materials}[$\mu$SR in
magnetic materials]
\author{P Dalmas de R\'eotier and A Yaouanc}
\address{Commissariat \`a l'Energie Atomique,\\
D\'epartement de Recherche
Fondamentale sur la Mati\`ere Condens\'ee,\\
Service de Physique Statistique, Magn\'etisme et Supraconductivit\'e,\\
F-38054 Grenoble cedex 9}
\today
\begin{abstract}
A review of the muon spin rotation and relaxation ($\mu$SR) studies 
on magnetic materials published from July 1993 is presented. It covers
the investigation of magnetic phase diagrams, of spin dynamics and the analysis 
of the magnetic properties of superconductors. We have chosen to focus on 
selected experimental works in these different topics. In addition, a list of 
published works is provided.\par
\end {abstract}

\tableofcontents

\section{Introduction}\label{intro}

The breadth, format and style of this article is intended to provide an
accessible and 
stimulating review of recent investigations of the physical properties of
magnetic materials by the $\mu$SR experimental technique. $\mu$SR is an
acronym for Muon Spin Rotation, Relaxation or Resonance. We shall only 
deal with the first two of these techniques, i.e. the two most commonly used 
in studies of magnetic materials. Measurements can be carried out with 
positive and negative muons. Since almost all the investigations were performed
using the
positive muon, we shall only discuss these. By no means have we attempted to 
write a comprehensive review. At the risk of being invidious, we discuss in 
three sections selected recent works which display the possibilities of the
technique. To achieve a balanced picture, we list the works published from July
1993. The material published before July 1993, including that presented at the 
$\mu$SR conference held in early 1993 on the island of Maui, Hawai, has already 
been nicely reviewed (Schenck and Gygax 1995). \par

The organisation of this article is as follows. In \sref{muon} we introduce 
the basic concepts of $\mu$SR. In \sref{diagram} we discuss magnetic phase 
diagram studies. \Sref{dynamique} presents two examples of the 
investigation of spin dynamics in magnets. The next section (\sref{mixed}) 
deals with the very successful studies of the magnetic properties of 
superconductors. In \sref{conclusion} we summarize the present status of
$\mu$SR and mention the scheduled technical developments at the $\mu$SR 
facilities. In the last section (\sref{list}) we give a list of published 
works. This review is completed by four appendices, the material of which is 
partly original.\par

\section{$\mu$SR: Muon Spin Rotation, Relaxation}\label{muon}

Since the $\mu$SR technique has been described in many reports (Seeger 1978,
Chappert and Grynszpan 1984, Schenck 1985, Chappert and Yaouanc 1986, Cox 1987,
Karlsson 1995, Schenck and Gygax 1995, Schatz and Weidinger 1995), we will 
only sketch it briefly. More information is provided in the appendices.\par

Currently, 
$\mu$SR experiments can be performed at three facilities located at i) 
TRIUMF (4004 Wesbrook Mall, Vancouver BC, Canada V6T 2A3), ii) the 
Paul Scherrer Institut (PSI $\mu$SR Facility, CH-5232 Villigen PSI, 
Switzerland) and iii) the Rutherford Appleton Laboratory 
(ISIS Facility, Chilton, Didcot, Oxon OX11 0QX, 
United Kingdom). In addition, measurements are carried out at the Meson Science
Laboratory of the Faculty of Science of the University of Tokyo (Bunkyo-ku,
Tokyo 113, Japan) and the Phasotron of the Joint Institute for Nuclear Research 
(Dubna, Head Post Office, P.O. Box 79, Moscow, Russia). The muon beams at the
ISIS Facility and Meson Science Laboratory being pulsed beams are well adapted
to study weak magnetic signals: this capability results from the virtual 
absence of background related to contamination of the beam with particles other 
than muons. However, their relatively low time-resolution means they are 
unsuitable for investigating systems exhibiting fast relaxation processes or 
appreciable local 
fields (larger than $\sim$ 50 mT) at the muon site. The latter limitations 
do not apply to the beams at the other institutions since they are 
quasi-continuous. In fact, the muon beams at the different institutions are 
complementary.\par
 
The $\mu$SR technique uses the positive muon as a probe. The muon may form a
bound state with an electron, called muonium, an exotic isotope of hydrogen.
However muonium has never been observed in metals and the probability of its
formation in oxides is small. We shall therefore consider only free muons.\par

In the $\mu$SR technique polarized muons are implanted into a sample where 
their polarization evolves in the local magnetic field until they decay
(the muon lifetime is $2.2\ \mu$s). Because of its positive charge, the muon 
localizes at an interstitial site. Due to the absence of quadrupolar electric 
moment (spin 1/2) the muon does not couple to electric field gradients. The 
decay positron is emitted preferentially along the muon spin direction; by 
collecting several million positrons, one can reconstruct the time dependence 
of the muon spin-depolarization function which, in turn, reflects the spatial 
and temporal distribution of magnetic fields at the muon site. 
Because of their large kinetic energy ($\approx$ 30 MeV), the positrons easily 
go through the sample. Fortunately they are weakly absorbed by cryostat or 
furnace walls so that complex sample-environment equipment may be used.\par

\begin{figure}
\centerline{\epsfbox{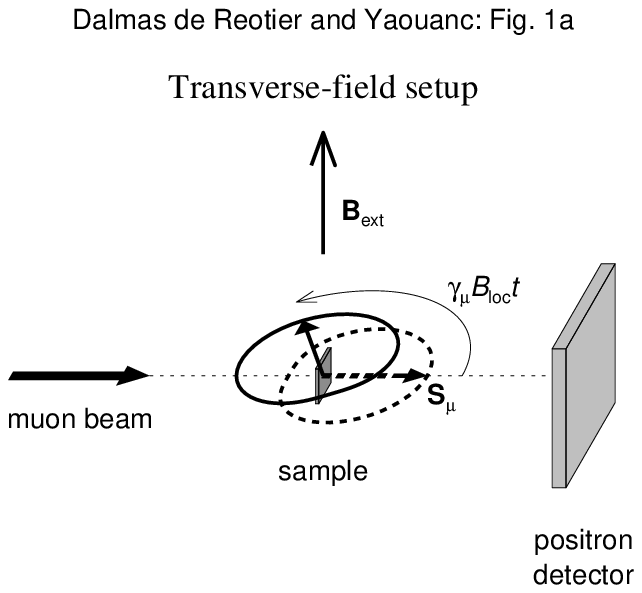}}
\centerline{\epsfbox{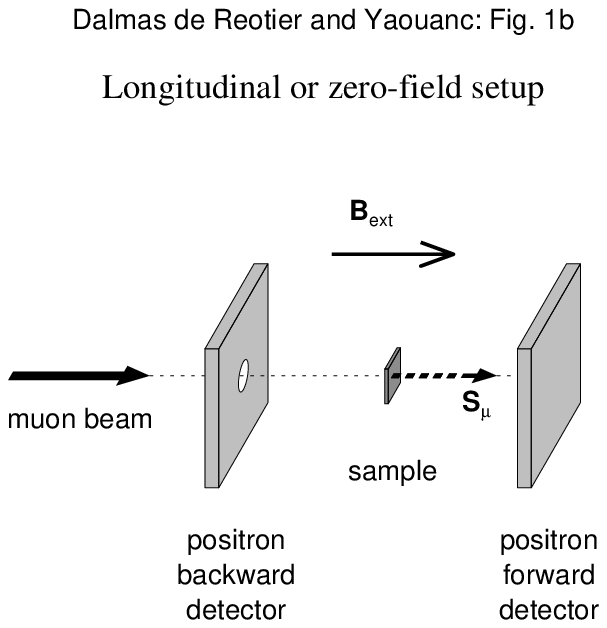}}
\caption{Principle of the two types of experimental geometries: the transverse 
and longitudinal set-ups are cartooned in the upper and lower panel, 
respectively. In order to produce drawings which are easy to understand, the 
muon beam momentum and the initial muon beam polarization have been taken 
parallel. In reality, these two vectors are antiparallel. The arrows 
originating from the sample sketch the muon spin direction: the dashed one at 
$t$ = 0 and the solid one at
$t$. The dashed and solid line cardioids drawn for the transverse field setup 
panel represent the probability of positron emission in a given direction 
relative to the muon spin position at the instant of the decay.}
\label{geometrie}
\end{figure}

Two types of experimental geometries are generally used, see 
\fref{geometrie}. In the longitudinal
geometry an external magnetic field $\bf {B}_{\rm ext}$ is applied along the 
initial muon beam polarization direction ${\bf S}_\mu$ and positron detectors 
are 
set parallel and antiparallel to ${\bf S}_\mu$. We then refer to the forward 
and backward direction, respectively. In the transverse geometry  
$\bf {B}_{\rm ext}$ is perpendicular to ${\bf S}_\mu$ and the positrons are 
detected perpendicular to $\bf {B}_{\rm ext}$. The zero-field 
measurements are performed with the longitudinal geometry.\par

The $\mu$SR technique is a local-probe hyperfine-method as are nuclear 
magnetic resonance (NMR), perturbed angular correlations (PAC) and M\"ossbauer
spectroscopy. Therefore the basic physical concepts developped for the latter
techniques can be transferred to $\mu$SR. One of its unique
characteristics is its ability to detect very small magnetic moments. This is
understood as follows. \par

We first recall that a magnetic moment precesses around 
a local magnetic field ${\bf B}_{\rm loc}$ at a pulsation frequency $\omega$ 
proportional to $B_{\rm loc}$: this is the Larmor precession. For the muon 
$\omega_{\mu}$ = $2 \pi \nu_\mu$ = $\gamma_\mu B_{\rm loc}$ where $\nu_\mu$ is 
the precession frequency and $\gamma_\mu$ is the muon gyromagnetic ratio 
($ \gamma_ \mu$ = 851.6 Mrad s$^{-1}$T$^{-1} $). Therefore, in a constant 
field, the moment rotates by an angle $\gamma_\mu B_{\rm loc} t$ in the elapsed 
time $t$. Since an angle of $0.5$ radian is measurable $(\cos 0.5 \simeq 0.88)$ 
and a $\mu$SR measurement can be routinely carried out up to $t$ = 15 $\mu$s, a
local field as small as 0.04 mT can be detected.
It might be produced by a nuclear magnetic moment which is 
$\sim 10^3$ times smaller than an electronic magnetic moment. \par

While the $\mu$SR technique can detect very small magnetic fields, very high
fields can not be measured: with a conventional spectrometer of a continuous 
muon source the maximum field is $\sim$ 3 T. Recent measurements at 6.5 T
have been performed by Riseman \etal 1995.\par

The muon spin depolarization function is written as $a P_\alpha(t)$. It is also
called the asymmetry. $a$ is the initial asymmetry (at $t = 0$) and 
$P_\alpha(t)$ the normalized depolarization function which will be referred to
as the depolarization function. The Cartesian label $\alpha$ denotes the 
direction along which the muon polarization is measured, i.e. it can be $X$,
$Y$ or $Z$. The value of $a$ depends on the experimental 
geometry; typically $a$ = $0.25$. We use an orthonormal laboratory reference 
frame. ${\bf B}_{\rm ext}$ is taken along the $Z$ axis and ${\bf S}_\mu$ 
parallel to $Z$ or $X$ in the longitudinal and transverse set-up, respectively. 
Whereas in a longitudinal set-up, only $P_Z(t)$ is of interest, in the 
transverse set-up, $P_X(t)$ and $ P_Y(t)$ can be measured. Since these two 
functions always contain the same information, we only consider $P_X(t)$.\par

$P_\alpha(t)$ monitors the properties of the magnetic field at the muon site. 
If all the muon spins precess in the same static magnetic field, oriented at an 
angle $\theta $ from ${\bf S}_\mu$, the Larmor equation yields
\begin{eqnarray}
P_\alpha(t) = \cos^2 \theta + \sin^2 \theta \cos(\omega_{\mu} t). 
\label{larmor}
\end{eqnarray}
This is the result on which the entire $\mu$SR technique is based. \par

\begin{figure}
\centerline{\epsfbox{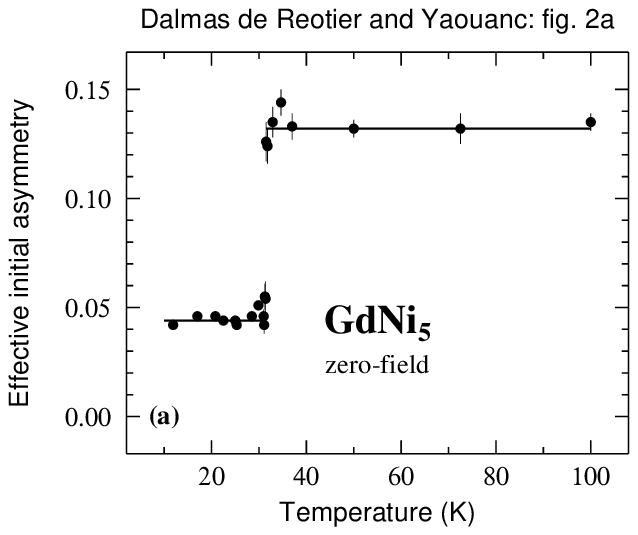}\hfill
\epsfbox{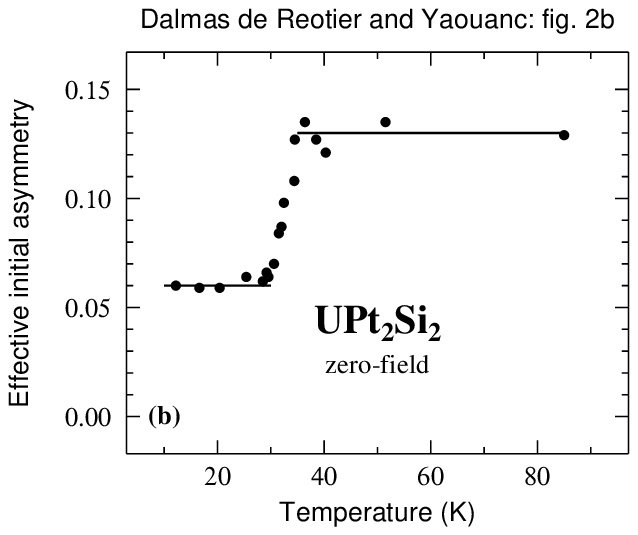}}
\caption{Temperature dependence of the effective initial asymmetry after
backgroung substraction measured in two polycrystalline samples at the ISIS 
Facility (Dalmas de R\'eotier \etal 1990 and Gubbens \etal 1996 ). 
Whereas this asymmetry drops at $T_C$ in a very 
small temperature range ($\sim$ 0.1 K) in the ferromagnet GdNi$_5$ (a), 
it changes gradually near $T_N$ (within $\sim$ 5 K) in the antiferromagnet 
UPt$_2$Si$_2$ (b). The solids lines are guide to the eyes.}
\label{assymetrie}
\end{figure}

We consider a magnet in zero external magnetic field. Then $P_Z(t)$ is of 
interest.
${\bf B}_{\rm loc}$ is usually not zero below the magnetic phase transition. 
For a magnet in polycrystalline form, the spatial average of \eref{larmor} has to 
be performed. If the sample is not textured, we obtain
\begin{eqnarray}
P_Z(t) = {1\over 3} + {2\over 3} \cos(\omega_{\mu} t).
\label{poudre}
\end{eqnarray}
The oscillating component reflects the magnetic order in the sample. This 
single component exists even in a polycrystalline sample because we suppose 
there is only one type of muon localization sites and for these sites 
${\bf B}_{\rm loc}$ is the same. If the magnet is disordered, i.e., if 
its correlation length is small, ${\bf B}_{\rm loc}$ can take a large number of
values. Then the oscillation can be strongly damped and even disappear.
If the muon spins precess too quickly relative to the time 
resolution of the spectrometer, the oscillation will not be observed and the 
resulting muon spin-polarization will be averaged out to 0. Then, at a magnetic 
phase transition, if no wiggles are observed in the muon signal, one expects a 
drop in the 
effective initial asymmetry from $a$ in the paramagnetic state to $a/3$ in the 
ordered state. In \fref{assymetrie} we present examples for two samples. \par

Whereas for GdNi$_5$ we do observe the expected behaviour, for UPt$_2$Si$_2$ 
the change at $T_N$ is smoother than expected and the ratio of the effective 
initial asymmetry in the paramagnetic and ordered state is smaller. We then 
conclude that the distribution of $T_N$ values is relatively large in the 
UPt$_2$Si$_2$ sample. In addition, this sample has a strong texture. The 
distribution is the signature of crystalline disorder: probably some Pt and Si
atoms interchange their atomic positions. Complementary measurements at ISIS on 
a single crystal sample of UPt$_2$Si$_2$ with ${\bf S}_\mu$ perpendicular to
${\bf B}_{\rm loc}$ (and therefore to the $c$ axis) show that the effective
initial asymmetry decreases to zero around $T_N$ in a temperature interval as 
large as $\sim$ 5 K (Gubbens \etal 1996). This result 
supports the conclusion deduced from 
the polycrystalline sample measurements. Note that the measurement of the 
effective initial asymmetry offers a stringent test of the sample quality. \par

\Fref{assymetrie} shows that the maximum initial asymmetry is smaller than 
expected. The difference is accounted for by the contribution from the 
background, i.e. muons which have not been stopped in the sample, but for
instance in the sample holder. We write 
$a P_Z(t)$ = $a_{\rm s} P_Z^{\rm s}(t)$ + $a_{\rm bg} P_Z^{\rm bg}(t)$
where the first and the second term describes the contribution from the 
sample and the background, respectively. $a_{\rm s}$ is plotted in 
\fref{assymetrie}. Usually the sample holder is a silver plate, 
$P_Z^{\rm bg}(t)$ is then taken as time independent 
($P_Z^{\rm bg}(t)$ = 1) because silver has no electronic moment and very small 
nuclear moments. On the other hand, if one analyses an extremely weakly damped
$\mu$SR signal, the damping due to the silver nuclear moments should be taken
into account.\par

The depolarization of the muon spins is caused either by a static distribution
of local fields (dephasing of the muon spins) or by fluctuating fields
(relaxation of the muon spins). The general framework needed to describe the
combination of the two mechanisms will be first given. 
In \sref{deptrans} and \sref{deplong}, respectively, the resulting $P_Z(t)$ and 
$P_X(t)$ 
will be presented. The relaxation induced by a purely dynamical process  
described in \sref{deplong}.\par

When measuring $P_\alpha(t)$, an ensemble average is made. Usually the muons 
do not all experience the same local field, i.e. there is a field distribution 
at the muon site, $D ({\bf B}_{\rm loc})$. The time evolution of the muon 
magnetic moment experiencing a field ${\bf B}_{\rm loc}$ can be computed by 
solving the Larmor equation in classical mechanics (or the Heisenberg equation
in quantum mechanics). We define 
$\hat S_{\mu,\alpha}(t,{\bf B}_{\rm loc})$, the projection along the $\alpha$
axis of the unit vector parallel to the muon moment. The depolarization
function is simply written
\begin{eqnarray}
P_\alpha (t) = \int \hat S_{\mu,\alpha}(t,{\bf B}_{\rm loc}) 
D ({\bf B}_{\rm loc}) \, d {\bf B}_{\rm loc}
\label{distribution}
\end{eqnarray}
If the field distribution is sharp, $D ({\bf B}_{\rm loc})$ is a Dirac
delta function and we recover \eref{larmor}. Different origins for 
$D ({\bf B}_{\rm loc})$ exist depending on the physical case. For a magnet the 
distribution is due to the nuclear and electronic magnetic moments. For a 
superconductor in the mixed phase, the nuclear magnetic moments and the 
flux quanta induce the distribution. First, for simplicity, we suppose 
that this distribution is static, isotropic and Gaussian centered at 0. 
Therefore the distribution is written as 
$D({\bf B}_{\rm loc})$ 
$\equiv$ $D(B^X_{\rm loc}, B^Y_{\rm loc}, B^Z_{\rm loc})$ 
where
\begin{eqnarray}
D({\bf B}_{\rm loc}) & = &
{\left( {\gamma_\mu \over \sqrt {2 \pi} \Delta} \right)}^3
\exp {\left[ -{ {(\gamma_\mu B^X_{\rm loc})}^2 \over 2 \Delta^2} \right]}
\exp {\left[ -{ {(\gamma_\mu B^Y_{\rm loc})}^2 \over 2 \Delta^2} \right]}\cr
& \times &
\exp {\left[ -{ \gamma_\mu^2 \left( B_{\rm ext} -B^Z_{\rm loc} \right)^2
\over 2 \Delta^2} \right]}. 
\label{distribution_gaussian}
\end{eqnarray}
$\Delta^2/ \gamma_\mu^2$ is the variance of the components of the field 
distribution. $\Delta$ can be expressed in terms of the interaction between the
muon spin and the spins of the compound and the characteristics of the latter
spins (see Schenck 1985 for details). In contrast to NMR where the field at the
probe of dipolar origin is negligible (except in the case of NMR on protrons), 
the dipolar
interaction dominates in the $\mu$SR case. This can have important concequences
as shown for example in \sref{dipolar}.\par

Below, we shall present the depolarization function for the two possible 
experimental geometries, and discuss the meaning of the magnetic field at the 
muon site.\par

\subsection{The transverse depolarization function}\label{deptrans} 

Up to the end of the eighties most of the experiments were performed with the 
transverse geometry (or in zero external field for the case of magnetically
ordered materials), mainly because the spectra are then less sensitive to the 
quality of the muon beam, e.g., contamination with other particles, 
than when recorded with the longitudinal geometry. 
Today the transverse geometry is mainly used for the investigation of 
the microscopic field distributions arising from either the vortex state in 
superconductors or magnetic domains in metals. In addition, a very active
field of investigation is the measurement of the local susceptibility at 
the muon site. This type of experiment yields information on the 
symmetry at the muon site and on the hyperfine coupling-constant.\par

The field distribution leads to a dephasing of the muon spins responsible for 
the depolarization. If $B_{\rm ext}$ is sufficiently large relative to 
$\Delta/\gamma_\mu$, i.e. if $B_{\rm ext} \geq 5(\Delta/\gamma_\mu)$ 
(Dalmas de R\'eotier and Yaouanc 1992), $P_X(t)$ probes only the field 
distribution along ${\bf B}_{\rm ext}$, i.e. along the $Z$ axis. Then, using 
\eref{distribution} and \eref{distribution_gaussian}, the following simple 
result is found:
\begin{eqnarray}
P_X(t) = \exp {\left( -{ \Delta^2 t^2 \over 2} \right)}\cos(\omega_\mu t).
\label{dep_gaussian_transverse}
\end{eqnarray}
If the argument of the Gaussian term is small, i.e. if $t \Delta$ is small, the 
envelope of $P_X(t)$ is well approximated by the parabolic form 
$(1 - \Delta^2 t^2/2)$.\par

Usually the field distribution is not static. A useful first approximation is 
to account for the dynamics with a single fluctuation rate, $\nu$, and to use
the stochastic theory of dynamical processes. Kehr \etal 1978 have used the 
strong collision model which supposes that $B_{\rm loc}$ takes a given value 
for a time $1/\nu$ followed by a new value not related to the previous one, 
i.e. they consider a Markov process. They have shown that the following 
analytical formula is a fair approximation:
\begin{eqnarray}
P_X(t) = \exp {\left\{ -{ \Delta^2 \over \nu^2} 
\left[ \exp(-\nu t) - 1 +\nu t \right] \right\} }\cos(\omega_\mu t).
\label{dep_abragam_transverse}
\end{eqnarray}
This formula, called the Abragam formula (in fact it was first derived by 
Anderson 
1954 as a model for the NMR line shape), interpolates between the static case, 
\eref{dep_gaussian_transverse}, and the fast fluctuation limit, the so-called 
motional-narrowing limit, for which the envelope is an exponential function:

\begin{eqnarray}
P_X(t) = \exp {\left( - \lambda_X t \right)}\cos(\omega_\mu t),
\ \ \ \ \ \  \lambda_X = \Delta^2/\nu.
\label{extra_trans}
\end{eqnarray}

Although the Gaussian field distribution model with the dynamics described as a
Markov process provides only a rough picture of the physics involved, it is 
useful since it clearly shows the physical origin of the measured damped 
oscillation. The limits of this model for the depolarization induced by 
nuclear magnetic moments is discussed by Dalmas de R\'eotier 
\etal 1992, Yaouanc and Dalmas de R\'eotier 1994 and Cameron and Sholl 1994. 
In a magnet, the 
functional form of $D({\bf B}_{\rm loc})$ reflects its magnetic structure. This
is discussed in \ref{appendix1}. In a type II superconductor, 
$D({\bf B}_{\rm loc})$ is usually not Gaussian. In fact its shape is
characteristic of the vortex state as shown in \sref{mixed}. Additional 
information is provided in \ref{appendix3}.\par

\subsection{The longitudinal depolarization function}\label{deplong} 

Nowadays, the longitudinal geometry is popular for the study of magnets since
it is well suited for the characterization of a magnetic phase transition.
The fact that the measurements can be performed in purely zero-field is a
definitive advantage of the $\mu$SR method.\par

\begin{figure}
\centerline{\epsfbox{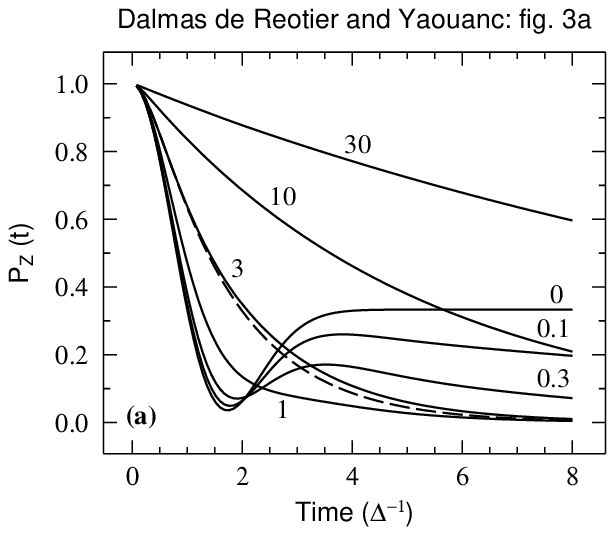}\hfill
\epsfbox{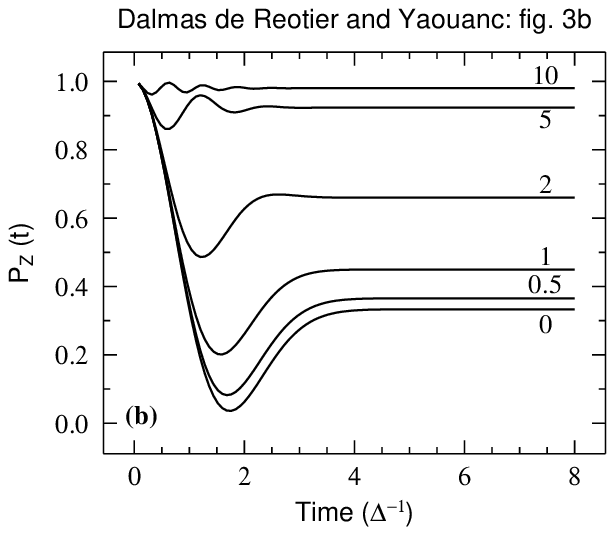}}
\caption{(a) Fluctuation-rate dependence of the zero-field depolarization 
function derived for a isotropic Gaussian field distribution and a Markov 
dynamical process. (b) Magnetic field dependence of the longitudinal 
depolarization function derived for a static isotropic Gaussian field 
distribution. The values of the fluctuation rate and magnetic field are 
respectively given in units of $\Delta$ and $\Delta/\gamma_\mu$. 
The dashed line in the (a) panel is the prediction
of the modified Abragam formula (\protect \eref{dep_gaussian_longitudinal})
for $\nu/\Delta$ = 3. The Kubo-Toyabe function (Eq. \protect 
\ref{KTfunction_full}) corresponds to the curves marked with 0 in both panels.}
\label{theorie_KT}
\end{figure}

In zero-field, an analytical formula is found for $P_Z(t)$ if the distribution
is isotropic and Gaussian. Using \eref{distribution} and 
\eref{distribution_gaussian} one derives
\begin{eqnarray}
P_Z(t)={1 \over 3} + {2 \over 3} (1-\Delta^2t^2)
\exp(-{1 \over 2} \Delta^2t^2).
\label{KTfunction_full}
\end{eqnarray}
This is the well-known Kubo-Toyabe function which is plotted in 
\fref{theorie_KT}. $P_Z(t)$ exhibits a dip at $t = {\sqrt 3}/\Delta$. If
$t \Delta $ is sufficiently large, it saturates to 1/3.
If $t \Delta $ is small, $P_Z(t)$
is well approximated by the parabolic form $P_Z(t) = 1 - \Delta^2t^2$. Relative 
to the transverse case, the initial depolarization is stronger in zero-field 
by a factor of two since for the latter geometry the two components of 
${\bf B}_{\rm loc}$ perpendicular in the $Z$ axis participate to
the depolarization.
\Eref{KTfunction_full} is strictly valid for an isotropic field distribution
at the muon site. Szeto 1987, Dalmas de R\'eotier 1990 and Solt 1995 have 
computed $P_Z(t)$ for different types of anisotropic field distributions. As 
expected, the value of $P_Z(t)$ for $t \Delta $ large reflects the type of 
anisotropy. \Eref{KTfunction_full} is derived with the hypothesis that the 
average field at the muon site is zero. An extension for a finite average 
field inside a polycrystalline sample has been given by 
Kornilov and Pomjakushin 1991.\par

As for the transverse geometry, the dynamics can be accounted for approximately
by the strong-collision model (Hayano \etal 1979). Examples are shown in  
\fref{theorie_KT}. Interestingly, at large $t$, the functional form of $P_Z(t)$ 
depends strongly on $\nu$ when the distribution is quasi-static, and 
the Kubo-Toyabe function can still be used but with the $1/3$ factor multiplied 
by $\exp(-2 \nu t/3)$. Therefore, slow dynamical processes can be characterized 
by zero-field measurements. If $\nu/\Delta$ is sufficiently large, $P_Z(t)$ is 
well approximated by the analytical formula (Dalmas de R\'eotier 
and Yaouanc 1992)
\begin{eqnarray}
P_Z(t) = \exp {\left\{ -{2 \Delta^2 \over \nu^2} 
\left[ \exp(-\nu t) - 1 +\nu t \right] \right\} }.
\label{dep_gaussian_longitudinal}
\end{eqnarray}
This approximation is compared in \fref{theorie_KT} to the exact numerical
solution for $\nu/\Delta = 3$. This analytical formula is the envelope of the 
transverse depolarization function (see \eref{dep_abragam_transverse}), except 
for the substitution $\Delta^2$ $\rightarrow 2 \Delta^2$. The factor 2 reflects 
again the fact that the depolarization is induced by the two components of 
${\bf B}_{\rm loc}$ perpendicular to the $Z$ axis. Further analysis of $P_Z(t)$
in a low field for time-dependent random magnetic fields is given by Shibata and
Shimoo 1995. In the motional narrowing limit, i.e. when the dynamics is 
fast, $\nu \ll 1/t$, $P_Z(t)$ is an exponential function
\begin{eqnarray}
P_Z = \exp(-\lambda_Z t),
\ \ \ \ \ \ \lambda_Z = 2 \Delta^2/\nu.
\label{explong}
\end{eqnarray}

In order to fully characterize a quasi-static process, it is important to
perform longitudinal field measurements. \Fref{theorie_KT} presents
the field dependence of $P_Z(t)$ for a static field distribution: $P_Z(t)$ is
strongly field dependent. Obviously, the strong collision model can be applied
to this depolarization functions to account for dynamics.
Recently Uemura \etal 1994 have pointed out that the
following scaling property holds:
\begin{eqnarray}
P_Z(\Delta, B_{\rm ext}, \nu, ft) = P_Z(f\Delta, fB_{\rm ext},f\nu,t).
\label{scaling}
\end{eqnarray}
Physically, $f$ can be the fraction ($f < 1$) of the time during which the 
muon experiences a non-zero magnetic field, i.e. in the time fraction $(1-f)$, 
this field is zero. Spin liquid systems could be an example of such a model
(Uemura \etal 1994, Bonville \etal 1997). 
Note that fitting a spectrum with a regular dynamical
Kubo-Toyabe function with a longitudinal field then yields 
reduced effective $f\Delta$, $fB_{\rm {ext}}$ and $f\nu$ values.\par

The field distribution at the muon site in disordered systems is Gaussian only 
when the spin concentration is large. Walstedt and Walker 1974 have shown that
the distribution is Lorentzian for a disordered dilute spin system. Kubo 1981 
has computed $P_Z(t)$ for such a distribution. In the motional narrowing limit
$P_Z(t)$ is then a square root stretched exponential function. $P_Z(t)$ for
disordered systems has been discussed recently 
by Berzin \etal 1993, Borgs \etal 1995 and Crook and Cywinski 1997.\par

Even if the Gaussian approximation is fair, as it is usually the case for the 
depolarization due to nuclear spins, the Kubo-Toyabe formula and its extension
to dynamical processes does not provide a sufficiently good model to
analyse high statistics spectra. The interaction between the muon spin and the
lattice spins has to be taken into account, at least in nuclear systems 
(Celio 1986, Dalmas de R\'eotier and Yaouanc 1992, Keren 1994, 
Yaouanc and Dalmas de R\'eotier 1995).\par

For a purely fast dynamical process, the relaxation of the muon spin leads to an
exponential depolarization function characterized by a relaxation rate
$\lambda_Z$; see \ref{appendix2bis}. 
This process is very similar to the NMR spin-lattice relaxation, 
and consists of an exchange of energy between the lattice spins and the two muon 
Zeeman levels. In that case, the name of relaxation function for $P_Z(t)$ 
seems more appropriate. We point out that $P_Z(t)$ is an exponential function 
only if no spatial average is necessary. Therefore, in general, $P_Z(t)$ is not 
an exponential function for a polycrystalline sample (see Bonville \etal 1996). 
In \ref{appendix2} we discuss the meaning of $\lambda_Z$ in terms of the spin 
correlation-functions of the magnet.\par

\subsection{The magnetic field at the muon site}\label{field} 

A detailed discussion of the field at the muon site, ${\bf B}_{\rm loc}$, 
is given by Schenck and Gygax 1995.\par

In general ${\bf B}_{\rm loc}$, is the sum of seven terms: 
\begin{eqnarray}
{\bf B}_{\rm loc} = {\bf B}_{\rm con} + {\bf B}_{\rm trans} + 
{\bf B}_{\rm dip} ^{\prime}
+ {\bf B}_{\rm L} + {\bf B}_{\rm dem} + {\bf B}_{\rm dia}+ {\bf B}_{\rm ext}.
\label{somme}
\end{eqnarray}
${\bf B}_{\rm con}$ is the contact hyperfine-field resulting from the spin 
density at the muon site which is induced by the polarization of the conduction 
electrons. Therefore it only exists in metals. ${\bf B}_{\rm trans}$ is the 
transferrred hyperfine field. In metals this field is due to the indirect 
Rudermann-Kittel-Kasuya-Yosida (RKKY) interaction. The next three terms in 
\eref{somme} reflect the muon spin 
interaction with the localized lattice spins through the dipolar interaction. 
This interaction gives rise to a dipolar field which is expressed as a 
lattice sum over the sample. This sum is split into two parts by separating the 
volume of the sample into a sphere around the muon (the Lorentz sphere) and the 
rest. ${\bf B}_{\rm dip} ^{\prime}$ is given 
by a lattice sum restricted to the Lorentz sphere; see \ref{appendix1}. 
Summing over the rest yields the Lorentz and the demagnetization fields, 
${\bf B}_{\rm L}$ ($= (\mu_0/3) {\bf M}_{\rm sat}$) and ${\bf B}_{\rm dem}$
($= -\mu_0 {\rm N} {\bf M}_{\rm bulk}$), respectively. ${\bf M}_{\rm sat}$ and
${\bf M}_{\rm bulk}$ are respectively the saturation and bulk magnetizations, 
$\mu_0$ the permeability of free space and ${\rm N}$ the demagnetization factor
tensor which depends only on the sample shape. The trace of ${\rm N}$ is 1.
For a sphere ${\bf B}_{\rm dem} = -\mu_0 {\bf M}_{\rm bulk}/3$. For an infinite 
plane, in practice for a disk with extremely small thickness relative to 
radius, ${\bf B}_{\rm dem}= 0$ if ${\bf B}_{\rm ext}$ is applied perpendicular
to the disk axis and ${\bf B}_{\rm dem} = -\mu_0{\bf M}_{\rm bulk}$ if 
${\bf B}_{\rm ext}$ is parallel to the disk axis. ${\bf B}_{\rm dia}$ in
\eref{somme} is the diamagnetic field which is important only for 
superconductors. In usual magnets, it can be neglected.\par

In a magnetized ferromagnet all the terms of \eref{somme}, except 
${\bf B}_{\rm dia}$, are to be taken into account. In a ferromagnet which is
not macroscopically magnetized, ${\bf B}_{\rm dem} = 0$ because 
${\bf M}_{\rm bulk} = 0$. For an antiferromagnet in zero field, 
${\bf B}_{\rm L}$ and ${\bf B}_{\rm dem}$ are necessarily zero. 
In a metal, ${\bf B}_{\rm con}$ and ${\bf B}_{\rm trans}$ are normally 
independent of the crystal direction. ${\bf B}_{\rm dip}^{\prime}$ depends 
strongly on the symmetry at the muon site; for examples, see Seeger 1978.\par

For a paramagnet or a superconductor in an external field, a shift of the muon
frequency is usually observed, i.e. ${\bf B}_{\rm loc}$ is different from 
${\bf B}_{\rm ext}$. The induced field $({\bf B}_{\rm loc} - {\bf B}_{\rm ext})$
is not necessarily parallel to ${\bf B}_{\rm ext}$. However, in general, 
$|{\bf B}_{\rm loc} - {\bf B}_{\rm ext}|$ $\ll |{\bf B}_{\rm ext}|$. Therefore
it is useful to characterize the observed shift by the projection of 
$({\bf B}_{\rm loc} - {\bf B}_{\rm ext})$ onto ${\bf B}_{\rm ext}$:
\begin{eqnarray}
K^{\rm exp} =  
{{\bf B}_{\rm ext} \cdot \left({\bf B}_{\rm loc} - {\bf B}_{\rm ext} \right)
\over B_{\rm ext}^2}.  
\label{knight_exp}
\end{eqnarray}
Since $K^{\rm exp}$ is not intrinsic to the compound (it depends on the sample
shape), one defines a new ratio, $K$, traditionally called the Knight shift:
\begin{eqnarray}
K =  
{{\bf B}_{\rm ext} \cdot  \left({\bf B}_{\rm con} + {\bf B}_{\rm trans} + 
{\bf B}_{\rm dip}^{\prime} \right) \over B_{\rm ext}^2}.  
\label{knight}
\end{eqnarray}
The two ratios can be related using \eref{somme}:
\begin{eqnarray}
K =  K^{\rm exp} - \left ({1\over 3} -n \right) \chi_p -
\left (1 -n \right) \chi_d,
\label{relation}
\end{eqnarray}
where $\chi_p$ is the paramagnetic susceptibility, $\chi_d$ the diamagnetic 
susceptibility produced by persistent currents in the mixed state of a 
superconductor and $n$ the appropriate component of ${\rm N}$ 
(Heffner \etal 1989). Obviously $\chi_d = 0$ if the compound is not 
superconducting at the temperature of the measurement. At first sight, one may 
be surprised that $\chi_p$ and $\chi_d$ are not multiplied by the same factor. 
This is understood if one remembers that ${\bf B}_{\rm L}$ responsible for the 
$1/3$ factor originates from the dipole moments in the Lorentz sphere. This 
concept is unsuitable for the superconducting electrons. The factor $1$ in
$(1-n)$ originates from ${\bf B}_{\rm dia}$, i.e. it does not concern the 
paramagnetic properties of the compound.\par

The concept of Knight shift is useful because one expects this quantity to
be field independent. In particular, in superconductors, its thermal behaviour
when crossing the superconducting temperature yields direct information on the
parity of the superconducting order parameter. However, recent measurements on
UBe$_{13}$ and UBe$_{12.91}B_{0.09}$ indicate that the Knight shift concept is 
not valid for these compounds since $[B_{\rm loc}(T) - B_{\rm loc}(T_c)]$ is 
field independent (Luke \etal 1991, Heffner \etal 1997), i.e. $B_{\rm loc}(T)$
does not scale with $B_{\rm ext}$. \par

Since $K$ depends linearly on the components of the susceptibitity tensor, for 
a given ${\bf B}_{\rm ext}$ direction, it is presented as a function of the 
corresponding susceptibility, the temperature being an implicit parameter. From 
such plots (the Clogston-Jaccarino plots) the value of the components of the 
dipolar tensor and of the hyperfine coupling constant can be determined 
(Schenck and Gygax 1995). \par

\begin{figure}
\centerline{\epsfbox{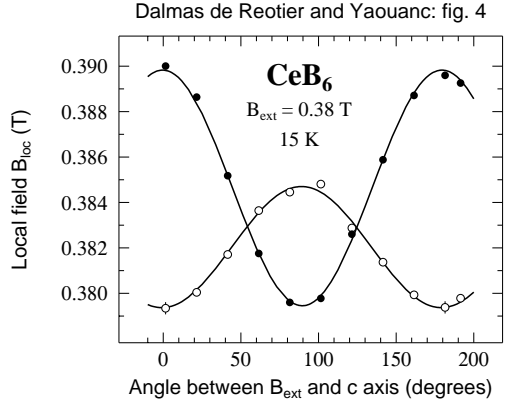}}
\caption{Angular dependence of the two local fields observed in CeB$_6$. 
${\bf B}_{\rm ext}$ was rotated around the $[1,1,0]$ axis. The two signals have 
an amplitude ratio 2:1 corresponding on one hand to the $(1/2, 0, 0)$ and 
$(0, 1/2,0)$ sites (open symbols) and on the other hand to the $(0,0,1/2)$ site 
(solid symbols). The solid lines are fit to the data 
(adapted from Amato \etal 1997).}
\label{site}
\end{figure}

The study of the angular dependence of the local field at the muon site
measured in transverse field on a single crystal in the paramagnetic phase 
gives the opportunity of determining the muon site. In \fref{site} such an 
angular dependence is shown for the cubic system CeB$_6$ 
(space group $Pm\bar{3}m$). The presence of two different fields (as seen in
the Fourier transform of the mesured signal) indicates the 
existence of two magnetically inequivalent sites for the muon. From the
analysis of this angular dependence and the value of the component of the
dipolar field deduced from the Clogston-Jaccarino plot, Amato \etal 1997
infer that the muon occupies the $d$ site 
$(0,0,1/2)$. In recent years this type of procedure has been used 
to determine the muon site in many intermetallic compounds. The results show
that it is not possible to predict reliably the muon site localization. For a
given crystal structure, this site depends on the chemical formula of the
compound. It may even depend on the temperature. Because of its electric 
charge, the muon may distort locally the crystal lattice and induce an electric
field gradient on its neighbor atoms.\par

\section{Magnetic phase diagrams}\label{diagram} 

The $\mu$SR technique has been popular in the condensed matter community 
for its 
success in detecting magnetic phase transitions in compounds with small 
magnetic moments. In this section we examine four such examples.
In \sref{organic} and \sref{uemura} we discuss recent results obtained for the
organic and low dimensional magnets, respectively. Whereas for the organic
compounds oscillating $\mu$SR signals are detected, pointing to the existence
of relatively well ordered magnetic strutures, in the 3-leg spin ladders one 
only observes wide field distributions, a fingerprint of disorder. 
In \sref{beryllium} we show that the $\mu$SR technique can be used to detect
magnetic domains in a diamagnetic metal such as beryllium. The appearance of the
domains is directly connected to the de Haas-van Alphen effect. Finally we
present in \sref{components} new results obtained for some heavy fermion metals 
in which two $f$ electron components exist.\par 

\subsection{Magnetic ordering in organic compounds}\label{organic}

The search for purely organic molecular ferromagnets which contain only light
elements (carbon, nitrogen, hydrogen and oxygen) is a subject of strong
current interest. The first such material to be found, the $\beta$ crystal
phase of {\sl para}-nitrophenyl nitronyl nitroxide 
({\sl p}-NPNN, C$_{13}$H$_{16}$N$_3$O$_4$), was reported to have a Curie 
temperature $T_C$ $\sim$ 0.6 K (Tamura \etal 1991). The unpaired spin is 
associated with the nitronyl nitroxide group (N-O group). The role of the 
rest of the molecule is to ensure the appropriate overlap of the correct 
orbitals on neighbouring molecules to produce 3D ferromagnetism. A whole 
series of materials which incorporate this N-O group has been synthesized.\par

\begin{figure}
\centerline{\epsfbox{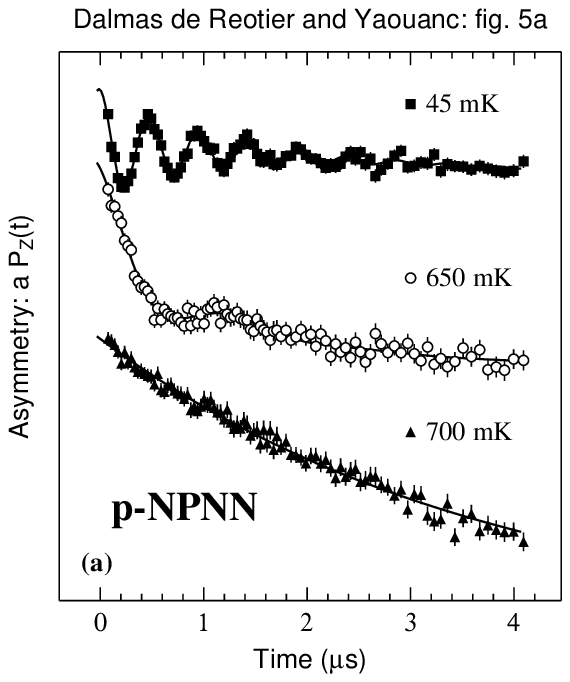}\hfill
\epsfbox{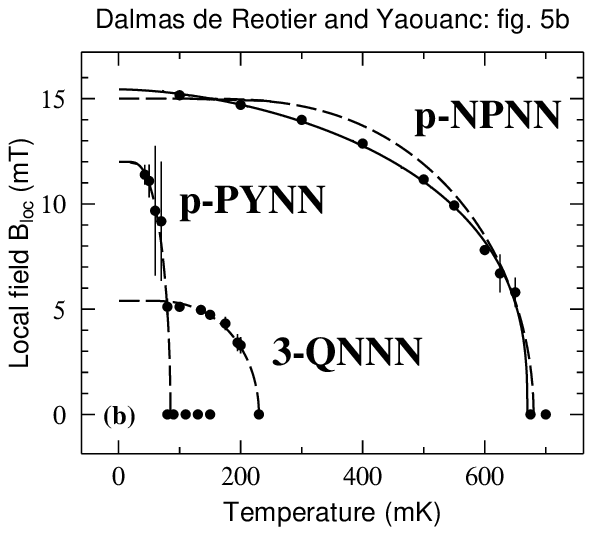}}
\caption{(a): some zero-field spectra recorded for 
{\sl p}-NPNN (adapted from Blundell \etal 1995). The solid lines are fits.
(b): temperature dependence of the local magnetic field at
the muon, $B_{\rm loc}$, in {\sl p}-NPNN, {\sl p}-PYNN and 3-QNNN 
(adapted from Blundell \etal 1994 and 1995). The solid line is a fit with 
$B_{\rm loc}(T)$ $\propto$ $[1- (T/T_C)^{\alpha}]^{\beta}$. The dashed lines 
are fits with the molecular field model with spin $S=1/2$.} 
\label{Blundell_spectre_diagramme}
\end{figure}

The first direct observation of spontaneous magnetic order in {\sl p}-NPNN was 
done using the $\mu$SR technique (Le \etal 1993b). It has been subsequently
confirmed by zero-field neutron diffraction (Zheludev \etal 1994). In 
\fref{Blundell_spectre_diagramme}a we present three zero-field $\mu$SR 
spectra. The mere observation of an oscillating signal at low temperature, 
i.e. of a spontaneous internal magnetic field $B_{\rm loc}$, is a clear 
signature of the existence of static magnetic correlations. Since a long-lived 
oscillation is detected rather than an increase in damping, {\sl p}-NPNN
orders with a well defined magnetic structure. Measurements on crystals 
indicate that ${\bf B}_{\rm loc}$, is nearly parallel to the $b$ crystal axis 
which is the easy axis (Le \etal 1993b). The shape of the spectra changes
drastically between 650 mK and 700 mK as seen in
\fref{Blundell_spectre_diagramme}a: 
the magnetic phase transition occurs between these temperatures. In 
\fref{Blundell_spectre_diagramme}b we present the temperature dependence of 
$B_{\rm loc}$. The solid line is a fit with $B_{\rm loc}(T)$
$\propto$ $[1- (T/T_C)^{\alpha}]^{\beta}$. This compact formula allows us to 
discuss the spin wave ($T \ll T_C$) and the critical regimes: 
for $T \ll T_C$, $[B_{\rm loc}(0)- B_{\rm loc}(T)]$ $\propto$ $T^{\alpha}$ and 
$B_{\rm loc}(T)$ $\propto$ $(T - T_C)^\beta$ near $T_C$. $B_{\rm loc}(T)$ is 
well described in the whole ferromagnetic state by the compact formula with 
$\alpha$ = 1.7 (4) and $\beta$ = 0.36 (5) (Le \etal 1993b and 
Blundell \etal 1995). These results are consistent with that of a 3D Heisenberg 
magnet. The weak magnetic anisotropy is accounted for by the dipolar 
interaction between the unpaired electron spins (Le \etal 1993b). The small
$B_{\rm loc}(T = 0 {\rm K})$ value found in the organic magnets is a strong
indication that the magnetic moment carried by the unpaired electrons is
small since it would be surprising that the dipolar field at the muon site 
accidentally cancels almost perfectly in all these compounds. An analysis of 
the behaviour of the oscillations in an applied longitudinal field, in terms of 
the magnetization process and demagnetising field, is given by 
Blundell \etal 1995.\par

The transition temperature and the nature of the magnetic ground state of the 
molecular crystals based on the nitronyl nitroxide radical depend strongly on 
their crystal structure. This is nicely seen in 
\fref{Blundell_spectre_diagramme}b:
$T_C \approx$ 90 mK for {\sl p}-PYNN , $T_C \approx$ 230 mK for 3-QNNN and 
$T_C \approx$ 670 mK for {\sl p}-NPNN. Other examples and a detailed discussion 
are given by Blundell \etal 1997b.\par

It is well known that neutron scattering can provide extensive information on
magnets, in particular on the magnetic structure and on the excitations. But in
compounds with small magnetic moments such as the organic magnets, the $\mu$SR 
technique is a sensitive and useful probe: it easily yields the value of the 
critical temperature and information on the thermal behaviour of the order
parameter. In addition, it should be possible to investigate the excitations
at a very small energy transfer (see \sref{dipolar} for an example for a
conventional ferromagnet).\par

\subsection{Magnetic ordering in spin chains and ladders}\label{uemura}

Hoping to gain insight into the mechanism of superconductivity, some 
experimentalists and theoreticians have recently turned their attention to 
systems as simple as one dimensional chains and ladders of copper and oxygen 
atoms with the aim to eventually apply their findings to the high $T_c$ 
superconductors where Cu-O planes play an important role. The $\mu$SR technique 
has been used to determine the magnetic properties of the ground state of some 
of these systems.\par

The ground state properties of a linear chain of antiferromagnetically coupled 
spins have been intensively studied because of the pronounced quantum 
effects. Both integer and half integer spin chain systems have a singlet ground 
state (Mermin and Wagner 1966, Haldane 1983a and 1983b). Because the spin 
excitations are gapless at momentum $q$ = 0 and $\pi$ for half integer spin
chains (des Cloizeaux and Pearson 1962), a magnetic ordering can be expected 
when interchain interactions are taken into account. However, for integer spin 
chains which have a so-called Haldane energy gap (Haldane 1983a and 1983b), no 
ordering should occur. The key parameters for half integer spin chains 
are the magnitude of the ratio $T_N/J$ 
($J$ is the intra-chain coupling constant) and the ordered magnetic moment at 
$T$ = 0 K, $M(T=0)$. $J$ is estimated from magnetic susceptibility and infrared 
light absorption measurements. The N\'eel temperature is easily determined by 
zero-field $\mu$SR in such systems characterized by small magnetic moments. 
Neutron diffraction is the most direct method to measure $M(T=0)$. 
Although the $\mu$SR technique cannot yield a precise $M(T=0)$ value for a 
given compound, it is well suited to measure accurately the relative size of 
moments for iso-structural materials. Therefore a combined  neutron 
diffraction and $\mu$SR study is expected to yield reliable results. 
Kojima \etal 1997 have done such a study for the quasi one-dimensional 
antiferromagnets Sr$_2$CuO$_3$ and Ca$_2$CuO$_3$. Plotting $M(T=0)$ versus 
$T_N/J$ and comparing with different models, they find that the chain mean 
field approach best explains the experimental results. This approach takes more 
quantum effects into account than the two different spin-wave 
approximations available. Probably, this is the reason why it provides a better 
description since the moment reduction is dominated by 
quantum spin fluctuations.\par  

\begin{figure}
\centerline{\epsfbox{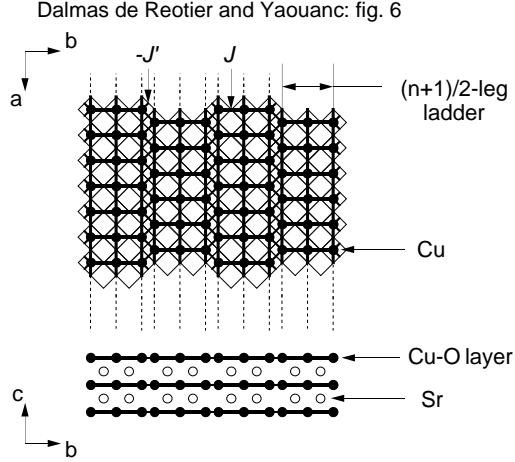}}
\caption{The 3-leg ladder structure (from Kojima \etal 1995a). Oxygen ions 
locate at each corner of the squares. The ferromagnetic interladder interaction
$J^\prime$ is much smaller than the antiferromagnetic intraladder interaction
$J$.}
\label{spinladder}
\end{figure}

As mentioned, the spin ladder systems have been studied because it is
thought that an understanding of their magnetic properties is a prerequisite
for a proper description of the magnetic properties of infinite CuO layer 
systems. A 3-leg ladder structure is displayed in \fref{spinladder}. The 
oxides Sr$_{n-1}$Cu$_{n+1}$O$_{2n}$ are realizations of such ladders. 
Indeed, one observes that the geometry of the ladder structure and of the CuO 
square lattice layer are related. A nice review of the physics of these systems 
is given by Goss Levi 1996. Note that the physics of the spin ladder systems 
and of the Haldane spin chains are closely related (Strong 1997).\par

A key prediction is that only ladders with even numbers of legs have a singlet 
ground state separated from the triplet state by a large spin gap 
(Rice \etal 1993). However, the odd-leg systems are expected to reach a 
magnetically ordered ground state in the presence of interladder interactions. 
Kojima \etal 1995a have performed zero-field and longitudinal 
field $\mu$SR measurements to test these theoretical predictions. \par

\begin{figure}
\centerline{\epsfbox{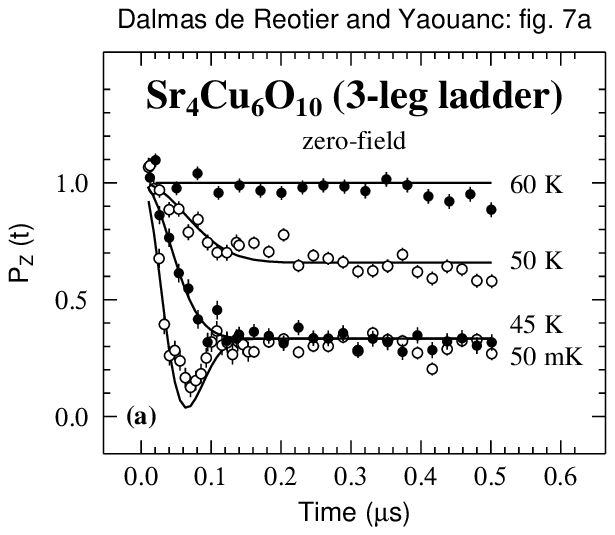}\hfill
\epsfbox{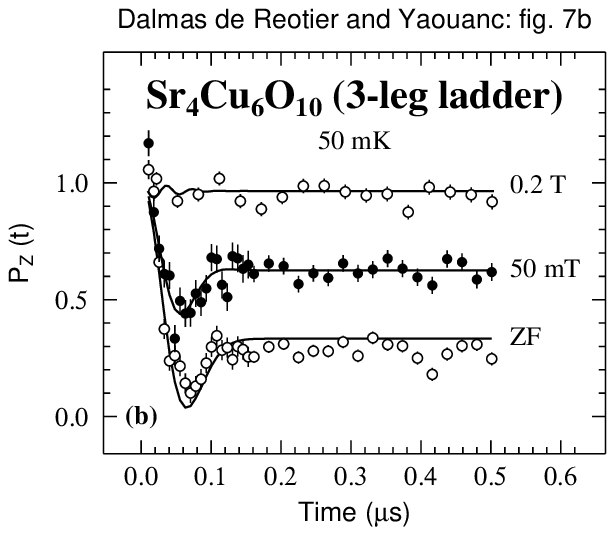}}
\caption{Some spectra recorded on Sr$_4$Cu$_6$O$_{10}$ which has a 3-leg spin 
ladder structure. The solid lines are fits (adapted from Kojima \etal 1995a).} 
\label{KOJIMA_3LEG}
\end{figure}

In \fref{KOJIMA_3LEG} spectra recorded on the 3-leg system are presented. 
The strong depolarization of the zero-field spectra at low temperature shows 
that the ground state is magnetic. Since no wiggles are detected (in contrast 
to the observations for the organic magnets; see \sref{organic}), the disorder
in the compound is important or the number of muon localization sites with
different local fields is large. 
Comparing the spectra recorded at 50 K and 60 K, 
we infer that a 3D magnetic phase transition occurs between these 
temperatures. The longitudinal field measurements confirm the interpretation of 
the zero-field spectra, i.e. the ground state of the 3-leg system with 
interladder interactions is a conventional static 
ordered state rather than a spin liquid system since $f$ $\simeq$ 1 (see
\eref{scaling}).\par
  
\begin{figure}
\centerline{\epsfbox{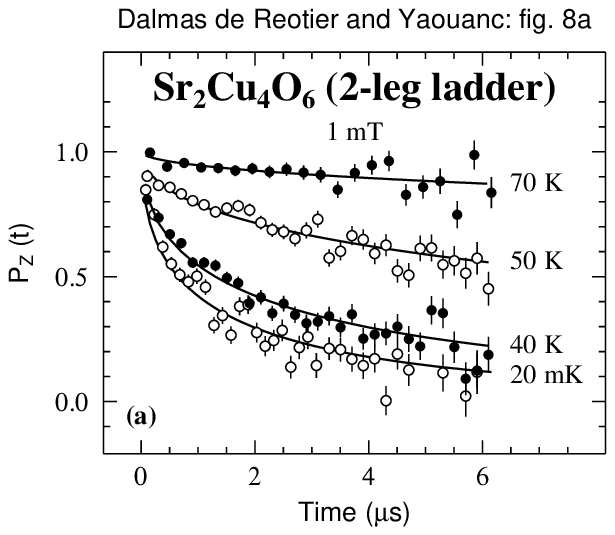}\hfill
\epsfbox{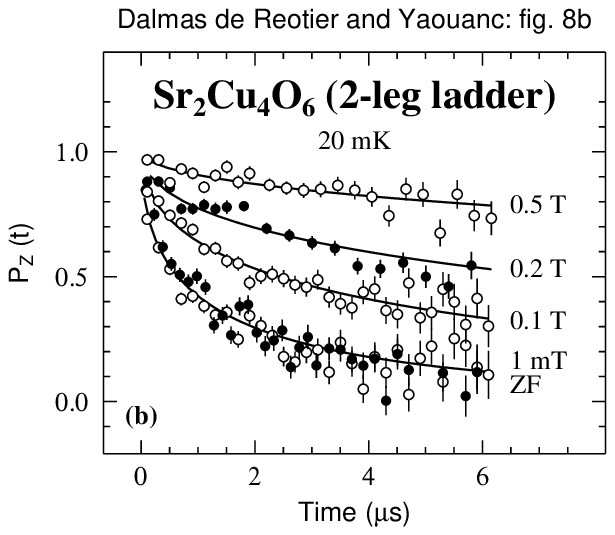}}
\caption{Some spectra recorded on Sr$_2$Cu$_4$O$_6$ which has a 2-leg spin 
ladder structure. The solid lines are fits (adapted from Kojima \etal 1995a).
Note that the horizontal scales are $\sim$ 10 times larger than in 
figure \protect\ref{KOJIMA_3LEG}.} 
\label{KOJIMA_2LEG}
\end{figure}

The magnetic behavior of the 2-leg and 3-leg ladder systems differ remarkably 
as seen in 
\fref{KOJIMA_2LEG}. The depolarization functions are described with a
square-root exponential function, appropriate for dilute fluctuating moments
(see \sref{deplong}). Therefore no static magnetic ordering is detected. The
depolarization originates from dilute unpaired spins which may be associated
with defects in the sample.\par

In conclusion, the work of Kojima \etal 1995a confirms the theoretical
predictions (Rice \etal 1993) that a 3-leg system becomes magnetic at low
temperature but not a 2-leg system.\par 

\subsection{Diamagnetic domains in beryllium}\label{beryllium}

In general terms, we first note that the relation between induction, external
field and magnetization for a given material is given by
\begin{eqnarray}
{\bf B} = {\bf B}_{\rm ext} -\mu_0 ({\rm N} -\openone){\bf M},
\label{constitutive}
\end{eqnarray}
where ${\rm N}$ is the demagnetization factor tensor and $\openone$ the unit
tensor.
Therefore if the external field ${\bf B}_{\rm ext}$ is applied perpendicular 
to an infinite plane, in practice to a platelike sample with extremely small 
thickness relative to radius, \eref{constitutive} yields (see \sref{field}) 
${\bf B}$ = ${\bf B}_{\rm ext}$, i.e. the induction should follow the 
external field.\par 

It is well known that the magnetic response of a non-magnetic metal can 
oscillate as a function of $B_{\rm ext}$. This is the de Haas-van Alphen 
(dHvA) effect which is used to study Fermi surfaces (Shoenberg 1984). Condon 
1966 noticed that if the oscillating amplitude is large within 
some part of each dHvA oscillation period, i.e. for a given $B_{\rm ext}$ 
range, the conduction electron states in this range are thermodynamically 
unstable and cannot follow the electrodynamics relation ${\bf B}$ = 
${\bf B}_{\rm ext}$. Therefore the electronic system should jump periodically 
over the forbidden intervals of $B$. A compromise could be the splitting of the
magnetic energy in alternating diamagnetic (with induction smaller than $B$) 
and paramagnetic (with induction larger than $B$) domains. With such a domain 
structure the relation ${\bf B}$ = ${\bf B}_{\rm ext}$ is fulfilled as an 
average over the sample. The so-called Condon domains are spectacular 
manisfestations 
of the collective behaviour of the electrons in quantized cyclotron orbits, 
i.e. Landau states.\par

The first direct observation of the Condon 
domain formation was made on silver by 
NMR measurements (Condon and Walstedt 1968). One had to wait 28 years for a
second report on the domain formation: using transverse field $\mu$SR
measurements, Solt \etal 1996a have reported the observation of domains in
beryllium. Interestingly, the prediction of Condon was made for beryllium.\par

\begin{figure}
\centerline{\epsfbox{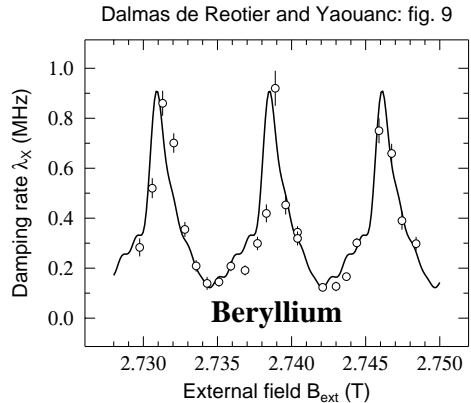}}
\caption{Exponential damping rate $\lambda_X$ of the $\mu$SR signal as a 
function of the intensity of the external field measured on beryllium. The
temperature is 0.8 K. The periodic sharp rises of 
$\lambda_X$ mark the onset of line splitting due to the domain formation. 
For $\lambda_X$ $\geq$ 0.4 MHz the broadened line turns out to be a well
resolved doublet. The solid line is a best fit to a truncated Fourier
series (adapted from Solt \etal 1996a).}
\label{solt_lambda}
\end{figure}

\begin{figure}
\centerline{\epsfbox{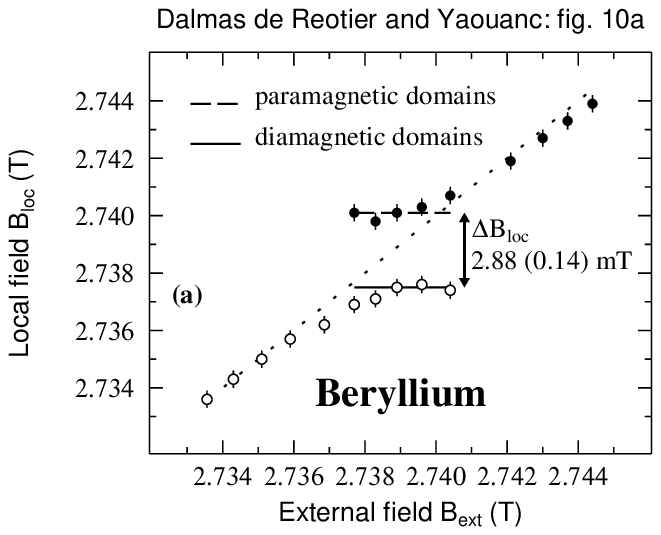}\hfill
\epsfbox{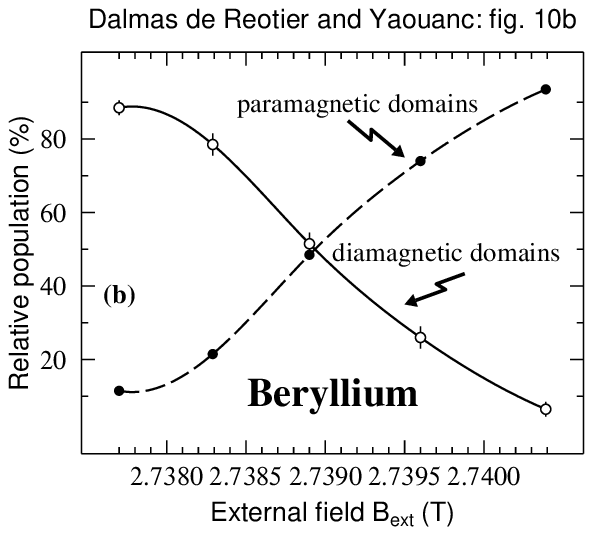}}
\caption{Analysis of the central peak of figure \protect\ref{solt_lambda}. 
The (a) panel displays the intensity of the local fields at the muon site
versus the intensity of the external field. The difference between the fields
in the paramagnetic and diamagnetic domains is clearly resolved. For reference
the line whose equation is $B_{\rm loc}$ = $B_{\rm ext}$ is plotted (dotted
line). The (b)
panel presents the relative populations of the paramagnetic and diamagnetic 
domains versus the intensity of the external field. The lines are guide to the
eyes. (adapted from Solt \etal 1996a).}
\label{frequence}
\end{figure}

As a first step, Solt \etal 1996a have analyzed their data using  
\eref{extra_trans}. The field dependence of $\lambda_X$ is presented in 
\fref{solt_lambda}. Any deviation from $\lambda_X (B_{\rm ext})$ $\approx$
constant must reflect the influence of the dHvA effect. The function
$\lambda_X (B_{\rm ext})$ rises periodically, reaching values about 10 times as
large as its minimum. The period is consistent with the expected dHvA value. The
broadening model is practical to visualize the dHvA
effect but not strictly justified. Solt \etal 1996a have then performed a 
Fourier
analysis. The results for the central peak region of \fref{solt_lambda}
are presented in \fref{frequence}. As expected, within a given field range, 
$B_{\rm loc}$, which is a measure of the induction, does not follow 
$B_{\rm ext}$. The diamagnetic and paramagnetic signals are clearly seen, their
respective populations changing smoothly as $B_{\rm ext}$ increases.
Interestingly, the domains could not be visualized by NMR on beryllium because
of the excessive linewidth produced by the electric field gradients. Since the 
muon does not have a quadrupolar moment, it does not couple to these gradients.
The Condon domains have been detected by a point-like probe such as the muon 
because the volume occupied by the domains is much larger that the
volume of the domain walls: at $B_{\rm ext}$ = 1 T, the domain and wall
thicknesses are $\sim$ 30 $\mu$m and $\sim$ 1 $\mu$m, respectively 
(Solt \etal 1996b). Note that the observation of the domains means that they
are pinned (static) for the time scale of the muon experiment 
($\sim$ 1 $\mu$s).\par

The Condon domains discovered in silver and beryllium correspond to the simplest
possibility. A variety of domain-like periodic structures have been predicted 
(for a review, see Solt \etal 1996b). We expect that with the improvement of
the high field transverse field $\mu$SR spectrometers, the investigation of
magnetic domains in the bulk of non-magnetic metals will attract much interest 
in the future. \par   

\subsection{Localized and itinerant $f$ electrons in heavy fermion materials}  
\label{components}

In recent years the analysis of $\mu$SR data recorded in some intermetallic 
compounds containing cerium or uranium atoms has suggested that two different 
substates of electrons of $f$ character are present. In this section the 
available experimental evidence found recently is presented. We discuss in 
some detail results obtained for the heavy fermion superconductor 
UPd$_2$Al$_3$ and the heavy fermion metal CeRu$_2$Si$_2$. We then mention the 
superconductor CeRu$_2$ and give a short discussion of the magnetic behaviour 
of these $f$ electron magnets, pointing out the complementary nature of the 
information obtained by $\mu$SR and inelastic neutron scattering.\par

UPd$_2$Al$_3$ belongs to the small family of heavy fermion 
superconductors; see for example Heffner and Norman 1996. It has a hexagonal 
crystal structure (space group $P6/mmm$) and exhibits a coexistence of  
superconductivity ($T_c$ $\simeq 1.5$ K) and antiferromagnetism 
($T_N$ $\simeq 14$ K). The uranium atom carries a relatively large magnetic 
moment of 0.85 $\mu_B$ as determined by neutron diffraction. The uranium 
moments are aligned ferromagnetically in the basal plane along the $a$ axis and 
stacked antiferromagnetically along the $c$ axis. The magnetic structure is not 
affected by the superconducting transition, which is intriguing in view of the 
unquestionable $5f$ character of both the heavy quasiparticles forming the 
Cooper pairs and the electrons responsible for the antiferromagnetic structure.
The transverse field $\mu$SR investigation of the superconducting phase by  
Feyerherm \etal 1994 provides information on the 5$f$ electrons involved in the
Cooper pair formation.\par

\begin{figure}
\centerline{\epsfbox{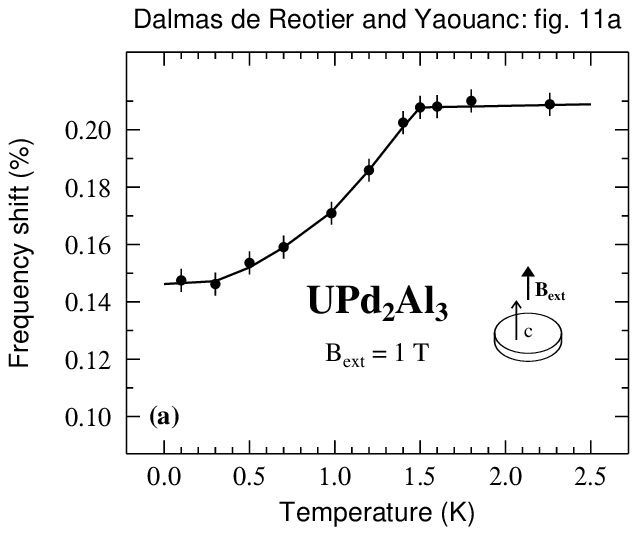}\hfill
\epsfbox{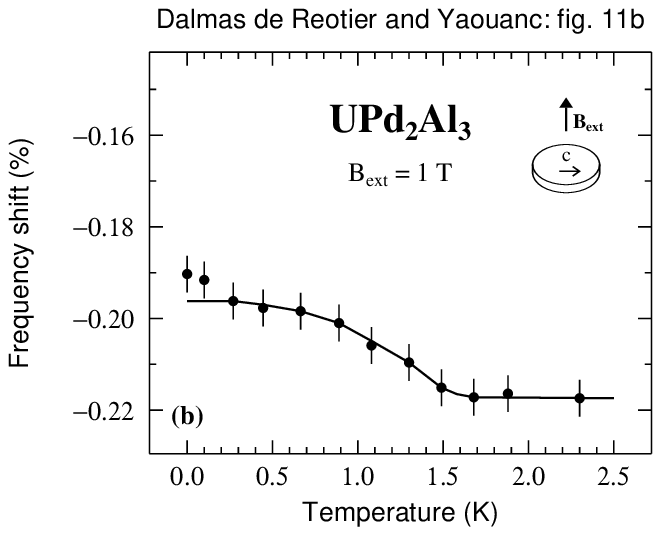}}
\caption{Temperature dependence of the muon frequency shift in UPd$_2$Al$_3$
measured in an external field of 1 T applied either along or perpendicular
to the $c$ axis. Note the strong reduction of the absolute value of the shifts 
below the superconducting transition ($T_c \simeq 1.5$ K). The solid lines are 
guides to the eyes. The data are not corrected for demagnetization and Lorentz 
fields. (adapted from Feyerherm \etal 1994).}
\label{upd2al3}
\end{figure}

Since the local field produced by the magnetic sublattices in the 
antiferromagnetic phase cancels at the muon localization site, a study of the 
muon frequency shift could be undertaken below $T_N$ and in particular in the 
superconducting phase. The key point is the observation that the positive
frequency shift along the $c$ axis as well as the negative frequency shift in 
the basal plane are decreasing in absolute values when the temperature decreases
below $T_c$ as seen in \fref{upd2al3}. In discussing the possible origins
for this decrease, Feyerherm \etal 1994 noticed that the diamagnetic shift due 
to the flux expulsion or a possible change of the hyperfine coupling cannot 
explain the data. Therefore, the observed partial reduction of the frequency 
shifts can only reflect a decrease of the susceptibility of the 5$f$ 
electrons. Taking into account the strong anisotropy of the bulk 
susceptibility, these authors infer that the $f$ electron 
susceptibility reduction due to superconductivity is isotropic. This, apparently
surprising, result can be undertood if the 5$f$ electrons are viewed as two 
essentially independent electron subsets. The decrease of the susceptibility 
is associated with the electron system formed by the heavy quasiparticles
condensing into Cooper pairs below $T_c$. Interestingly, this reduction
suggests singlet pairing of the Cooper pairs. The residual 
susceptibility is ascribed to the electron subsystem associated with the local 
antiferromagnetism which is not affected by superconductivity. We mention 
that the analysis of NMR Knight shift data leads to the same physical picture 
if one assumes that the hyperfine coupling constant is temperature independent,
in contrast to the $\mu$SR analysis which does not need this assumption
(Kyogaku \etal 1993). \par

The picture of an isotropic itinerant 5$f$ electron subset is supported by the
$\mu$SR observation of an almost isotropic London penetration depth 
(Feyerherm \etal 1994). It is noteworthy that specific heat measurements under 
pressure indicate that the itinerant subset accounts for 80 \% of the linear 
coefficient of the specific heat (Caspary \etal 1993).

\begin{figure}
\centerline{\epsfbox{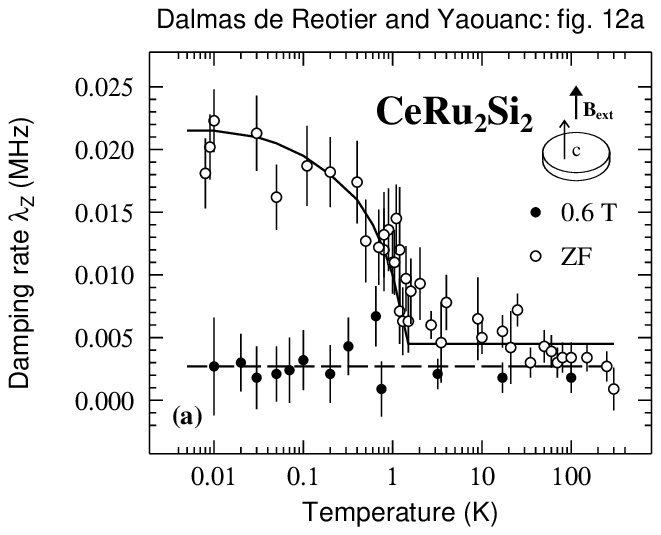}\hfill
\epsfbox{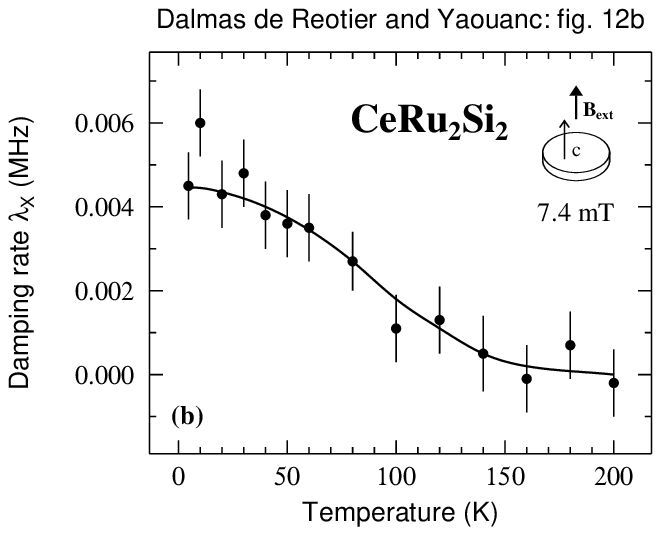}}
\caption{ Temperature dependence of the damping rates measured with a single 
crystal of CeRu$_2$Si$_2$. On the (a) and (b) panels are presented the rates 
recorded with the longitudinal and transverse geometry, respectively. These
measurements show that CeRu$_2$Si$_2$ exhibits, in addition to a static
magnetic field distribution below $\sim$ 2 K, a dynamical magnetic component
persisting up to $\sim$ 150 K. The solid lines are guide to the eyes
(adapted from Amato \etal 1993)}
\label{ceru2si2}
\end{figure}

CeRu$_2$Si$_2$ has attracted considerable interest because it lies at the 
borderline of
a magnetic instability, between long-range magnetic order and paramagnetic
ground state (for a review, see Kambe \etal 1996). \par

In \fref{ceru2si2} is presented the temperature dependence of the 
zero-field exponential damping rate measured on a CeRu$_2$Si$_2$ crystal 
(Amato \etal 1993). The rate increases significantly below $\sim$ 2 K, 
corresponding to an enhancement of the field spread at the muon site of the 
order of 20 $\mu$T. The static nature of the local field is proved by the 
strong reduction of the damping rate measured in a longitudinal field as shown 
in \fref{ceru2si2}. The exponential shape of $P_Z(t)$ in zero-field is 
surprising as one would expect to observe a parabolic function for such small 
damping rates. Amato \etal 1994 argue that the enhancement of the field 
distribution originates from a static electronic magnetic moment carried 
by the Ce atoms. The Ce magnetic moment is estimated to be 
$\simeq$ 10$^{-3} \mu_B$. Because of this extremely small value, this 
magnetism is probably itinerant in nature. Since the value is found using a 
localized magnet model, there is quite a large uncertainty on the value of the 
magnetic moment deduced from the data. The small longitudinal 
damping rate which seems to exist at high magnetic field suggests the presence 
of fast electronic spin fluctuations. The existence of such fluctuations is 
confirmed by the transverse field measurements shown in \fref{ceru2si2}. 
These fluctuations, which are observed up to $\sim$ 150 K, have been 
related to the short-range dynamical magnetic correlations detected below 
$\sim$ 60 K by inelastic neutron scattering (Regnault \etal 1988). They 
involve a relatively large Ce magnetic moment 
(0.6 $\mu_B$, see Amato \etal 1994) and an extremely short correlation time of 
$\sim 10^{-13}$ s (Amato \etal 1993).\par 

Recently, zero-field and longitudinal field measurements on a polycrystalline
sample of the superconductor CeRu$_2$ ($T_c = 6.1$ K) have shown that this 
compound presents a magnetic phase transition at $\simeq 40$ K characterized by 
an extremely small magnetic moment of $\sim$ 10$^{-4}$ $\mu_B$ 
(Huxley \etal 1996). Because of this extremely small value, the 
magnetism is itinerant in nature. We note that a high energy neutron scattering 
study of CeNi$_2$, a compound  which should have the same magnetic 
characteristics as CeRu$_2$, has revealed a strong inelastic paramagnetic 
response of the Ce ion with a large Kondo temperature 
(Murani and Eccleston 1996). The combination of the $\mu$SR results on CeRu$_2$
and the neutron data on CeNi$_2$ suggests again the existence of a two 
component 4$f$ electronic system: in this case, a 4$f$ weakly polarized 
itinerant electron system and a paramagnetic 4$f$ localized electron system.\par

The picture of two different substates of electrons of $f$ character 
which emerges from the analysis of the UPd$_2$Al$_3$, CeRu$_2$Si$_2$ and 
CeRu$_2$ data was already suggested for UCu$_5$ by Schenck \etal 1990. Some
years ago inelastic neutron scattering measurements on CeCu$_6$ and 
CeRu$_2$Si$_2$ were analyzed in terms of two magnetic contributions 
(Aeppli \etal 1986 and Regnault \etal 1988). Amato \etal 1993 have shown that
the two contributions observed by neutron and $\mu$SR can be nicely related in
the case of CeRu$_2$Si$_2$. Clearly, it would be of interest to carry on this 
type of comparison keeping in mind that, contrary to the inelastic neutron
scattering technique, the $\mu$SR method probes the spectral weight of the
modes at extremely small energy transfer, i.e., quasi-static modes.\par

The two component picture which is proposed from the analysis of the data is 
in qualitative agreement with the duality model for heavy fermion 
(Kuramoto and Miyake 1990) which introduces two coupled electron components: 
itinerant and localized. The physical basis of such a picture is attributed to
the fact that the one-particle density of states has a triple peak structure in
strongly correlated fermion systems. This structure consists of two
broad peaks corresponding to the upper and lower Hubbard bands and a narrow
quasiparticle peak at the Fermi level. Interestingly, this type of structure is 
found by recent numerical works using the 
$d = \infty$ technique, $d$ being the space dimension 
(see for example, Georges and Kotliar 1992). A recent theoretical discussion is 
given by P\'epin and Lavagna 1997.\par

A high resolution photoemission study of CeRu$_2$ has found a substantial 4$f$ 
electron density at the Fermi level, in agreement with the interpretation of 
the $\mu$SR result on this compound and the duality picture 
(Yang \etal 1996). \par

\section{Spin dynamics in magnets}\label{dynamique} 

The study of the spin dynamics in magnets has been a traditional subject of 
$\mu$SR. But it is only recently that quantitative information has been 
extracted, thanks to greatly improved experimental conditions and better data 
analysis. In \sref{dipolar} and \sref{glass}, spectra recorded respectively in 
the critical regime of a ferromagnet and for a spin-glass are analyzed in 
terms of spin-spin correlation-functions. The first study has taken
advantage of the possibility to carry out the measurements in truely zero field 
and the second one, of the large spectrum of fluctuations which can be probed.
The discussion is partly based on the material presented in \ref{appendix2bis}
and \ref{appendix2}.\par

\subsection{Critical and low temperature spin dynamics in ferromagnets}
\label{dipolar}

The first detailed analysis of data recorded in the critical regime has been 
given for Ni and Fe by Yaouanc \etal 1993a and 1993b and for Gd by Dalmas de 
R\'eotier and Yaouanc 1994. These works were restricted to the critical 
paramagnetic regime. In this section we present an analysis of the spin 
dynamics of the Gd$^{3+}$ ion spins in the dipolar axial ferromagnet GdNi$_5$ 
covering the whole temperature range (Yaouanc \etal 1996a). This zero-field 
study reveals the effect of the dipolar interaction on the physical properties 
of the spin-spin correlation-tensor $\tilde\Lambda^{ \alpha \beta}({\bf q})$, 
both in the paramagnetic and ferromagnetic state. This effect is expected to be 
strong since the measurements probe the fluctuation modes in the zone center 
where the long range nature of the dipolar interaction dominates the dynamics. 
This interaction introduces non-conserving terms in the Hamiltonian which 
prevents the slowing down of the longitudinal (to the wave vector) modes near 
the Curie temperature, $T_C$ (K\"otzler 1986). Therefore, if the measured 
relaxation rate reflects only these modes, it should saturate as the 
temperature approaches $T_C$.\par

GdNi$_5$ crystallizes in the hexagonal CaCu$_5$ crystal structure and exhibits 
a ferromagnetic phase transition at $T_C$ $\simeq$ 32 K characterized by a 
small magnetic dipolar anisotropy field as determined by magnetization 
measurements: $B_a$($T$ = 5 K) $\simeq $ 0.21~T. The muon localization site(s) 
is unknown.\par

\begin{figure}
\centerline{\epsfbox{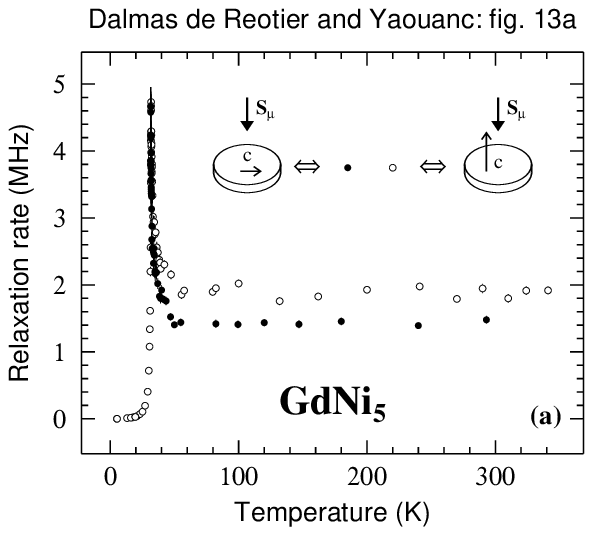}\hfill
\epsfbox{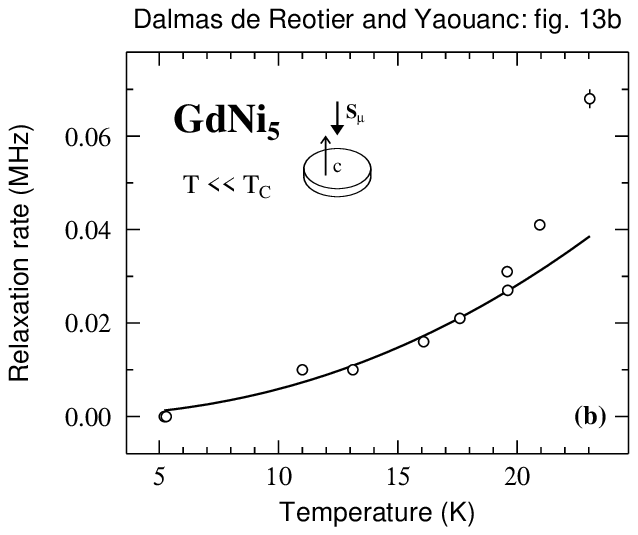}}
\caption{(a): an overview of the zero-field relaxation rate of 
GdNi$_5$ (Yaouanc \etal 1996a).
The two sets of data refer to measurements performed on two crystals
which differ by the orientation of ${\bf S}_\mu$ relative to the $c$ axis.
(b): zero-field relaxation rate of GdNi$_5$ at low 
temperature (Yaouanc \etal 1996b). The full line is the 
prediction for 
the relaxation induced by the Raman spin wave process 
(equation (\protect\ref{raman})).}
\label{gdni5_general_raman}
\end{figure}

An overview of the zero-field relaxation rate $\lambda_Z$ is presented in
\fref{gdni5_general_raman}a. This rate reflects the fluctuations of the 
Gd$^{3+}$ ion spins. No signal is found in the ferromagnetic state when 
${\bf S}_\mu$ is perpendicular to the crystal $c$ axis 
since the spontaneous muon spin rotation is too fast to be resolved at the 
pulsed source (ISIS) where the data were recorded. Four temperature regions 
can be distinguished: far above or below the Curie temperature $T_C$ and the 
critical paramagnetic and ferromagnetic regions. \par

Far above $T_C$, the wave vector dependent spin-spin correlation-tensor, 
$\tilde\Lambda^{ \alpha \beta}({\bf q})$, is expected to be isotropic and 
independent of the wave vector: $\tilde\Lambda^{ \alpha \beta}({\bf q})$ = 
$\tilde \Lambda \delta^{ \alpha \beta}$. The observed anisotropy of 
$\lambda_Z$, which reflects the anisotropy of the coupling tensor between the 
muon spin and the Gd$^{3+}$ spins, can be explained if the muon occupies the 
interstitial site of coordinates $(1/2, 0, 0)$.\par

Far below $T_C$, the spin-lattice relaxation rate increases with temperature 
as shown in \fref{gdni5_general_raman}b. Its thermal variation is interpreted 
as due to the relaxation of the muon spins by the magnons. 
Since the minimum magnon energy is much larger than the Zeeman muon energy 
$\hbar \omega_\mu$, a one magnon process cannot flip the muon spin because 
energy-conservation requirement would not be ensured. Therefore, the 
perturbation 
operator which induces the muon spin flip cannot contain the $J^{x}$ and 
$J^{y}$ operators ($J$ denotes the Gd$^{3+}$ spin). In terms of the 
correlation-tensor 
$\tilde\Lambda^{ \alpha \beta}({\bf q})$, this means that one does not 
have to consider the correlations with $\{ \alpha \beta \}$ = $\{x,y \}$. On 
the other hand, a two-magnon, or Raman, process does not present problems as 
regards energy conservation because the only requirement is that the energies 
of the annihilated and created magnons must be equal (we neglect 
$\hbar \omega_\mu$). Therefore the energy principle tells us that only the 
parallel (to the easy axis $z$ ; the $Z$ and $z$ axes are parallel) 
fluctuations contributes to the relaxation, i.e. the measurements only probe 
the correlation-function $\tilde  \Lambda^{ zz}({\bf q})$ 
(Yaouanc and Dalmas de R\'eotier 1991).\par 

While in NMR the Raman process is almost never observed because the
hyperfine interaction between the spin probe and the lattice spins is 
isotropic (therefore a perturbation operator such as $\sigma_\pm J^z$ does not
exist; $\sigma$ is the Pauli operator of the spin probe),
this process is expected to be active in $\mu$SR since the interaction between
the spins is mainly due to the dipolar interaction which is spatially 
anisotropic.\par

The key parameter which determines if the muon spin is relaxed by the Raman
process is the ratio of the minimum magnon energy over the thermal 
energy. The relaxation is effective only if this ratio is sufficiently small. 
This means that $\lambda_Z$ = 0 at very low temperature as observed
experimentally in \fref{gdni5_general_raman}b. For a small ratio, i.e. at high
temperature, we expect $\lambda_Z$ $\propto$ $T^2$ since two magnons are
involved. This quadratic law is a robust result in the sense that it does not
depend on the details of the model.\par  

Using the simple magnon dispersion relation $\hbar\omega_{\bf q}$ = 
$ D_m q^2+ \Delta$, where $D_m$ is the magnon stiffness constant and $\Delta$
the anisotropy energy which is of dipolar origin, and the fact that 
$\Delta$ = $g_L \mu_B B_a$ 
$\ll k_B T$ for GdNi$_5$, the following approximate result holds 
(Dalmas de R\'eotier and Yaouanc 1995):
\begin{eqnarray}
\fl
\lambda_ Z & = &{{\cal C}g^2_L T^{2}\over D_m^3}
\left [1 +{15 \over 2} \left[ {( C^{xz}({\bf q =0}))^2
+ ( C^{yz}({\bf q =0}))^2 }  \right] \right] 
\ln \left({k_BT/\Delta} \right). 
\label{raman} 
\end{eqnarray}
${\rm C}$ is the analytical part of the tensor describing the dipolar 
interaction 
between the muon spin and the Gd$^{3+}$ spins; see \eref{appen113bis}. 
\Eref{raman} accounts for the effect of the spin waves near the zone center. 
Note that $\lambda_Z$ is independent of the characteristics of the muon 
localization site(s) if $ C^{zx} ({\bf q}=0) $ = $ C^{zy} ({\bf q}=0) $ = 0 as 
it is the case for the possible muon sites in GdNi$_5$. $g_L$ is the Land\'e 
factor and $ k_B$ the Boltzmann constant. $ {\cal C}$ is a universal constant 
($ {\cal C}$ = 129.39 (meV)$^3$.\AA$^6$.s$^{-1}$.K$^{-2}$). The fit to 
\eref{raman} presented in \fref{gdni5_general_raman}b yields 
$D_m$ = 3.2 (1) meV.\AA$^2$. This estimate allows one to extract the value of 
an important parameter: namely
the dipolar wave vector, $ q_D $, which determines the 
relative strength of the dipolar and exchange interactions. The analysis gives 
$ q_D$ = 0.19 \AA$^{-1} $. The simple linear magnon theory cannot describe the 
data above $\sim$ 20 K because the magnon-magnon interactions are neglected.\par

\begin{figure}
\centerline{\epsfbox{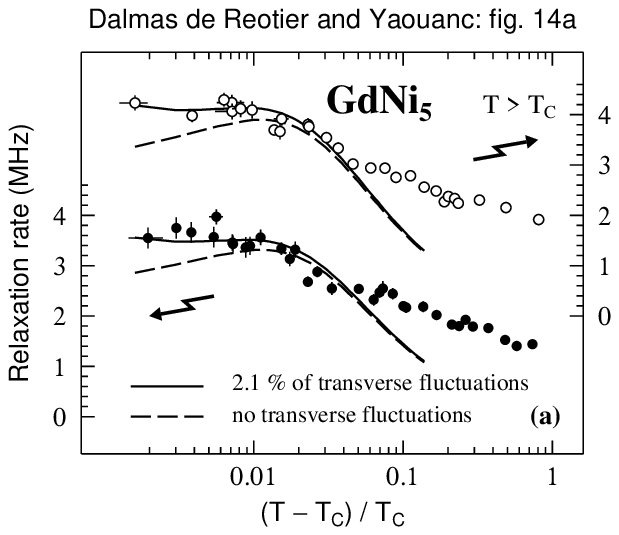}\hfill
\epsfbox{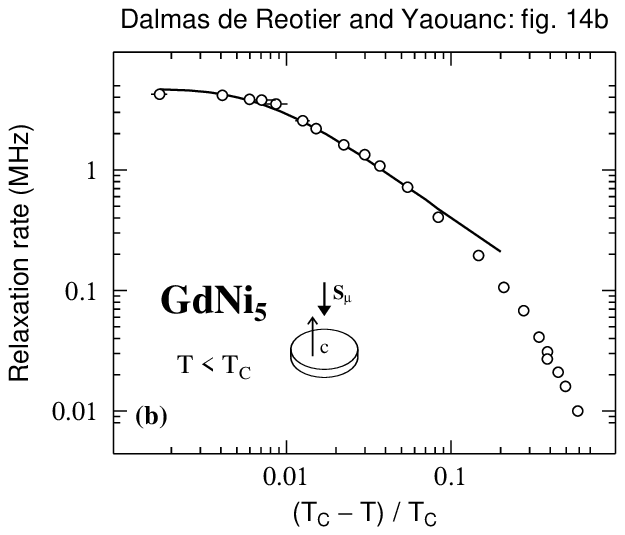}}
\caption{(a): zero-field relaxation rate measured in the critical 
paramagnetic state given as a function of the temperature relative to the Curie
temperature and the orientation of ${\bf S}_\mu$ relative to the $c$ axis
(same symbol convention as in figure \protect\ref{gdni5_general_raman}a). The 
full
and dashed lines are predictions of the mode coupling theory for the critical 
behaviour of $\lambda_Z$(T) in a dipolar Heisenberg ferromagnet 
(Frey and Schwabl 1988, 1989 and 1994). (b): zero-field relaxation rate 
measured in the ferromagnetic state near the Curie temperature. The full line 
is the prediction for the critical paramagnetic fluctuations. The relative 
weight of the longitudinal and transverse fluctuations is taken as given by 
the analysis of the paramagnetic fluctuations. Both figures are from 
Yaouanc \etal 1996a.}
\label{gdni5_para_ferro}
\end{figure}

We now analyze the paramagnetic critical behaviour of $\lambda_Z$. As shown in 
\fref{gdni5_para_ferro}a, $\lambda_Z$ is almost isotropic. This is expected for 
a dipolar Heisenberg paramagnet for which the correlation-tensor is
\begin{eqnarray} 
\tilde  \Lambda^{ \beta \gamma} ( {\bf q} )  =
\tilde  \Lambda^T(q)P^{\beta \gamma}_ T({\bf q} ) +
\tilde  \Lambda^L(q)P^{\beta \gamma}_ L({\bf q} ) 
\label{decomposition_heisenberg}
\end{eqnarray}
where $P^{\beta \gamma}_ T( {\bf q} )$ and $P^{\beta \gamma}_ T( {\bf q} )$ are 
the transverse and longitudinal (relative to $\bf q$) projector operators, 
respectively. This decomposition reflects the symmetry of the Hamiltonian at 
small ${\bf q}$ value, i.e. of the long range dipolar interaction between the 
Gd$^{3+}$ ions (Frey and Schwabl 1988, 1989 and 1994). Since the dynamics near 
$T_C$ is driven by the modes at small ${\bf q}$, we need to consider the 
expansion of the muon-lattice coupling tensor $G^{\alpha \beta}( {\bf q})$ 
only near ${\bf q} = 0$: $G^{\alpha \beta}( {\bf q})  =  -4\pi \left[
P^{\alpha \beta}_ L( {\bf q} )-C^{\alpha \beta} (0)
- {H^{\alpha \beta} (0)\over 4\pi}\right]$. ${\rm H}$ is the hyperfine tensor. 
These functional forms of $\tilde  \Lambda$ and ${\rm G}$, together 
with the expression for 
$\lambda_Z$ given in \ref{appendix2}, leads to the simple result
\begin{eqnarray} 
\lambda_ Z & = &{\cal W} [a_LI^L(\varphi) + a_TI^T(\varphi)],
\label{lambda_heisenberg}
\end{eqnarray}
where ${\cal W}$ is a nonuniversal constant, $a_{L,T}$ depends only on the 
muon localization site(s) and $ I^{L,T}(\varphi ) $ are universal fluctuation 
functions of the temperature through the angle $\varphi$. $ I^L(\varphi ) $ and 
$ I^T(\varphi ) $ account for the longitudinal and transverse fluctuations, 
respectively. We have $ \varphi$ = $\arctan (q_D\xi)$ with $\xi$ = 
$\xi_0 t^{-\nu}$. $\xi$ is the correlation length, $\xi_0$ the correlation 
length extrapollated to $ T = 2T_C$, $t$ $\equiv$ $|T-T_C|/T_C$ and $\nu$ the 
correlation length critical exponent ($\nu$ $\simeq $ 0.69). Therefore, in
general, $\lambda_Z$ is a weighted sum of $I^L(\varphi )$ and $I^T(\varphi)$.
Experimentally, a saturation is observed when approaching $T_C$ (see
\fref{gdni5_para_ferro}a). This is understood if $a_LI^L(\varphi)$ $\gg$ 
$a_TI^T(\varphi)$ since $I^L(\varphi)$ saturates as $T_C$ is approached.\par

\begin{figure}
\centerline{\epsfbox{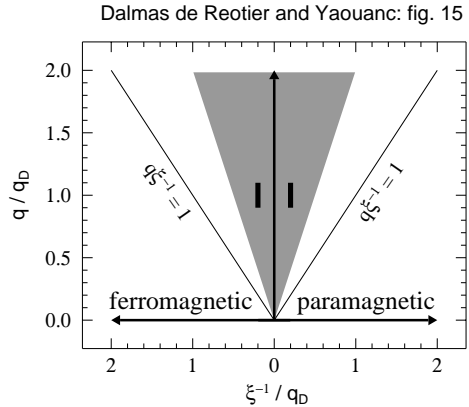}}
\caption{Halperin-Hohenberg diagram for the $\mu$SR measurements of GdNi$_5$ at
$|T_{\rm C} - T|/T_{\rm C}$ = $10^{-2}$ (from Yaouanc \etal 1996b). 
Their locations are indicated by the two bars. They have been clearly performed 
in the critical regime delimited by the shaded region.}
\label{gdni5_halperin}
\end{figure}

An interesting result of this study is the observed similarity between the 
paramagnetic and ferromagnetic longitudinal critical fluctuations as 
discovered when comparing \fref{gdni5_para_ferro}a and 
\fref{gdni5_para_ferro}b. 
Using the dynamical scaling theory of Halperin and Hohenberg (Halperin and 
Hohenberg 1967), it can be understood as follows. The basic quantity which 
distinguishes the different regions in the ($q, \xi^{-1}$) plot is the product 
$q \xi$. Since the measurements are mostly sensitive to longitudinal
fluctuations, they probe the modes with $q$ $\sim$ $q_D$ 
(Dalmas de R\'eotier \etal 1994). 
Therefore the relevant quantity is $q_D \xi$. Taking $\xi_0$ = 1 \AA\ one finds 
$q_D \xi$ $\simeq$ 5 at $t$ = $10^{-2}$. Despite this rough estimate for the 
correlation length, the measurements are still clearly in the critical regime 
of the paramagnetic and ferromagnetic dynamics as shown in 
\fref{gdni5_halperin}. Thus the continuity of the dynamical behavior when 
crossing $T_C$ and therefore the observed similarity is understood. 
Nevertheless this argument calls for a detailed theoretical justification : in 
a dipolar magnet two scaling variables are needed instead of one for the
isotropic model of Halperin and Hohenberg.\par

We note that the relaxation rate of the points very close to $T_C$ (namely the 
points which have been used for the determination of $T_C$ and 
correspond to $t \leq 0.0015$) is significatively larger than the saturation
value obtained for 0.002 $\leq t \leq$ 0.02. This can be seen by comparing the 
values of the damping rate in \fref{gdni5_para_ferro}a and 
\fref{gdni5_para_ferro}b 
with \fref{gdni5_general_raman}a. This increase of the damping rate near $T_C$
could be due to the Ising crossover that has been observed in metallic Gd 
(Dalmas de R\'eotier and Yaouanc 1994).\par

We point out that systematic zero-field investigations of the dynamics in 
rare-earth metals would be of great interest: it could provide essential 
information for the determination of their universality class through the
measurement of the dynamical critical exponent.\par

Recently Heffner \etal 1996 have measured $\lambda_Z$ in the ferromagnet
La$_{0.67}$Ca$_{0.33}$MnO$_3$ which has attracted much interest because of its 
colossal magnetoresistance. \Eref{raman} fails to predict the order of 
magnitude of $\lambda_Z$. This may not be surprising since this oxide is an 
inhomogeneous mixed-valence compound. The results of Heffner \etal 1996 reveal 
the 
presence of a density of states for magnetic excitations larger than expected,
which has a strong influence on the correlations along the easy axis. Its 
origin could be related to the existence of a second $d$ electron component in 
addition to the localized $d$ component responsible for the conventional spin 
waves detected by inelastic neutron scattering 
(Perring \etal 1996, Moussa \etal 1996). This picture is identical to the one
used to describe results obtained on some metallic compounds containing Ce or 
U atoms (see \sref{components}). In order to characterize these
unexpected excitations further, 
it would be of interest to extend the measurements to
crystals and investigate the possible field dependence of $\lambda_Z$ over the
whole temperature range.\par 

\subsection{The correlation-function in spin-glasses}\label{glass}

The intensive investigation of glass and spin-glass forming systems in the 
eighties has shown that the key to a comprehensive understanding of the 
transition lies in the dynamics (Fisher and Hertz 1991). The spin-spin self 
correlation-function is the most important 
quantity in the spin glass systems since cross correlations are 
zero, i.e. one can neglect the wave vector dependence of the 
correlation-function. 
Therefore for an isotropic spin glass system one should only consider 
the correlation-function $\Lambda (t)$ $\equiv$ $\Lambda (t, {\bf q = 0})$
(see \ref{appendix2}).\par

Below the spin glass transition temperature $T_g$, a $\mu$SR study has shown 
that $\Lambda (t)$ decays as a power law after some microscopic time, of the 
order of 10$^{-14}$ s (MacLaughlin \etal 1983). However it is only recently
that the form of $\Lambda (t)$ above $T_g$ has been established 
(Keren \etal 1996b). In this section we highlight this work.\par

The goal of the work by 
Keren \etal 1996b was to distinguish, using the longitudinal
field method, between three possible forms of $\Lambda (t)$ above $T_g$: the 
power law ($d \cdot |t|^{- \alpha}$), the stretched exponential 
($d \exp[-(\zeta |t|)^{\beta}]$) and the cutoff power law 
($d \cdot |t|^{- \alpha} f(\zeta |t|)$) which is often approximated by the 
Ogielski form ($d \cdot |t|^{- \alpha} \exp[-(\zeta |t|)^{\beta}]$). 
The exponents $\alpha$ and $\beta$ are positive by definition. 
The difference between these forms is fundamental: the power law is 
time-invariant, the stretched exponential has a well-defined time scale given 
by $1/ \zeta$ and the cutoff power law is time-invariant only at times shorter 
than $1/ \zeta$. \par

Since the longitudinal field method does not probe directly $\Lambda (t)$ but
only $P_Z(t)$, one has to determine the relation between these two quantities.  
Because the magnetic impurities are randomly distributed, the spin environment
of each muon is different. We first consider an expression of $P_Z(t)$ for a 
given environment. As written in \eref{explong}, for a crystal in the 
fast-fluctuation 
limit, $P_Z(t)$ is an exponential function with a relaxation rate 
$\lambda_Z$ = $2 \Delta^2/\nu$. The field dependence of $1/\nu$ is given in 
terms of the correlation-function:
\begin{eqnarray}
{1 \over \nu(B_{\rm ext})} = {1 \over 2 \Lambda(0)} \int^{\infty}_{-\infty} 
\Lambda(t) \cos \left (\gamma_\mu B_{\rm ext} t \right) dt.
\label{campbell2}
\end{eqnarray}
This formula, valid for an isotropic system, can be deduced from 
\ref{appendix2bis}.
$\Lambda(t)$ does not depend on $B_{\rm ext}$ if the electronic Zeeman energy 
is much smaller than the spin-spin coupling energy. The parameters $\Delta$ and 
$\nu$ can vary from one muon site to another, hence their average values 
should be considered. Most treatments have assumed that  $\nu$ is site 
independent. In order to understand their data, Keren \etal 1996b are forced to 
use a weaker assumption. They write $\nu(B_{\rm ext})$  = $c \ l(B_{\rm ext})$ 
where the site dependence enters through the prefactor $c$. Thus the measured 
depolarization, which is an average, is given by 
\begin{eqnarray}
P_Z(t) = \int \int \rho (\Delta, c) 
\exp \left[{- 2 \Delta^2 \over c \ l(B_{\rm ext})} t \right] dc\ d\Delta,
\label{campbell3}
\end{eqnarray}
where $\rho (\Delta, c)$ is the probability of occurence of the values
$\Delta$ and $c$. For the three possible forms of $\Lambda(t)$, $P_Z(t)$ obeys 
asymptotically the scaling relation
\begin{eqnarray}
P_Z(t, B_{\rm ext}) = P_Z(t/B_{\rm ext}^{\gamma}),
\label{campbell4}
\end{eqnarray}
where $\gamma = 1 - \alpha$ for the power law and the cutoff power law, and 
$\gamma = 1 + \beta$ for the stretched exponential form. Asymptotically means
for the stretched exponential, $\gamma_\mu B_{\rm ext} \gg \zeta$, and for the
cutoff form, $\zeta |t| \ll 1$. \par

It is interesting to notice that \eref{campbell4} should be valid with 
$\gamma$ =1 for a conventional magnet if the spin dynamics is slow because  
$\Lambda(t)$ is an exponential function for such a magnet.\par 

\begin{figure}
\centerline{\epsfbox{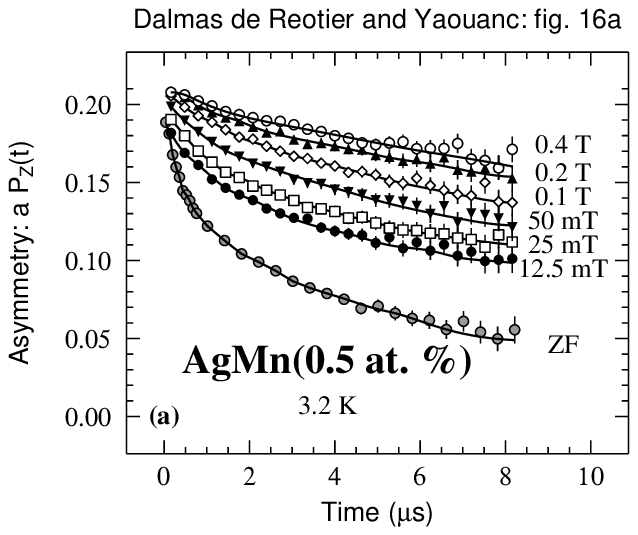}\hfill
\epsfbox{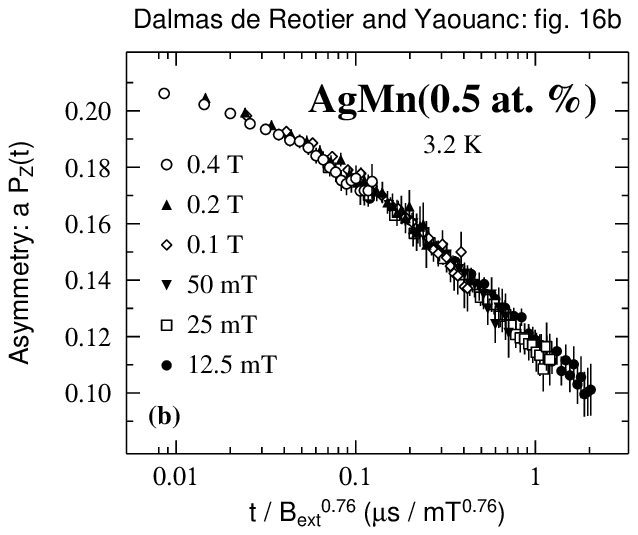}}
\caption{(a) Dependence of the $\mu$SR spectra on the intensity of 
the longitudinal magnetic field recorded on the spin glass AgMn(0.5 at. $\%$). 
The measurements have been performed at T = 
3.2 K, i.e. above the spin glass transition temperature ($T_g$ = 2.95 K). 
The solid lines are guides for the eyes.
(b) The same spectra as presented on figure 
\protect\ref{campbell_spectre_champ_universel}a but plotted
as a function of the scaling variable $t/B_{\rm ext}^{0.76}$ and for various
values of $B_{\rm ext}$. $t$ is the time and $B_{\rm ext}$ the external 
magnetic field. Both figures are adapted from Keren \etal 1996b. 
}
\label{campbell_spectre_champ_universel}
\end{figure}

The measurements have been done on the canonical Heisenberg spin glass system
AgMn(0.5 at. $\%$). In \fref{campbell_spectre_champ_universel}a 
we present the field dependence of the spectra recorded at 3.2 K, i.e. just 
above $T_g$. In \fref{campbell_spectre_champ_universel}b the validity of the 
scaling law (\ref{campbell4}) is demonstrated for $\gamma$ = 0.76 (5) over 
three orders of magnitude in $t/B_{\rm ext}^{\gamma}$. The  stretched
exponential form is inconsistent with the measured $\gamma$ value. According to 
\eref{campbell4}, an instantaneous relaxation should occur as 
$B_{\rm ext}$ $\rightarrow 0$ since then $P_Z(t)$ $\rightarrow 0$ at any $t$. 
This is not observed as confirmed 
in \fref{campbell_spectre_champ_universel}a. But 
as $B_{\rm ext}$ $\rightarrow 0$ the exponential term of the cutoff power law 
provides a cutoff. Therefore, only the cutoff power law describes the field 
dependence of $P_Z(t)$. This conclusion is achieved without assuming a specific 
functional form of $P_Z(t)$. \par

\begin{figure}
\centerline{\epsfbox{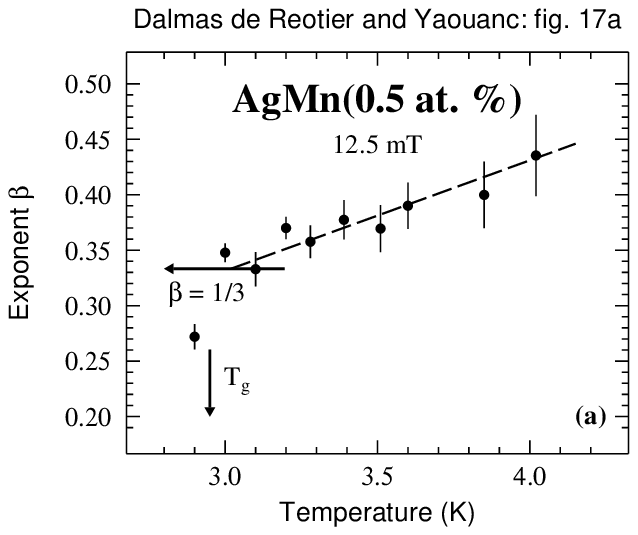}\hfill
\epsfbox{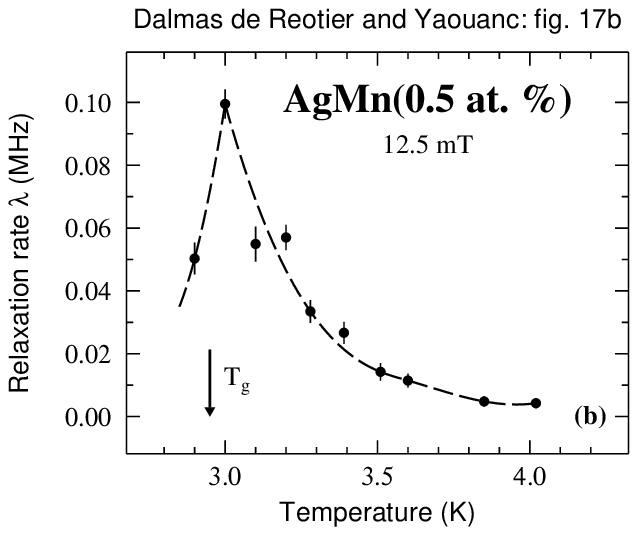}}
\caption{ Temperature dependence of the two physical parameters characterizing 
the stretched exponential used to fit spectra recorded on AgMn(0.5 a. $\%$) 
at various temperatures around the spin glass transition temperature ($T_g$ =
2.95 K) by cooling the sample in a longitudinal field of 12.5 mT 
(adapted from Keren \etal 1996b). The dashed lines are guides for the eyes.}
\label{campbell_parametre}
\end{figure}

A complementary approach is to test a functional form for $P_Z(t)$. One
possibility, which has been tested with success at a 
high impurity concentration 
(Campbell \etal 1994), is the stretched exponential function: 
\begin{eqnarray}
P_Z(t) = \exp \left[- (\lambda t)^{\beta} \right].
\label{campbell5}
\end{eqnarray}
Keren \etal 1996b show that this function provides a very good description of 
the AgMn(0.5 at. $\%$) spectra recorded 
at various temperatures around the spin glass transition temperature by cooling 
the sample in a longitudinal field of 12.5 mT. In \fref{campbell_parametre}
are plotted the two parameters characterizing the stretched exponential. The 
maximum of $\lambda$ indicates the vicinity of the phase transition. As the 
temperature is lowered towards $T_g$, the exponent $\beta$ saturates to 1/3. 
This result holds for a wide range of concentration in AgMn.\par

Keren \etal 1996b have been able to distinguish between three possible forms of 
the spin-spin self correlation-function with only a scaling argument. Using 
the neutron spin echo technique, Mezei and Murani 1979 could not achieve this 
result. Interestingly, the success of the $\mu$SR work lies in the use of the 
spin correlation-function concept for the data analysis. It would be of 
interest
to check if the limit $\beta \rightarrow 1/3$ as $T \rightarrow  T_g$ is a 
universal characteristics of spin glasses.\par

\section{Probing the magnetic properties of superconductors}\label{mixed}

$\mu$SR spectroscopy has been intensively used in recent years to probe the
magnetic properties of superconductors. Below we present three experimental
works which have greatly benefited from the availability of high quality 
crystals. In these works the physical properties of the mixed state are
probed but with quite different results. In \sref{sonier} 
and \sref{fusion} the field distribution due to the vortices is visualized for
YBa$_2$Cu$_3$O$_{6.95}$ and Bi$_{2 +x}$Sr$_{2 - x}$CaCu$_2$O$_{8+ \delta}$,
respectively. Whereas the observation of a conventional 3D flux-line lattice
for YBa$_2$Cu$_3$O$_{6.95}$ 
leads to the analysis of the data in terms of the symmetry properties
of the superconducting order-parameter, the strong temperature and field
dependence of the Bi$_{2 +x}$Sr$_{2 - x}$CaCu$_2$O$_{8+ \delta}$ field 
distribution offers the possibility of investigating its
phase diagram in the temperature-field plane. In \sref{upt3}, the measurements 
in UPt$_3$ show that, in addition to the analysis of the field distribution in 
terms of the symmetry of its superconducting order-parameter, one is able to 
unravel some basic magnetic properties of the compound and their interplay with 
superconductivity.\par

\subsection{The symmetry of the superconducting order-parameter in 
YBa$_2$Cu$_3$O$_{6.95}$}\label{sonier}

The symmetry of the superconducting order-parameter of the cuprate 
superconductors is the subject of ongoing reseach. A possible way of obtaining 
information on this symmetry is to investigate the excitation spectrum from the
temperature dependence of the superconducting condensate density, $n_s$. The 
flux line lattice in YBa$_2$Cu$_3$O$_{6.95}$ is expected to be conventional,
i.e. not to melt (at least for a usual magnetic field intensity; see 
\sref{fusion}). Therefore the field distribution due to the flux line lattice is
characterized by the coherence length and the London penetration depth,
$\lambda$, which is directly related to $n_s$: namely $1/\lambda^2$ $\propto$ 
$n_s$. An effective method to measure the field distribution is the muon spin 
rotation 
technique (see \ref{appendix3}). It probes the bulk of the material.\par
 
Early $\mu$SR measurements of the field distribution in sintered powders and 
low quality crystals concluded that $1/ \lambda^2$ has a weak temperature
dependence for $T$ $\ll$ $T_c$, suggesting the existence of an energy gap in 
the spectrum of excitations, in contradiction to the interpretation of other 
measurements. In particular, microwave data of Hardy \etal 1993 on high quality 
crystals show that $1/ \lambda_{ab}^2$ depends linearly on temperature. 
$\lambda_{ab}$ is defined as $\left(\lambda_a \lambda_b \right)^{1/2}$ where 
$\lambda_a$ and $\lambda_b$ are the penetration depths for currents flowing 
along the $a$ and $b$ axis, respectively. This is consistent with the expected
line of nodes for the order-parameter from singlet $d_{x^2 - y^2}$ wave 
pairing. 

\begin{figure}
\centerline{\epsfbox{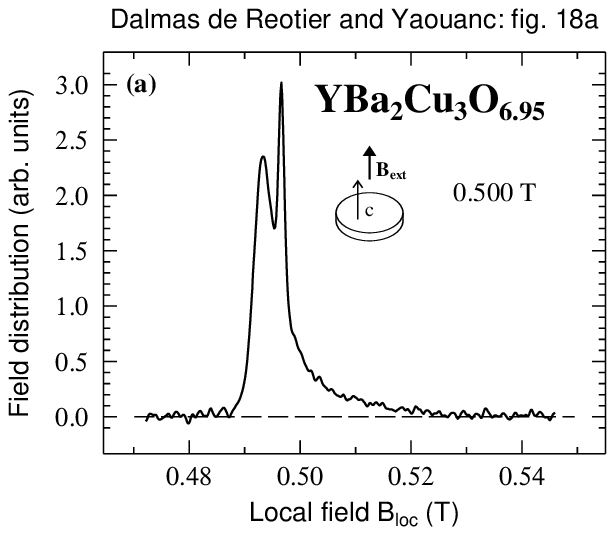}\hfill
\epsfbox{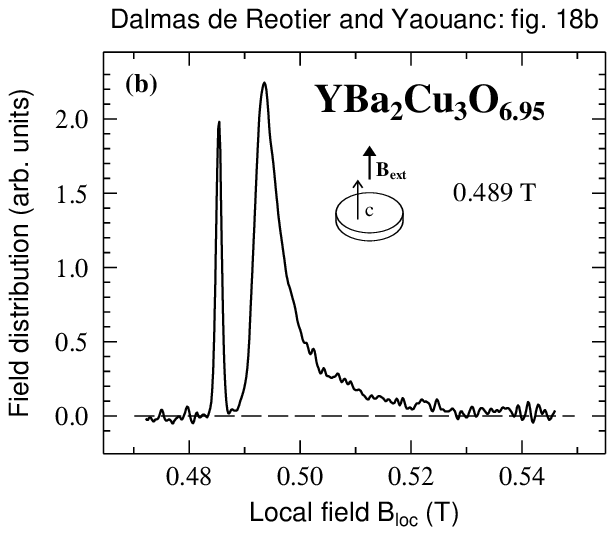}}
\caption{ Typical field distributions recorded in YBa$_2$Cu$_3$O$_{6.95}$ after
cooling the sample in a field of 0.500 T (adapted from Sonier \etal 1994).
While the distribution of figure \protect\ref{sonier_DISTRIBUTION}a
has been recorded at 5.4 K with that field value, the distribution of 
figure \protect\ref{sonier_DISTRIBUTION}b was measured at 6 K after decreasing 
the field to 0.489 T. Because of the high superconducting temperature of this
compound, 
the difference in temperature for the two measurements is expected not to be
significant. The sharp lines below 0.50 and 0.49 T in figure 
\protect\ref{sonier_DISTRIBUTION}a and \protect\ref{sonier_DISTRIBUTION}b
respectively represent the background contribution, i.e. muons which are not
stopped in the sample. The other part of the distribution arises from the
superconductor and is not affected by the field decrease. 
These measurements prove that the vortices are pinned.}
\label{sonier_DISTRIBUTION}
\end{figure}

Although the microwave method has a high precision, it is not sensitive to the
absolute value of $\lambda_{ab}(0)$ and probes only the skin depth at the
surface. Sonier \etal 1994 and 1997 and Riseman \etal 1995 have performed 
$\mu$SR measurements in a mosaic of single crystals. 
In \fref{sonier_DISTRIBUTION} we present the Fourier transformation of two 
spectra which have been recorded using a field cooling procedure with 
$B_{\rm ext}$ = 0.500~T. But while the distribution on the left has been 
recorded with that field value, the distribution on the right was taken in a 
field of 0.489 T after initially 
cooling the sample in a field of 0.500 T. Whereas the 
signal from muons in the sample remains practically unchanged due to the strong 
pinning of the vortex lattice, the sharp peak attributed to the background 
(muons stopped in the sample holder and cryostat walls) shifts down by $\sim$ 
11.2 mT.\par

\begin{figure}
\centerline{\epsfbox{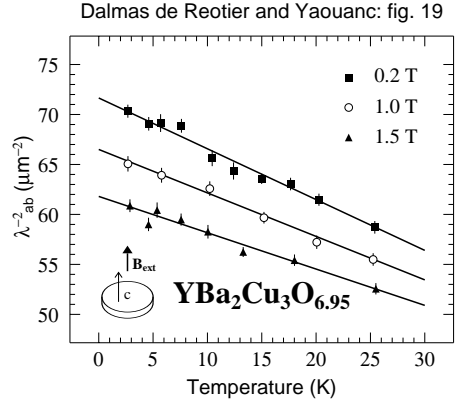}}
\caption{ Temperature dependence of $1/ \lambda_{ab}^2$ measured on
YBa$_2$Cu$_3$O$_{6.95}$ for three values of the external field applied along 
the $c$ axis (adapted from Sonier \etal 1997). The measured linear temperature
dependence is consistent with a $d$-wave superconducting order-parameter.
The observed field dependence is probably explained if a proper proper cutoff 
function is used (Yaouanc \etal 1997a).}
\label{sonier_densite}
\end{figure}

From the measured field distribution, Sonier \etal 1994 and 1997 have extracted 
$\lambda_{ab}$ using the theory explained in \ref{appendix3}, with the
restriction that they do not use a proper cutoff function to account for
the finite size of the vortex cores (Yaouanc \etal 1997a). In 
\fref{sonier_densite} we present $1/ \lambda_{ab}^2 (T)$ for $B_{\rm ext}$ 
= 0.2 T, 1.0 T and 1.5 T. The linear dependence of $1/ \lambda_{ab}^2$ vs 
$T$, confirms the zero-field microwave measurements.
We note that the weak field-dependence of $1/ \lambda_{ab}^2 (T = 0)$ can 
probably be explained using a proper cutoff function (Yaouanc \etal 1997a). In 
their analysis, Sonier \etal 1997 take into account the in-plane anisotropy of 
the penetration depth discovered by $\mu$SR (Tallon \etal 1995, Bernhard \etal 
1995b) and infrared measurements (Basov \etal 1995).\par

Measurements carried out by Riseman \etal 1995 for 
$1.9$ T $\leq$ B$_{\rm ext}$ $\leq$ $6.5$ T show that the Ginzburg-Landau 
paramater $\kappa$ is constant between 30 K and 75 K with a value $\kappa$ 
$\approx$ 70. In addition, these authors finds an upper critical field 
$B_{c2}$ = 90 (10) T.

As pointed out by Sonier \etal 1994, the sample quality seems to have a
significant influence on the experimental results. This is understandable since
$\lambda$ is in fact expressed in terms of an integrated excitation spectrum
over parts of the Fermi surface 
(see Gross \etal 1986 and Gross-Alltag \etal 1991). Therefore, additional 
methods of investigation of the symmetry of the order-parameter are highly 
desirable. A recent example is given by Bernhard \etal 1996 who have determined 
by $\mu$SR the depression of $n_s$ as a function of the level of Zn doping. 
They argue that the initial decrease of $n_s$ is inconsistent with $s$-wave 
pairing and magnetic scattering but points rather towards $d$-wave pairing, in 
agreement with the microwave data of Hardy \etal 1993, the $\mu$SR data of 
Sonier \etal 1994 and 1997 and the micro-SQUID results of Tsuei \etal 1994. 
However, this conclusion is disputed by Nachumi \etal 1996. Therefore, untill 
the dispute is settled, it is not possible to give a definite conclusion 
concerning the interpretation of the effect of Zn doping on $n_s$.
\par

Since the discovery of the high $T_c$ superconductors, trends in the 
relations between some of their parameters have been looked for. Some years 
ago a remarkable empirical relationship between $T_c$ and the low-temperature 
variance of the vortex field distribution was found for some high 
$T_c$ oxides (Uemura \etal 1991). This experimental result was taken as 
evidence for a high-energy-scale pairing mechanism, consistent with the picture 
of real-space paired bosons. It was suggested that this scaling was valid for 
all the high $T_c$ materials. This suggestion has triggered an 
important experimental activity. The recent reports (Uemura \etal 1993, 
Niedermayer \etal 1993, Weber \etal 1993, Bernhard \etal 1995b, 
Zimmermann \etal 1995, Tallon \etal 1995) show that the simple universal 
scaling-law is only partially valid. For example, $T_c$ is not solely a 
function of the hole concentration: the oxides with and without Cu chains do 
not belong to the same class. \par
We note that practically all these works have 
been performed in powder samples and the $\mu$SR spectra have been analyzed 
supposing a Gaussian field distribution. Although this methodology may not be 
completely safe (Sonier \etal 1994 and Harshman and Fiory 1994), it has 
nevertheless been able to provide the first proof of the in-plane anisotropy of 
the London penetration depth in YBa$_2$Cu$_3$O$_{z - \delta}$ 
(Tallon \etal 1995).\par

From the $\mu$SR technical point of view, the works of Sonier \etal 1994 and
1997 and Riseman \etal 1995 
are remarkable since they have been performed both at high fields 
and with small samples (one of the Sonier's sample covers an area of 
$5 \times 5$ mm$^2$ perpendicular to the $c$ axis). In a near future one may 
foresee routine $\mu$SR measurements on even smaller samples and therefore on 
samples of even higher quality.\par

\subsection{The vortex state in highly anisotropic high $T_c$ superconductors}
\label{fusion}

The state of the vortices in some of the highly anisotropic superconducting 
oxides is still a subject of discussion (see Bishop 1996a and 1996b, 
Nelson 1997 and 
Crabtree and Nelson 1997 for reviews). In these compounds such as
Bi$_{2 +x}$Sr$_{2 - x}$CaCu$_2$O$_{8+ \delta}$, the vortices are best described 
as layered systems of 2D pancake vortices interacting via a combination of 
tunneling Josephson currents and electromagnetic interactions. Increasing the 
field or the temperature, one expects to observe changes in the typical 3D flux
line lattice field distribution due to either disordering of the vortex
lattice, a reduction of its dimensionality or its melting.\par

In this section we focus on results obtained by Lee \etal 1993, 1995 and 1997,
Aegerter \etal 1996 and Bernhard \etal 1995a.\par

\begin{figure}
\centerline{\epsfbox{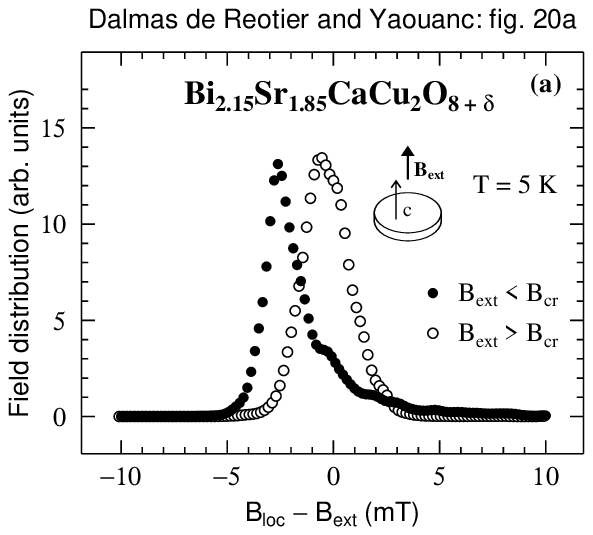}\hfill
\epsfbox{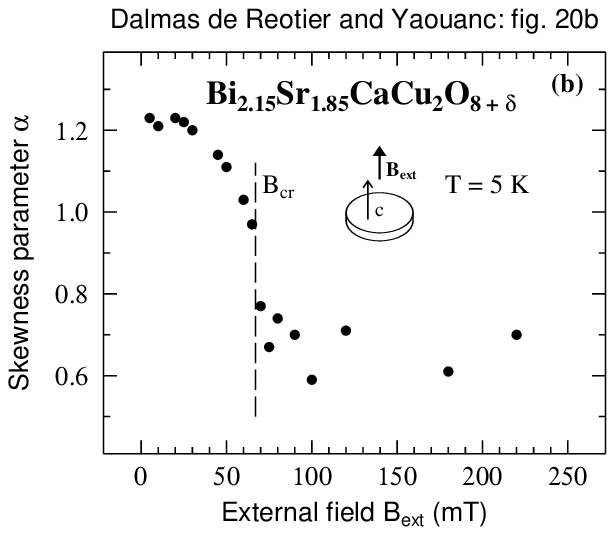}}
\caption{The vortex state in the highly anisotropic material 
Bi$_{2.15}$Sr$_{1.85}$CaCu$_2$O$_{8+ \delta}$ as the intensity of the 
external magnetic field is increased. The temperature is kept fixed far below
$T_c$. As shown in figure \protect\ref{AEGERTER_champ}a, the shape of the field 
distribution becomes more symmetric at high field. The field dependence of the 
skewness parameter, which allows one to quantify the change in the shape of the
distribution, is presented in figure \protect\ref{AEGERTER_champ}b 
(Aegerter 1997 and adapted from Aegerter \etal 1996). Note the sharpness of the 
crossover at $B_{\rm cr}$.}
\label{AEGERTER_champ}
\end{figure}

At low temperature and for a field less than the crossover field $B_{\rm cr}$, 
a lattice of extended flux lines is observed. As seen in
\fref{sonier_DISTRIBUTION} and \fref{AEGERTER_champ}, its $\mu$SR 
signature is a typical strongly asymmetric field distribution with a pronounced 
tail towards high fields, showing that some of the muon spins precess in the 
local field caused by flux cores that are extended in the $c$ direction. This 
interpretation is supported by the occurence of Bragg peaks in the neutron 
scattering experiments (Cubitt \etal 1993). Drastic changes occurs when 
$B_{\rm ext}$ exceeds $B_{\rm cr}$. The field distribution becomes 
more symmetric 
as shown in \fref{AEGERTER_champ}. Simultaneously the neutron Bragg peaks 
disappear indicating that the long-range coherence of the flux lattice is 
destroyed. The change of the shape of the field distribution is quantified by 
the skewness parameter $\alpha$ (see \ref{appendix3}). Its field dependence is 
presented in \fref{AEGERTER_champ}. $\alpha$ is definitively smaller above 
$B_{\rm cr}$ than below, reflecting the truncation of the high-field tail. 
This reduction is either due to the motion of the vortices 
(Harshman \etal 1991, Inui and Harshman 1993), a reduced dimensionality of the 
vortex structure 
(Brandt 1991 and Harshman \etal 1993) or a transition to a glass phase 
(Ryu \etal 1996, Ryu and Stroud 1996 and Gingras and Huse 1996). The last two 
statements have 
common features and it may not be easy to distinguish between them
(Gingras and Huse 1996). Lee \etal 1993 suggest that their data points to a 
reduction of the dimensionality of the vortex system: in a high field, the 
system 
consists of an array of pancake vortices that are uncorrelated in the $c$ 
direction but ordered two-dimensionally within each stack of CuO$_2$ planes. 
This interpretation is supported by the numerical results of Schneider \etal
1995. Such a dimensional crossover has been predicted to occur at $B_{\rm cr}$ 
(Vinokur \etal 1990 and Glazman and Koshelev 1991). Additional measurements 
(Bernhard \etal 1995a) on underdoped and strongly overdoped crystals support 
this interpretation. Recently Aegerter \etal 1996 have discovered that 
$B_{\rm cr}$ = $\Phi_0/\lambda^2_{ab}$ where $\Phi_0$ = 
2.07 $\times$ 10$^{-15}$ Tm$^2$ is the quantum of flux. This latter result is 
understood in terms of a system of vortices controlled predominantly by 
electromagnetic interactions. This is in contrast to other materials, such as 
HgBa$_2$Ca$_3$Cu$_4$O$_{10 + \delta}$, where Josephson coupling plays a more 
significant role.\par

\begin{figure}
\centerline{\epsfbox{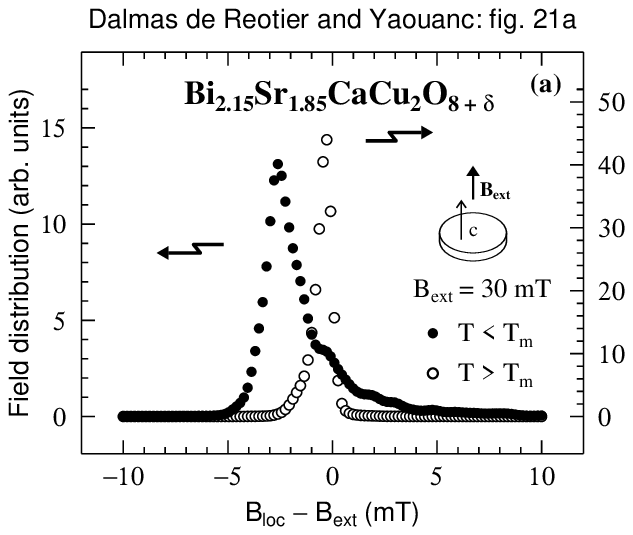}\hfill
\epsfbox{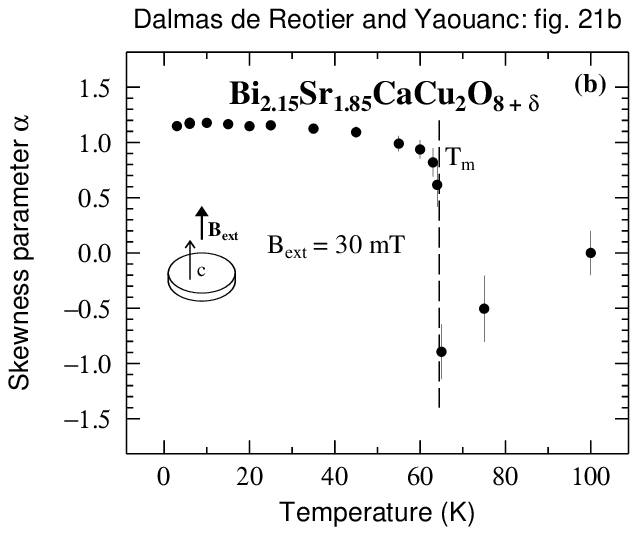}}
\caption{The vortex state in the highly anisotropic material
Bi$_{2.15}$Sr$_{1.85}$CaCu$_2$O$_{8+ \delta}$ as the temperature is increased.
The intensity of the external magnetic field is kept fixed. As shown in 
figure \protect\ref{AEGERTER_temperature}a, the shape of the field distribution 
changes and becomes narrow at high temperature. 
The temperature dependence of the skewness parameter is presented in figure 
\protect\ref{AEGERTER_temperature}b (Aegerter 1997 and adapted from 
Lee \etal 1997). Note the sharp change of $\alpha$ at $T_m$ and the negative 
value of $\alpha$ for $T > T_m$.}
\label{AEGERTER_temperature}
\end{figure}

There is much interest in investigating the vortex state as a function of the
temperature since it is expected that, at sufficiently high temperature, the 
vortex lattice should melt (Bishop 1996). In \fref{AEGERTER_temperature} we 
compare field distributions recorded at low and high temperature for a given 
field intensity. The high temperature distribution is very narrow and the 
shapes at low and high temperature are distinctly different. The temperature 
dependence of the skewness parameter indicates a sharp change at $T_m$. 
$\alpha$ is even negative above $T_m$, whereas in \fref{AEGERTER_champ} it 
is always positive. Comparing these results with the numerical calculations of 
the field distribution by Schneider \etal 1995, $T_m$ is found to be 
the fusion temperature of the vortex lattice. A thorough analysis is 
presented by Lee \etal 1997 which supports this interpretation. In addition, 
this analysis points out the determinant role of the electromagnetic coupling 
between the superconducting layers, In fact, it is this coupling rather than 
the Josephson coupling which determines the phase diagram below $\sim$ 70 K, a 
rather high temperature relative to the superconducting temperature 
($T_c$ $\sim$ 80 K).\par

The detailed characterization of the vortex state in the highly anisotropic 
oxides is a complex problem. The microscopic methods like $\mu$SR and 
small angle neutron scattering, in combination with magnetization measurements,
can lead to an improved understanding of the physics involved. For
example, in Bi$_{2}$Sr$_{2}$CaCu$_2$O$_{8+ \delta}$, the ``second peak'' 
observed in the hysteresis loops and the sharp change in the field distribution 
detected by $\mu$SR both occur at the same field, the crossover field 
$B_{\rm cr}$, allowing the origin of the ``second peak'' to be understood
(Bernhard \etal 1995a).\par 

\subsection{Anisotropy of the magnetic response in UPt$_3$}\label{upt3}

The hexagonal heavy fermion superconductor UPt$_3$ has the unique physical 
property of showing two Meissner phases, at $T_{c1}$ $\simeq$ 0.48 K and 
$T_{c2}$ $\simeq$ 0.53 K, and three flux phases. In addition, neutron and
magnetic X-ray diffractions have indicated an antiferromagnetic phase 
transition at $T_N$ $\sim$ 6 K characterized by a tiny magnetic moment of
0.02 (1) $\mu_B$/U-atom at low temperature, lying in the basal plane along the
$b$ axis. This 
transition has never been detected by macroscopic measurements. The origin of 
the magnetic and superconducting phases of UPt$_3$ is one of the most debated 
subjects in condensed matter physics. The superconducting multiphases identify 
UPt$_3$ has a candidate for unconventional superconductivity. The term 
``unconventional'' refers to the fact that the order-parameter has a lower 
rotational symmetry in the superconducting phases than in the normal state. 
More information on UPt$_3$ can be found in the reviews of Sauls 1994 and of 
Heffner and Norman 1996. A review on unconventional superconductivity has 
been recently published (Muzikar 1997).\par

\begin{figure}
\centerline{\epsfbox{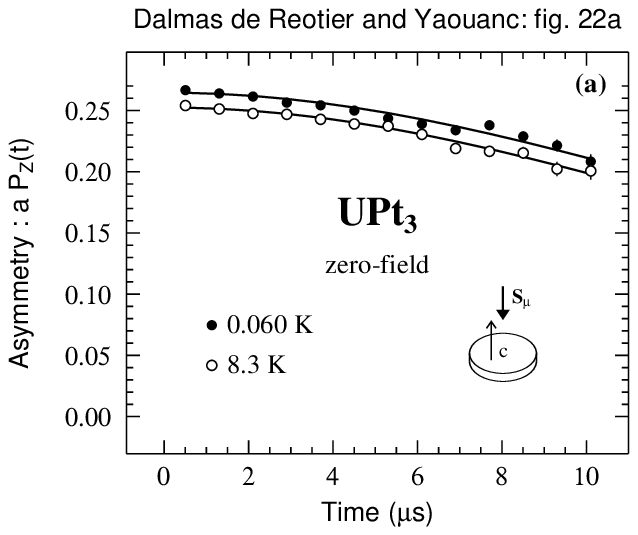}\hfill
\epsfbox{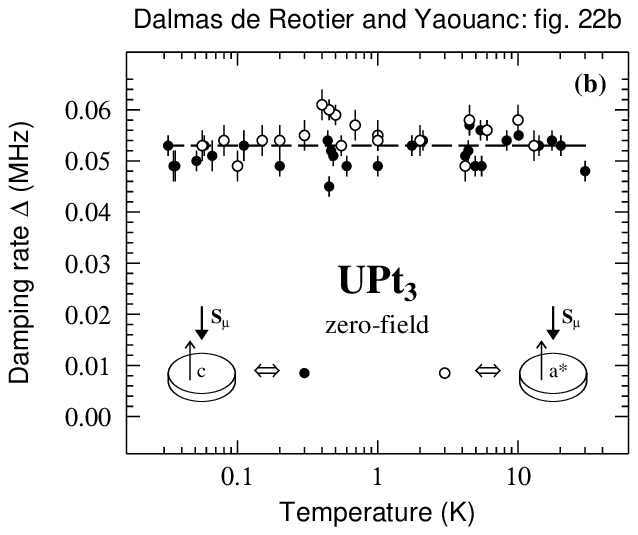}}
\caption{(a): Typical zero field spectra measured for UPt$_3$. The 
full lines are fits. This figure shows that there is no additional local field 
at the magnetic phase transition and in the low temperature superconducting 
phase.
(b): Temperature dependence of the Kubo-Toyabe damping rate 
$\Delta$. The dashed straight line indicates its average. These results show 
that $\Delta$ is independent of the temperature and orientation of the crystal 
axes relative to ${\bf S}_\mu$. Both figures are from Dalmas de R\'eotier
\etal 1995.}
\label{upt3_fig1}
\end{figure}

We first present some recent zero-field measurements (Dalmas de R\'eotier
\etal 1995). They have been performed with two purposes: to detect the
magnetic phase transition at $T_N$ and a signature of an eventual internal
magnetic field induced by the Cooper pairs in the low temperature 
superconducting B phase (below $T_{c1}$). The results are presented in 
\fref{upt3_fig1}. For 
the large temperature range probed and the two crystals investigated 
(${\bf S}_\mu$ either parallel or perpendicular to the $c$ axis), 
$P_Z(t)$ is simply described by a Kubo-Toyabe function. The values of the 
damping rates are consistently explained as arising from the $^{195}$Pt nuclear 
magnetic moments. Therefore neither the phase transition at $T_N$, 
nor an additional magnetic field in the superconducting B phase is observed.\par

Since neutron diffraction measurements performed on some of the slices of the 
$\mu$SR samples show that the antiferromagnetic phase transition still exists, 
one must conclude that either the electronic dipolar fields at the 
muon site cancel out exactly or the electronic uranium magnetic moments 
fluctuate too fast to be detected, i.e. their characteristic time is shorter
than $\sim$ 10$^{-7}$ s. From neutron diffraction (Aeppli \etal 1988) we know
that this time is longer than $\sim$ 10$^{-11}$ s.\par

The zero-field measurements (Dalmas de R\'eotier \etal 1995) indicate that a 
possible change in $B_{\rm loc}$ induced by magnetism or superconductivity, if 
any, has to be smaller than approximately 3 $\mu$T. Recently high precision 
magnetization measurements have shown that the possible change in bulk 
magnetization (rather than in the local field as observed by $\mu$SR) in the 
superconducting B phase is smaller than 0.2 $\mu$T (Kambara \etal 1996).
A Cooper pair produces an orbital magnetic field at the muon site. If it has a 
spin, a spin density has to be added to the orbital density. Since the orbital 
moment is expected to be much smaller than the spin moment, the measurements 
put a limit on the possible value of the spin density at the muon site. Taking
into account the available theoretical estimates for the spin density, 
Dalmas de R\'eotier \etal 1995 conclude that their results do not
support models predicting a triplet spin state for the Cooper pair.\par

Lussier \etal 1996 have investigated the magnetic field response of UPt$_3$ by
single crystal neutron diffraction. Their results show that a field in the basal
plane of up to 3.2 T has no effect on the magnetic Bragg peaks. Since the
intensity of these peaks is extremely small, the precision of the measurements
is limited: it is only known that the angle of rotation of the magnetic
moment in the hexagonal plane is smaller than 26$^\circ$.
Taking into account that the $\mu$SR technique is 
well adapted to study small moment systems, Yaouanc \etal 1997c have 
performed transverse high field $\mu$SR 
mesurements. They detect the magnetic phase transition if a large field is 
applied in the basal plane. As expected, the phase transition is not observed 
for a field applied along the $c$ axis. Therefore the magnetic properties of 
UPt$_3$ are found to be field dependent if probed at high field by a 
sufficiently sensitive method.\par

The magnetic properties of UPt$_3$ are remarkably sensitive to alloying. In the
case of the pseudobinaries U$\left({\rm Pt}_{1-x}{\rm Pd}_x \right)_3$, large 
uranium magnetic moments have been detected for $0.02 \leq x \leq 0.07$. 
Neutron diffraction experiments on 
U$\left({\rm Pt}_{0.95}{\rm Pd}_{0.05} \right)_3$ show that the size of ordered 
magnetic moment equals 0.6 (2) $\mu_B$/U-atom, much larger than in the pure
sample (0.02 (1) $\mu_B$/U-atom). In the zero-field $\mu$SR measurements of 
deVisser \etal 1997 on this compound, two frequencies (or one frequency and a
strongly damped Kubo-Toyabe signal; Amato 1997) are 
observed, indicating 
two magnetically inequivalent muon stopping sites. On the other hand, the 
zero-field signal in U$\left({\rm Pt}_{0.998}{\rm Pd}_{0.002} \right)_3$ and 
in the pure sample are the same, i.e. no electronic magnetic signal is 
observed.\par

The preliminary data of deVisser \etal 1997 demonstrate a salient difference 
in the $\mu$SR response between compounds with small and large magnetic moments 
in the U$\left({\rm Pt}_{1-x}{\rm Pd}_x \right)_3$ series. It is still not 
possible to extract reliable physical information from the limited amount of 
available experimental $\mu$SR data. An important issue that could be resolved
by combining these measurements with neutron measurements on the same samples 
concerns origin of the large difference between the intensity of the low 
temperature magnetic moment for UPt$_3$ and 
U$\left({\rm Pt}_{0.95}{\rm Pd}_{0.05} \right)_3$.\par

\begin{figure}
\centerline{\epsfbox{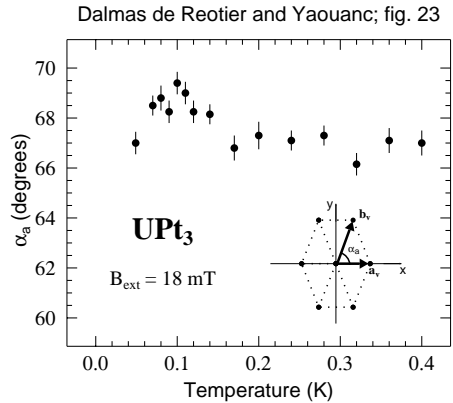}}
\caption{Temperature dependence of the angle characterizing the vortex lattice 
of UPt$_3$ for the external field applied along the $a$ axis. The angle of the
real space vortex lattice is defined in the insert.
(adapted from Yaouanc \etal 1997b).}
\label{angle_vortex}
\end{figure}

The temperature dependence and the anisotropy of the magnetic field penetration 
lengths in the superconducting phases of UPt$_3$ have been studied by muon 
spin rotation measurements (Yaouanc \etal 1997b). 
The analysis of the temperature 
dependence of the penetration length parallel and perpendicular to the $c$ 
axis has shown that the superconducting order-parameter in the B phase cannot
just have a line of nodes in the equatorial plane of the Fermi surface. The 
analysis supports an hybrid order-parameter with point nodes at the poles and
a line of nodes at the equatorial plane. A remarkable result of the
measurements is the observation of a strong axial anisotropy of the penetration 
length. As explained at \ref{appendix3}, this anisotropy is directly related to
the opening angle of the vortex lattice. In \fref{angle_vortex} we present the 
temperature dependence of this angle for ${\bf B}_{\rm ext}$ applied along the
$a$ axis. If the penetration lengths were isotropic, it would be equal to
60$^\circ$. It has already been measured by small angle neutron scattering 
(Kleiman \etal 1992).  But, because the neutron cross-section decreases 
dramatically as the temperature is increased, it could only be measured at 
0.05 K. We note that the angle is pratically temperature independent. When 
${\bf B}_{\rm ext}$ is applied along the $c$ axis of the crystal lattice, 
the angle is found to be temperature independent with a value of $\simeq$ 
60$^\circ$, i.e. the vortex lattice is a traditional hexagonal lattice. \par

\section{Summary and perspectives}\label{conclusion}

In this review we have attempted to detail the possibilities offered by the 
$\mu$SR technique for the study of the magnetic properties of materials.
A quick look at all the examples described in this work shows that the
prominently used properties of the technique are its high sensitivity to local
fields, and its capability to characterize the time scale of their
fluctuations. Interestingly, 
information can be extracted from compounds which do not display long range 
magnetic correlations. In fact, many $\mu$SR investigations concern disordered 
or frustrated magnetic compounds. Two examples are given in this review. The 
typical spot size of a muon beam is some millimetres at a continuous 
source. Although this means that extremely small samples can not be studied, 
this macroscopic size allows one to probe magnetic structure with nanometre or 
micrometre scales. We have presented four examples of such studies.\par

The $\mu$SR technique has many attractive feactures which explain its present
fast development. Although we have not explained in any detail how a $\mu$SR 
experiment is performed (this information is found in the general references 
provided at the beginning of \sref{muon}), relative to other non local 
microscopic techniques, a $\mu$SR experiment is relatively straightforward.
The $\mu$SR spectroscopy is of general use: the sample does not need to contain 
special nuclei. A relatively involved sample environment can be used as 
illustrated, for example, by the high number of measurements performed in the 
30 mK - 1 K range using dilution refrigerators. Measurements have been 
performed up to $B_{\rm ext}$ = 6.5 T (Riseman \etal 1995).
The recording of a $\mu$SR spectrum requires in most cases less than an hour. 
Therefore, in a reasonable amount of time, it is possible to map out the 
temperature and magnetic field $\mu$SR response of a compound.\par

As does any experimental technique, the $\mu$SR method has its own limitations.
One of
the major drawbacks is that the sample has to be sufficiently big. If a disk is
used and if the muon beam is perpendicular to its plane, in the most favourable 
cases, the diameter of the disk should not be smaller that $\sim$ 5 mm. A 
larger diameter is unfortunately required at a pulsed source such as ISIS, 
where the diameter has to be at least 20 mm. For the more widely used
beamlines, the so-called surface muon beams (i.e. relatively slow muon beams), 
150 mg of material 
per cm$^2$ of beam cross section is required to stop the muons. This means that 
the sample thickness needs not to be much larger than $\sim$ 1 mm for organic
materials (density $\grsim$ 1) and even much less (one to three hundred 
micrometres) for denser material such as oxides or metals. \par

{\sl A priori}, the muon diffusion and
localization site(s) properties are unknown. This can turn out to be a problem 
to extract detailed information from the measurements. In this respect, 
it has recently been shown that frequency shift measurements may be of great 
help (see \sref{field}).\par

In order to respond to the demand of the users, major technical developments 
are scheduled at the $\mu$SR facilities. \par

The investigation of the vortex state in superconductors is at present one of 
the main subject of $\mu$SR research. Optimized spectrometers in transverse 
geometry with intense magnetic fields at relatively high temperature are needed 
for this type of work. They should be available in a few years time
both at TRIUMF and PSI. \par

Surface science is a major topic of modern physics which, so far, is untouched 
by $\mu$SR. The reason is that the momentum of the muons in presently
available beam is too large and therefore the muons are stopped at best in a 
few hundred micrometres. Ultra slow muon beams are being developed at
facilities. The challenge is to keep a high degree of polarization in the 
slowing down process together with a reasonable efficiency. The possibility to
moderately reaccelerate the ultra slow muons provides the opportunity of fine 
tuning the stopping range of the muons. For instance, such a beam will allow 
the study of magnetic multilayers which are of interest both for basic research 
and technological applications.\par

The possibility of building an intense pulsed neutron facility in Europe, the
European Spallation Source, is also
being considered. As at the ISIS Facility, where pulsed neutrons and 
muons are available, many muon beams are scheduled to be introduced 
at this future facility. The
experience gained at ISIS has shown that having beams of both particles
at the same
institution is a real advantage: it provides opportunities for the neutron and 
muon communities to interact. \par

\section{List of works published from July 1993}\label{list}

The tables provide a list of published works. When a given work has been 
described in short
articles and later on in an extended paper, we only refer to the latter 
report.\par
\begin{table}
\caption{A selection of work carried out on conventional magnets}
\label{conventional}
\footnotesize\rm
\begin{tabular}{@{}lll}
\br
Chemical formula&Comments&References\\
\mr
\centre{3}{Magnetic correlations in magnets}\\
\mr
(Mn$_{1-x}$Fe$_x$)Pt$_3$
  &dynamical crossover from ferromagnetism to 
    &Barsov \etal 1993 and 1994 \\
  &asperomagnetism & \\
\ms
MnSi
  &strong correlation between the $^{55}$Mn electric 
    &Kadono \etal 1993 \\
  &field gradient and magnetic susceptibility & \\
\ms
Ag$_{1-x}$Mn$_x$
  &spin correlation-function in a spin-glass
    &Campbell \etal 1994 \\
  &see \sref{glass} 
    &Keren \etal 1996b\\
\ms
Gd
  &critical paramagnetic fluctuations and 
    &Dalmas de R\'eotier and \\
  &correlation-functions; effect of dipolar interaction &Yaouanc 1994 \\
\ms
Ni and Fe
  &evaluation of the Brillouin zone probed by $\mu$SR
    &Dalmas de R\'eotier \etal 1994 \\
  &study of critical paramagnetic fluctuations & \\
\ms
Gd$_{0.696}$Y$_{0.304}$
  &magnetic phase diagram investigation
    &Eccleston \etal 1995 \\
\ms
CeB$_6$
  &field response in disagreement with the
    &Feyerherm \etal 1995c\\
  &hypothesis of an antiferroquadrupolar phase & \\ 
\ms
PrCo$_2$Si$_2$
  &spin dynamics in an axial magnet
    &Gubbens \etal 1995\\
\ms
Y$_6$(Mn$_{1-x}$Fe$_x$)$_{23}$
  &static and dynamical magnetism in competition
    &Kilcoyne and Telling 1995 \\
\ms
CeSb
  &slow spin dynamics below $T_N$
    &Klauss \etal 1995\\
\ms
EuO, EuS
  &theoretical prediction of the thermal behaviour 
    &Lovesey and Engdahl 1995 \\
  &of the relaxation rate; see Yaouanc \etal 1993b &\\
\ms
RbMnF$_3$
  &theoretical prediction of the thermal behaviour
    &Lovesey \etal 1995 \\
  &of the relaxation rate &\\
\ms
Y$_{0.97}$Sc$_{0.03}$Mn$_2$
  &a spin liquid ground state is suggested
    &Mekata \etal 1995\\
\ms
Al-Mn-Si quasicrystal
  &study of the spin-glass state 
    &Noakes \etal 1995\\
\ms
UNiGa
  &magnetic phase diagram investigation 
    &Prokes \etal 1995\\
\ms
YMn$_2$
  &comparison between neutron and $\mu$SR results
    &Rainford \etal 1995a\\
\ms
Ni
  &temperature and pressure dependence of the 
    &Stammler \etal 1995\\
  &Fermi contact field &\\
\ms
Cr$_{1-x}$Fe$_x$
  &spin correlation-function in a spin-glass
    &Telling and Cywinski 1995\\
\ms
$\beta$-(NH$_4$)$_2$FeF$_5$
  &combined M\"ossbauer and $\mu$SR analysis 
    &Attenborough \etal 1996\\
\ms
MnF$_2$
  &first proof of Ising dynamical critical behaviour
    &Brown \etal 1996\\
\ms
CeRh$_3$B$_2$
  &investigation of the magnetic state
    &Cooke \etal 1996\\
\ms
La$_{0.67}$Ca$_{0.33}$MnO$_3$
  &ferromagnet with colossal magnetoresistance; 
    &Heffner \etal 1996\\
  &unusual spin dynamics; see \sref{dipolar}&\\
\ms
YMn$_2$D$_x$
  &effect of deuterium on spin dynamics 
    &Latroche \etal 1996\\
  &and magnetic phase diagram &\\
\ms
ReGa$_6$
  &combined $\mu$SR and neutron study of these 
    &Lidstr\"om \etal 1996a\\
  &quasi-two-dimensional magnets &\\
\ms
GdNi$_5$
  &effect of the dipolar interaction on the critical 
    &Yaouanc \etal 1996a \\
  &ferromagnetic fluctuations; see \sref{dipolar} &\\
\ms
Ho
  &study of the incommensurate helicoidal structure
    &Krivosheev \etal 1997 \\
\br
\end{tabular}
\end{table}

\begin{table}
\caption{A selection of work carried out on conventional magnets
(continuation of Table \protect \ref{conventional})}
\label{continuation1}
\footnotesize\rm
\begin{tabular}{@{}lll}
\br
Chemical formula&Comments&References\\
\mr
\centre{3}{Magnetic phase diagrams in organic magnets}\\
\mr
(TMTSF)$_2$-X 
  &collective low-energy spin-density-wave 
    &Le \etal 1993a\\
X = PF$_6$, NO$_3$, ClO$_4$
  &in addition to single-particle excitations & \\
\ms
{\sl p}-NPNN
  &3D Heisenberg ferromagnet with $T_C$ $\simeq$ 0.67 K;
    &Le \etal 1993b\\
  &effect of a longitudinal field; see \sref{organic} 
    &Blundell \etal 1995 \\
\ms
{\sl p}-PYNN
  &magnetic phase transition at $\simeq$ 0.09 K
    &Blundell \etal 1994\\
  &see \sref{organic}\\
\ms
3-QNNN
  &magnetic phase transition at $\simeq$ 0.21 K
    &Pattenden \etal 1995\\
  &see \sref{organic}\\
\ms
$\alpha$-(BEDT-TTF)$_2$KHg(SCN)$_4$
  &detection of two spin density wave transitions 
    &Pratt \etal 1995\\
\ms
1-NAPNN, 2-NAPNN
  &1-NAPNN has a magnetic transition below 0.1 K
    &Blundell \etal 1996\\
  &and 2-NAPNN does not order magnetically \\
\ms
MEM(TCNQ)$_2$
  &spin dynamics in this spin-Peierls system
    &Blundell \etal 1997a\\
\ms
{\sl p}-CNPNN, 4-QNNN
  &{\sl p}-CNPNN has a magnetic transition at $\simeq$ 0.17 K
    &Blundell \etal 1997b\\
  &and 4-QNNN does not order magnetically
    & \\
\mr
\centre{3}{Borocarbide materials}\\
\mr
YNi$_2$B$_2$C
  &characterization of the vortex state of this 
    &Cywinski \etal 1994\\
  &conventional superconductor &\\
\ms
YNi$_4$BC$_{0.2}$
  &proof that it is not a bulk superconductor
    &S\"ullow \etal 1994\\
\ms
TmNi$_2$B$_2$C
  &coexistence of magnetism and superconductivity
    &Cooke \etal 1995\\
  &
    &Le \etal 1995\\
\ms
ErNi$_2$B$_2$C
  &observation of one frequency below $T_N$ 
    &Le \etal 1995\\
\ms
SmNi$_2$B$_2$C
  &exhibits a magnetic phase transition 
    &Prassides  \etal 1995\\
YNi$_2$B$_2$C
  &no magnetic correlations  &\\
\ms
HoNi$_2$B$_2$C
  &observation of two magnetic transitions : one 
    &Le \etal 1996a\\
  &commensurate and one incommensurate &\\
\br
\end{tabular}
\end{table}

\begin{table}
\caption{A selection of work carried out on strongly correlated 
electronic systems}
\label{correlated}
\footnotesize\rm
\begin{tabular}{@{}lll}
\br
Chemical formula&Comments&References\\
\mr
\centre{3}{Non-superconducting compounds}\\
\mr
CeRu$_2$Si$_2$ 
  &a two component $4f$ system; see \sref{components}
    &Amato \etal 1993 and 1994\\
\ms
Y$_{1-x}$U$_x$Pd$_y$
  &competition between Kondo, RKKY and crystal field
    &Wu \etal 1994\\
  &interactions: effect on the magnetic phase diagram &\\
\ms
CeCu$_{5.9}$Au$_{0.1}$
  &no magnetic ordering in this non-Fermi-liquid 
    &Amato \etal 1995\\
  &system; Kondo disorder is maybe negligible 
    &Bernal \etal 1996\\
  & 
    &MacLaughlin \etal 1996\\
\ms
CeCu$_{5.5}$Au$_{0.5}$
  &phase transition at $T_N$ = 0.95 K
    &Chattopadhyay \etal 1995\\
\ms
CeNiSn
  &magnetic correlations with surprising temperature 
    &Kalvius \etal 1995a\\
  &dependence (the analysis yields a too small Kondo 
    &\\
  &temperature; see Dalmas de R\'eotier \etal 1996) &\\ 
\ms
CeTrSn, Tr = Pt, Pd
  &CePdSn exhibits a simple second order transition;
    &Kalvius \etal 1995b and 1995c\\
  &the transition in CePtSn is unusual &\\
\ms
CePt$_2$Sn$_2$
  &investigation of magnetic correlations; see 
    &Luke \etal 1995\\
  &comments of Dalmas de R\'eotier \etal 1996   &\\
\ms
UNi$_4$B
  &study of the magnetic phase diagram 
    &Nieuwenhuys \etal 1995\\
\ms
CeRhSb
  &spin correlations below $\approx$ 0.6 K
    &Rainford \etal 1995b\\
\ms
CeCu$_{5-x}$Al$_x$
  &magnetic phase diagram investigation
    &Wiesinger \etal 1995\\
\ms
UCu$_{5-x}$Pd$_x$
  &non-Fermi-liquid alloy; strong Kondo disorder 
    &Bernal \etal 1996\\
  &
    &Maclaughlin \etal 1996\\
\ms
YbAuCu$_4$
  &effect of crystal field on Kondo-type fluctuations;
    &Bonville \etal 1996\\
  &possible muon induced crystal field effect &\\
\ms
Ce$_{1-x}$Re$_x$Ni$_{1-y}$Tr$_y$Sn
  &effect of doping on the electonic ground state
    &Flaschin \etal 1996\\
\ms
CeTr$_2$Sn$_2$
  &Investigation of the magnetic correlations in 
    &Lidstr\"om \etal 1996b\\
Tr = Cu, Pt, Pd
  &heavy fermion antiferromagnets &\\
\ms
YbPdSb
  &a spin liquid system
    &Bonville \etal 1997\\
\ms
CePt$_2$Si$_2$
  &Kondo lattice compound with no magnetic ordering%
    &Dalmas de R\'eotier \etal 1997\\
\ms 
CeRuSi$_2$
  &ferromagnetic phase transition  at $T_C$ = 11.6 K; 
    &Duginov \etal 1997 \\
  &very small ordered Ce magnetic moment \\
\mr
\centre{3}{Superconductors}\\
\mr
UPd$_2$Al$_3$
  &London penetration depth approximately isotropic; 
    &Feyerherm \etal 1994\\
  &a two component $5f$ system; see \sref{components} &\\
\ms
Ce$_{1+x}$Cu$_{2+y}$Si$_2$
  &competition between superconductivity and magnetism
    &Luke \etal 1994\\
  &
    &Feyerherm \etal 1995a\\
\ms
UPt$_3$          
  &absence of electronic magnetic signal in zero-field;
    &Dalmas de R\'eotier \etal 1995\\
  &study of London penetration depths; see \sref{upt3}
    &Yaouanc \etal 1997b \\
\ms
U$_{1-x}$Re$_x$Ru$_2$Si$_2$ 
  &magnetic phase diagram investigation
    &Cywinski \etal 1995\\
Re = La, Y  
   &&Park \etal 1996\\
\ms
CeRu$_2$         
  &detection of a magnetic phase transition with 
    &Huxley \etal 1996\\
  &very small magnetic moments; see \sref{components} &\\
\ms
U(Pt$_{1-x}$Pd$_x$)$_3$
  &study of effect of doping on magnetism; 
    &de Visser \etal 1997\\
  &see \sref{upt3} &\\
\br
\end{tabular}
\end{table}

\begin{table}
\caption{A selection of miscellaneous studies}
\label{divers}
\footnotesize\rm
\begin{tabular}{@{}lll}
\br
Chemical formula&Remarks&Reference\\
\mr
AMo$_6$S$_{8-x}$Se$_x$, A= Sn, Pb
  &Penetration depth measurements
    &Birrer \etal 1993\\
\ms
$\kappa-$[BEDT-TTF]$_2$Cu[NCS]$_2$
  &Penetration depth measurement and fluxon dynamics
    &Harshman \etal 1994\\
\ms
Bi
  &theoretical investigation of line broadening due to
    &Solt 1994 \\
  &inhomogeneity of the Landau orbital magnetization & \\
\ms
$\alpha$-O$_2$
  &characterization of the magnetic phase transition 
    &Storchak \etal 1994 \\
  &below the $\alpha$-$\beta$ transition temperature & \\
\ms
Sb
  &investigation of the anomalous Knight shift
    &Birrer \etal 1995\\
\ms
PrNi$_5$
  &observation of muon induced crystal field effects
    &Feyerherm \etal 1995b\\
\ms
RbC$_{60}$
  &magnetic phase diagram investigation 
    &Cristofolini \etal 1995\\
  & 
    &MacFarlane \etal 1995\\
  &
    &Uemura \etal 1995\\
\ms
Be
  &diamagnetic domains; see \sref{beryllium}
    &Solt \etal 1996a\\
\br
\end{tabular}
\end{table}

\begin{table}
\caption{A selection of work carried out on superconducting and
non-superconducting oxides}
\label{oxyde1}
\footnotesize\rm
\begin{tabular}{@{}lll}
\br
Chemical formula&Remarks&Reference\\
\mr
\centre{3}{Superconducting oxides}\\
\mr
Bi$_{2 +x}$Sr$_{2 - x}$CaCu$_2$O$_{8+ \delta}$
  &flux-line lattice study ; possible observation
    &Lee \etal 1993, 1995 and 1997\\
  &of melting; see \sref{fusion} 
    &Bernhard \etal 1995a\\
  &
    &Aegerter \etal 1996\\
\ms
HgBa$_2$Ca$_3$Cu$_4$O$_{10 + \delta}$
  &test of the universal behaviour;
    &Niedermayer \etal 1993\\
  &see \sref{sonier}  & \\
\ms
Bi$_2$Sr$_2$Ca$_{1-x}$Y$_x$Cu$_2$O$_{8 +\delta}$
  &results unconsistent with universal behaviour;
    &Weber \etal 1993\\
Bi$_{2-x}$Pb$_x$Sr$_2$CaCu$_2$O$_{8 +\delta}$
  &see \sref{sonier}  & \\
\ms
YBa$_2$Cu$_3$O$_y$
  &theoretical analysis of the damping of the 
    &Aristov and Maleyev 1994\\
  &precession component & \\
\ms
YBa$_2$Cu$_3$O$_{6.6}$
  &investigation of effect of sulfur substitution
    &Cloots \etal 1994\\
  &through measurement of the penetration depth & \\
\ms
YBa$_2{(\rm{Cu}_{0.96}\rm{Zn}_{0.04})}_3$O$_x$
  &mapping of the magnetic phase diagram for 
    &Mendels \etal 1994\\
  &$6.00 \leq x \leq 6.92$ & \\
\ms
La$_2$CuO$_4$
  &computation of the hyperfine field
    &Shukri B Sulaiman \etal 1994\\
\ms
YBa$_2$Cu$_3$O$_{6.95}$
  &study of the flux-line lattice field distribution 
    &Sonier \etal 1994 and 1997\\
  &for a mosaic of single crystals; see \sref{sonier} 
    &Riseman \etal 1995\\
\ms
La$_{2-x}$Sr$_x$CuO$_4$
  &NMR and $\mu$SR study of the weakly doped region; 
    &Borsa \etal 1995\\
  &possible evidence for phase separation &\\
\ms
La$_2$Cu$_{1-x}$Zn$_x$O$_4$  
  &NMR and $\mu$SR study of the effects of substitution 
    &Corti \etal 1995\\
  &of magnetic Cu$^{2+}$ with diamagnetic Zn$^{2+}$    & \\
\ms
YBa$_2$Cu$_4$O$_8$H$_x$
  &study of the antiferromagnetic order 
    &Gl\"uckler \etal 1995\\
\ms
YBa$_x$Cu$_y$O$_{z - \delta}$
  &in-plane London penetration depth anisotropic due
    &Tallon \etal 1995 and 1996\\
  &to superconductivity in chains; see \sref{sonier} 
    &Bernhard \etal 1995b\\
\ms
YBa$_2$Cu$_3$O$_y$
   &sintered samples; unconsistent with results of Sonier 
     &Zimmermann \etal 1995\\
   &\etal 1994 and 1997 and Riseman \etal 1995 
     &\\
   &for $y$ = 6.95; see \sref{sonier} &\\
\ms
Y$_{0.8}$Ca$_{0.2}$Ba$_2{(\rm{Cu}_{1-y}\rm{Zn}_y)}_3$O$_{7 - \delta}$
  &dependence of the condensate density on Zn doping
    &Bernhard \etal 1996 \\
  &points towards $d$-wave pairing; see \sref{sonier} \\
\ms
La$_{2-x}$Sr$_x$NiO$_{4+ \delta}$
  &magnetic phase diagram investigation
    &Chow \etal 1996\\
\ms
Nd$_{2-x}$Ce$_x$CuO$_4$
  &no magnetic order but spin fluctuations for $x =0.02$; 
    &Hillberg \etal 1997\\
  &ordering of the Nd moments at low T for $x =0$ \\
\ms
YBa$_2{(\rm{Cu}_{1-y}\rm{Zn}_y)}_3$O$_x$
  &effect of Zn doping on the superconducting electron 
    &Nachumi \etal 1996 \\
La$_{2-x}$Sr$_x$(Cu$_{1-y}$Zn$_y$)O$_4$
  &density; see \sref{sonier}& \\
\ms
La$_2$CuO$_{4+ y}$
  &investigation of the spin-glass state
    &Pomjakushin \etal 1996\\
\br
\end{tabular}
\end{table}

\begin{table}
\caption{A selection of work carried out on superconducting and
non-superconducting oxides (continuation of Table \protect \ref{oxyde1})}
\label{continuation2}
\footnotesize\rm
\begin{tabular}{@{}lll}
\br
Chemical formula&Comments&References\\
\mr
\centre{3}{Non superconducting oxides}\\
\mr
Ca$_{0.86}$Sr$_{0.14}$CuO$_2$, Sr$_2$CuO$_3$
  &magnetic order in infinite-layer and chain
    &Keren \etal 1993 and 1995\\
\ms
CuGeO$_3$ and Cu$_{1-x}$Zn$_x$GeO$_3$
  &possible spin-glass ground state of the doped 
    &Lappas \etal 1994 \\
  &compound; see comments of Kadono 1997
    &Garc\'ia-Mu\~noz \etal 1995b \\
  &
    &Sohma \etal 1995 \\
  & 
    &Tchernyshyov \etal 1995 \\
\ms
SrCr$_8$Ga$_4$O$_{19}$
  &a spin-liquid ground state is proposed;
    &Uemura \etal 1994\\
  &see comments of Dunsiger \etal 1996 & \\
\ms
Y$_2$Cu$_2$O$_5$
  &magnetic phase diagram investigation
    &Duginov \etal 1995\\
\ms
RNiO$_3$
  &characterization of the magnetic order in the 
    &Garc\'ia-Mu\~noz \etal 1995a \\
  &low-temperature insulating phase & \\
\ms
Sr$_{n-1}$Cu$_{n+1}$O$_{2n}$
  &magnetic phase diagram of spin ladder 
    &Kojima \etal 1995a\\
  &systems; see \sref{uemura} & \\
\ms
$\left( {\rm Y}_{2-x}{\rm Ca}_x \right)$Ba$\left( {\rm Ni}_{1-y}{\rm Mg}_y 
\right)$O$_5$   
  &Haldane system; chain length controlled 
    &Kojima \etal 1995b\\
  &by doping; effect of doping on the ground state &\\
\ms
La$_2$Co$_x$Cu$_{1-x}$O$_{4+ \delta}$
  &magnetic phase diagram investigation
    &Lappas \etal 1995\\
\ms
Y$_2$Mo$_2$O$_7$, Tb$_2$$_2$Mo$_2$O$_7$
  &spin dynamics of geometrically frustrated 
    &Dunsiger \etal 1996 \\
  &magnets; large density of states near zero energy &\\
\ms
CuO
  &magnetic phase diagram investigation
    &Grebinnik \etal 1996\\
\ms
LaCuO$_{2.5}$
  &the ground state is magnetically ordered 
    &Kadono \etal 1996 \\
  &rather than spin-liquid &\\
\ms
KTr$_3$$\left( {\rm OH} \right)_6 \left( {\rm SO}_4 \right)_2$ 
  &long range ordering in the Fe compound;
    &Keren \etal 1996a\\
Tr = Cr, Fe
  &no such ordering in the Cr compound &\\
\ms
La$_2$Cu$_{1-x}$Li$_x$O$_4$
  &magnetic phase diagram; formation of a singlet 
    &Le \etal 1996b \\
  &ground state at large doping &\\
\ms
Ca$_2$CuO$_3$, Sr$_2$CuO$_3$
  &infinite-chain cuprates; see \sref{uemura} 
    &Kojima \etal 1997\\
\ms
LiV$_2$O$_4$
  &heavy fermion oxide with no magnetic ordering
    &Kondo \etal 1997\\
\ms
SrCuO$_2$
  &magnetic phase transition at $\sim$ 2 K 
    &Matsuda \etal 1997 \\
  &in this zigzag chain compound & \\
\br
\end{tabular}
\end{table}

\clearpage
\appendix
\section{The magnetic field at the muon site}
\label{appendix1}
 
We first express the measured local magnetic field in terms of the field
components in the crystal reference frame. Then we write these field components
in terms of their spatial Fourier components. Finally, we complete this 
appendix by 
providing a method of computing the field distribution at the muon site for a 
given magnetic structure.\par 

\subsection{The magnetic field in the laboratory and crystal 
reference frames}\label{appendix1_1}

Whereas we are interested in the characterization of the magnetic field
components at the muon site in terms of parameters of the crystal under study,
the measurements are done in the laboratory reference frame. Here
we express the linear relation between the field components in the two frames.
\par

For simplicity we suppose that the magnet has only one type of localized 
magnetic moment and that the muon occupies only one interstitial site. We must 
consider two orthonormal reference frames: the laboratory reference frame 
(${\bf X,Y,Z}$) where ${\bf X}$, ${\bf Y}$, and ${\bf Z}$ are unit vectors and 
a reference frame (${\bf x,y,z}$) attached to the crystal axes. Its unit 
vectors are chosen parallel to the crystal axes according to the symmetry of 
the compound. ($\bf X,Y,Z$) are defined in the ($\bf x,y,z$) frame through the 
Euler angles $\theta$, $\varphi$ and $\psi$. Note that $\theta$ and $\varphi$ 
are also the polar angles of the $\bf Z$ axis in the  (${\bf x,y,z}$) frame. The
components of the local field in the laboratory frame are written as a function 
of the components in the crystal frame: $B^\varrho_{\rm loc}$ = 
$\sum^{ }_\alpha R_{\varrho \alpha} (\theta, \varphi,\psi) 
\tilde B^\alpha_{\rm loc}$. We list the nine $R_{\varrho \alpha}$ components: 
\begin{eqnarray}
\fl
R_{x x} = \cos\psi\cos\varphi\cos\theta -\sin\psi\sin\varphi, 
R_{x y} = \cos\psi\sin\varphi\cos\theta + \sin\psi\cos\varphi, 
\label{appen101}
\end{eqnarray}

\begin{eqnarray}
\fl
R_{y x} =  -\sin\psi\cos\varphi\cos\theta -\cos\psi\sin\varphi, 
R_{y y} = -\sin\psi\sin\varphi\cos\theta + \cos\psi\cos\varphi, 
\label{appen102}
\end{eqnarray}

\begin{eqnarray}
\fl
R_{x z} = -\cos\psi\sin\theta, 
R_{y z} = \sin\psi\sin\theta,
\label{appen103}
\end{eqnarray}

\begin{eqnarray}
\fl
R_{z x} = \cos\varphi\sin\theta,
R_{z y} = \sin\varphi\sin\theta, 
R_{z z} = \cos\theta.
\label{appen104}
\end{eqnarray}

Since $R$ is a unitary matrix, it is easy to express
$\tilde  B^\alpha_{\rm loc}$ in terms of $B^\varrho_{\rm loc}$ using the matrix 
elements listed above. \par 

\subsection{The Fourier components of the magnetic field}\label{appendix1_2}

This appendix is based on an extension of the work of Yaouanc \etal 1993b.
The reader will consult with profit the paper of Lovesey and Engdahl 1995.\par

\begin{figure}
\centerline{\epsfbox{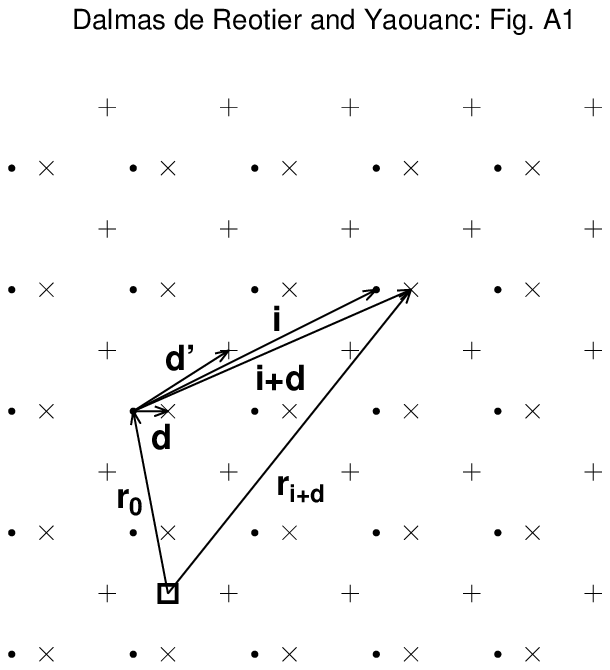}}
\caption{Definition of vectors relative to the crystallographic 
($+$, $\times$) and muon ($\Box$) sites. The points ($\bullet$) specify the
origin of each cell. The point at the extremity of ${\bf r}_0$ 
is the origin of the crystal lattice. The drawing is done for a two
dimensional square lattice with two magnetic atoms per cell. See the text for
the definitions of the different vectors.}
\label{cristal}
\end{figure}

We express the local field in the crystal frame as a function, on one hand, of
a tensor which describes the coupling between the localized spins of the magnet 
and the muon spin and, on the other hand, of the localized spins components 
themselves. The $\alpha$ component of the field is given by
\begin{eqnarray}
\tilde B^\alpha_{\rm loc}   ={\mu_ 0 \over 4\pi}{ g_L\mu_ B \over {n_d v}}
\sum^{ }_{ \beta =x,y,z}  \sum^{ }_ {i,d}
G^{\alpha \beta}_{ {\bf r}_{i+d}}J_{i+d}^\beta.
\label{appen110}
\end{eqnarray}
$v$ is the volume per magnetic atom, $n_d$ the number of magnetic atoms in the
cell used for the description of the magnet, $g_L$ the Land\'e factor of spin 
$ {\bf J}_{i+d}$ which is at the distance vector ${\bf r}_{i+d}$ from the muon, 
$\mu_ B$ the Bohr magneton, $\mu_ 0$ the permeability of free space and 
$ \{ \alpha ,\beta\} $=$ \{ x,y,z\} $. ${\bf r}_0$ defines the location of the
muon relative to the origin of the crystal lattice; see \fref{cristal}. {\sl 
Note
that the equivalent crystallographic sites allowed by the point group symmetry
at the muon may give rise to
different $G^{\alpha \beta}_d ({\bf q})$ and therefore to inequivalent magnetic
sites, each characterized by a given ${\bf r}_0$}. The index $i$ runs over the 
cells and $d$ over the sites inside a cell. Since one of the most important 
characteristics of a magnetic structure is its periodicity, we introduce the 
spatial-Fourier component $J^{\alpha}_d({\bf q})$. Using the vectors defined in 
\fref{cristal} we derive
\begin{eqnarray}
\fl
\tilde B^\alpha_{\rm loc} = {\mu_ 0 \over 4\pi}{g_L\mu_ B \over n_d v} 
\sum^{ }_{ \beta =x,y,z} \sum^{ }_ d \sum^{ }_{\bf q}{ 
G^{\alpha \beta}_d ({\bf q}) \exp(-i {\bf q} \cdot {\bf r}_0) 
J^{\beta}_d ({\bf q})},
\label{appen111}
\end{eqnarray}
where we have defined
\begin{eqnarray}
G^{\alpha \beta}_d ({\bf q})  = \sum^{ }_ i G^{\alpha \beta}_{{\bf r}_{i+d}}
\exp (i {\bf q} \cdot {\bf r}_{i+d}),
\label{appen112}
\end{eqnarray}
and
\begin{eqnarray}
J^{\beta}_d ({\bf q}) = \sum^{ }_ i J^{\beta}_{i+d}
\exp [- i {\bf q} \cdot ({\bf i} + {\bf d} )].
\label{appen112extra}
\end{eqnarray}
\par

The tensor $G^{\alpha \beta}_d ({\bf q})$ is the sum of two tensors:
$G^{\alpha \beta}_d ({\bf q})$ = $D^{\alpha \beta}_d ({\bf q})$ +
$H^{\alpha \beta}_d ({\bf q})$. Whereas $D^{\alpha \beta}_d ({\bf q})$
describes the dipolar interaction between the spins of the magnet and the muon 
spin, $H^{\alpha \beta}_d ({\bf q})$ accounts for their hyperfine interaction.
It is convenient to rewrite
\begin{eqnarray}
D^{\alpha \beta}_d( {\bf q}) & = &
-4\pi \left[P^{\alpha \beta}_ L( {\bf q} )-C^{\alpha \beta}_d ( {\bf q} )
\right],
\label{appen113}
\end{eqnarray}
where $P^{\alpha \beta}_ L( {\bf q} )$ is the longitudinal projection
operator and C$^{\alpha \beta}_d ( {\bf q} )$ a symmetric tensor which reflects
the point group symmetry at the muon site. The expression of 
C$^{\alpha \beta}_d ( {\bf q} )$ is computed using Ewald's method:
\begin{eqnarray}
\fl
C^{\alpha \beta}_d ( {\bf q} )  = 
{ q_\alpha q_\beta \over q^2} \left[1-\exp \left({-q^2
\over 4\varrho^ 2} \right) \right] \cr
\fl
-  {1 \over 4\varrho^ 2} \sum^{}_{ {\bf K} \not= {\bf 0}}
\left(K_\alpha +q_\alpha \right) \left(K_\beta
+q_\beta \right)\varphi_ 0 \left({( {\bf q} + {\bf K})^2 \over 4\varrho^ 2}
\right)\exp \left(-i {\bf K} \cdot {\bf r}_{0+d} \right)   \cr
\fl
+  {n_d v \varrho^ 3
\over 2(\pi)^{3/2}} \sum^{ }_ {i} 
\left[2\varrho^ 2r_{i+d,\alpha} r_{i+d,\beta}
\varphi_{ 3/2} \left(\varrho^ 2r^2_{i+d} \right)-
\delta^{\alpha\beta}\varphi_{1/2} \left(\varrho^ 2r^2_{i+d} \right) 
\right]\exp \left(i {\bf q} \cdot {\bf r}_{i+d} \right),
\label{appen113bis}
\end{eqnarray}
where we use the Misra functions:
\begin{eqnarray}
\varphi_ m(x)  = \int^{ \infty}_ 1 d\beta  \beta^ m \exp(-\beta x).
\label{appen113tri}
\end{eqnarray}
$\bf K$ is a vector of the reciprocal lattice.
Expression (\ref{appen113bis}) gives the same result for all values of the 
Ewald parameter $\varrho$, but for numerical applications a value of $\varrho$ 
is chosen which ensures that both series of \eref{appen113bis} converge 
rapidly.\par

Whereas $P^{\alpha \beta}_L( {\bf q} )$ is only piecewise continous at 
${\bf q = 0}$, reflecting the long range nature of the dipolar interaction, 
$C^{\alpha \beta}_d ( {\bf q} )$ is analytical. The presence of the projection 
operator in \eref{appen113} is important for understanding the critical 
behaviour of the relaxation rate in ferromagnets and the temperature dependence 
of this rate in the spin wave regime (see \sref{dipolar}).\par

In \sref{field} we have noted that ${\bf B}_{\rm loc}$ can be written as a sum
of seven terms. We can identify the first five terms. 
$({\bf B}_{\rm con} +{\bf B}_{\rm trans})$ results from the polarized 
conduction electrons and the transferred fields. To first approximation, these 
two fields are isotropic in metals. The 
$({\bf B}_{\rm con} +{\bf B}_{\rm trans})$ contribution to the local field is 
obtained by substituting ${\rm H}$ for ${\rm G}$ in \eref{appen111}. In the 
same way, the
$({\bf B}_{\rm dip} ^{\prime} + {\bf B}_{\rm L})$ and ${\bf B}_{\rm dem}$
contributions are derived from \eref{appen111} by substituting 
$4 \pi {\rm C}$ and 
$- 4 \pi {\rm P}_L$ for ${\rm G}$, respectively. \par

Although the spatial-Fourier transform may seem a complicated procedure for a 
result which can be obtained by a simple lattice sum, \ref{appendix1_3} and 
\ref{appendix2} will show its unique possibilities.\par

\subsection{The field distribution at the muon site}\label{appendix1_3}

Even for a simple magnetic structure, crystallographically equivalent sites may
not be magnetically equivalent, i.e. more than one local magnetic field may
exist. This is nicely explained
for the case of Fe by Seeger 1978. If the number of fields is
sufficiently large and their values are close, the muons will probe a field 
distribution which may be far from Gaussian as supposed in \sref{muon}. 
The relation between a magnetic model and the field distribution at the muon 
site, $D({\bf B}_{\rm loc})$, was first investigated by Szeto 1987. \par

In an experiment the several million implanted muons may enter different
crystal sites which are related by the lattice translation periodicity. 
Therefore an average over these $\bf r$ sites is required: ${\bf r}$ =
${\bf r}_0+ l {\bf x} + m {\bf y} + n {\bf z} $ where
$\{l,m,n\}$ are integers. We recall here that if the point group symmetry of
the muon site gives several equivalent positions, one has to consider each of
these as they may give rise to different coupling tensors 
${\rm G}_d ({\bf q})$ i.e. to non equivalent magnetic sites.\par

For the sake of simplicity, we consider a magnetic structure described by a 
single propagation vector $\bf k$. This represents the majority of 
structures found in nature, and the extension to multi-$\bf k$ structures is
easily performed. Then the Fourier components of the magnetic moment may be
written as
\begin{eqnarray}
{\bf M}_d({\bf q}) = \sum_p {{\bf M}_d (p{\bf k}) \delta({\bf q}- p{\bf k})},
\label{appen114}
\end{eqnarray}
where $\delta$ is the Kronecker symbol. The sum is over the harmonics of the 
magnetic structure; $p= \pm 1$ for a sinusoidal modulation. 
Note that ${\bf M}_d({\bf q})$ is related to ${\bf J}_d({\bf q})$ 
via the relation
${\bf M}_d({\bf q})$ = $g_L\mu_ B {\bf J}_d({\bf q})$.
Using \eref{appen114} and \eref{appen111}, we derive
\begin{eqnarray}
\fl
\tilde B^\alpha_{\rm loc} ({\bf r})   ={\mu_ 0 \over 4\pi} {1\over n_d v}
\sum^{ }_ d \sum_p \sum^{ }_{ \beta =x,y,z} 
G^{\alpha \beta}_d (p{\bf k}) M^{\beta}_d (p{\bf k}) 
\exp(-i p{\bf k} \cdot {\bf r}) 
\label{append115}
\end{eqnarray}
For a commensurate magnetic structure ${\bf k}$ = 
$(r/s) {\bf G}$ where $r$ and $s$ are integers with $r \leq s$ and ${\bf G}$ a 
reciprocal lattice vector. From the well known properties of ${\bf G}$, we 
derive ${\bf k} \cdot {\bf r}$ = ${\bf k} \cdot {\bf r}_0$ + $ 2 \pi (r/s) u$ 
where $u$ is an integer. Using this result, the exponential term of 
\eref{append115} is rewritten as
\begin{eqnarray}
\exp[-i p{\bf k} \cdot {\bf r} ] =
\exp(-i p{\bf k} \cdot {\bf r}_0 )
\left[\exp \left(i { 2 \pi p r \over s} \right)\right]^u.
\label{appen116}
\end{eqnarray}
During a measurement, an ensemble average is made, so that the integer $u$ takes
an enormous number of values. But, since
$\left\{\exp \left[i ( 2 \pi p r)/ s) \right]\right\}^u$ as a function of $u$ 
is a periodic function of period $s$, this phase factor takes a maximum of 
$s$ different values.\par 

Since ${\bf B}^\alpha_{\rm loc} ({\bf r})$ depends linearly on 
$\tilde B^\varrho_{\rm loc}({\bf r})$ and this latter field is a sum of Fourier 
components, ${\bf B}_{\rm loc} ({\bf r})$ is a sum Fourier components. Taking
into account the property of the phase factor, we deduce that each component
can take many different values. \par

We discuss this important result. For simplicity we consider a primitive
Bravais lattice
($n_d = 1$), a sinusoidal modulation and suppose that the laboratory 
and crystal frames are aligned with $\bf Z$ and $\bf z$ parallel. 
For instance, if $r$ = 1 and $s$ = 3, up to three different fields can exist.
If $s$ is big, a large number of fields can result and in the limit, a 
quasi-continuous distribution is generated. \par

To proceed further analytically we consider this limit and write 
${\bf B}_{\rm loc}({\bf r})$ = $\cos[2\pi w({\bf r})]{\bf B}_m$ with 
$0 \leq w \leq 1$. This form is strictly valid for an incommensurate magnetic
structure. Using the well known formula for a distribution  
\begin{eqnarray}  
D \left({\bf B}_{\rm loc} \right) = \left \langle \left \langle \delta 
\left[{\bf B}_{\rm loc} -{\bf B}_{\rm loc} ({\bf r})  \right] 
\right \rangle  \right \rangle,
\label{appen117}
\end{eqnarray}
where $\langle \langle ... \rangle  \rangle$ stands for the spatial average 
over ${\bf r}$, we derive (Le \etal 1993a)
\begin{eqnarray}  
D \left({\bf B}_{\rm loc} \right) = {2 \over \pi}
{1 \over \sqrt{B_m^2 -B_{\rm loc}^2}},
\label{appen118}
\end{eqnarray}
for $0 \leq B_{\rm loc} \leq B_m$ and $D \left({\bf B}_{\rm loc} \right)= 0$
otherwise. If we suppose that ${\bf B}_{\rm loc}$ is perpendicular to $Z$, we 
find using \eref{distribution} 
\begin{eqnarray}
P_Z(t) = J_0(\gamma_\mu B_m t),  
\label{appen119}
\end{eqnarray}
where $J_0$ is a Bessel function. When $\gamma_\mu B_m t$ is large, the model 
predicts $P_Z(t)$ =
$\sqrt{2/ (\pi \gamma_\mu B_m t)}\cos(\gamma_\mu B_m  t - \pi/4)$ 
instead of the usual $P_Z(t)$ = $\cos(\gamma_\mu B_m t)$, i.e. the
depolarization function presents at large $t$ a damped oscillation shifted by 
45$^\circ$. \par

If the muon site is known, using the $\tilde B^\alpha_{\rm loc} ({\bf r})$ 
expression \eref{append115} it is possible to compute numerically the field 
distribution for any magnetic structure. While with the $\mu$SR technique it is
not possible to determine in detail a magnetic structure, one can test proposed 
structures.\par

\section{Muon spin relaxation in a longitudinal field}
\label{appendix2bis}

When the dynamics of the magnetic field at the muon site is sufficiently rapid, 
$P_Z(t)$ takes the form
(McMullen and Zaremba 1978 and Dalmas de R\'eotier and Yaouanc 1992)
\begin{eqnarray}
P_Z(t) = \exp \left[ - \psi_Z(t) \right],
\label{extra1}
\end{eqnarray}
with
\begin{eqnarray}
\psi_Z(t) = {\gamma^ 2_\mu} \int^{ t}_{0} & d\tau & \ (t-\tau)\left\{
\cos (\omega_\mu \tau) \left[ \Phi^{ XX}(\tau)+\Phi^{ YY}(\tau) \right] \right.
\cr
 & + &
\left.
\sin (\omega_\mu \tau) \left[ \Phi^{ XY}(\tau)-\Phi^{ YX}(\tau) \right]
\right\} .
\label{extra2}
\end{eqnarray}
$\Phi(\tau)$ is the symmetrized correlation-tensor of the magnetic field at
the muon site: 
\begin{eqnarray}
\Phi^{ \alpha \beta} (\tau )  ={1 \over 2} \left[
\left\langle B^\alpha_{\rm loc} (\tau ) B^\beta_{\rm loc} \right\rangle +
\left\langle B^\beta_{\rm loc} B^\alpha_{\rm loc} (\tau ) \right\rangle \right] 
\label{extra3}
\end{eqnarray}
$\left\langle ... \right\rangle$ stands for the thermal average.\par

If the fluctuations are sufficiently rapid, one neglects $\tau$ in the 
$(t-\tau)$ factor and extends the integral to infinity. Then $P_Z(t)$ is an
exponential function characterized by the relaxation rate $\lambda_Z$ = 
$\psi_Z(t)/t$.\par

In zero-field $\Phi^{ \alpha \beta}(\tau)$ is an even function of $t$. In an
applied field, this may not be true because time reversal symmetry is broken. 
However, in many cases the breaking
terms in the Hamiltonian are so small that the effect of the field is 
negligible. Then, even in an applied field, $\Phi^{ \alpha \beta}(\tau)$ is an
even function of $t$. We will suppose that this property holds.
This enables the integration over $\tau$ in
\eref{extra2} to be extended from $-\infty$ to $\infty$. \par

In zero field $\lambda_Z$ is given by
\begin{eqnarray}
\lambda_ Z  = {\gamma^2_\mu \over 2} \int^{ \infty}_{-\infty}
d\tau  \left[\Phi^{ XX}(\tau )+\Phi^{ YY}(\tau ) \right].
\label{extra4}
\end{eqnarray}
This simple result can be derived from the Fermi golden rule 
(Lovesey \etal 1992). 

In a longitudinal field, a term proportional to $\sin(\omega_\mu \tau)$ is
present as seen in \eref{extra2}. In most cases this term is zero either 
because time reversal symmetry is not broken or $\omega_\mu \tau_c \ll 1$ where 
$\tau_c$ is a characteristic time of the fluctuations. Then we have
\begin{eqnarray}
\lambda_ Z  = {\gamma^2_\mu \over 2} \int^{ \infty}_{-\infty}
d\tau \cos \left(\omega_\mu \tau \right)
\left[\Phi^{ XX}(\tau )+\Phi^{ YY}(\tau ) \right].
\label{extra5}
\end{eqnarray}

As stressed by Dalmas de R\'eotier \etal 1996, $P_Z(t)$ is, in general, an 
exponential function only if no spatial average of the depolarization function 
is needed. This means that $P_Z(t)$ has, in general, no reason to be an 
exponential function for measurements on a polycrystalline sample.\par 

\section{Longitudinal relaxation rate and correlation-functions}
\label{appendix2}

In this section we analyze the relaxation rate in terms of the spin-spin
correlation-tensor of the magnet.\par

Using the results presented in the two previous sections, $\Phi(\tau)$ can be
written in terms of correlation-functions of the spatial Fourier components of 
the lattice spins. It is useful to introduce the time-Fourier transform of a 
function $ f(\tau ) $:
\begin{eqnarray}
f(\omega )  = {1 \over 2\pi} \int^{ \infty}_{ -\infty} d\tau
{\bf \ } \exp(-i\omega \tau )f(\tau ).
\label{extra6}
\end{eqnarray}
Using \eref{appen111} and \eref{extra5},
$\lambda_Z$ can be expressed in terms of the spin-spin correlation-tensor 
between spins belonging to sublattices $d$ and $d'$, 
$\Lambda_{d,d'} ( {\bf q} ,\omega )$. With the definition
\begin{eqnarray}
\fl
\Lambda^{ \gamma \gamma^\prime}_{dd'}( {\bf q}, \omega) =  
{1 \over 2} \left[
\left\langle J^{\gamma}_d ({\bf q}, \omega ) 
J^{\gamma^\prime}_{d^\prime}({\bf -q}) \right\rangle +
\left\langle J^{\gamma^\prime}_{d^\prime}({\bf -q})
J^\gamma_d ({\bf q}, \omega)\right\rangle \right],
\label{extra7}
\end{eqnarray}
the $\lambda_Z$ expression writes
\begin{eqnarray}
\fl
\lambda_ Z  = {\pi{\cal D} \over V} 
\sum^{ }_ { \beta ,\alpha} L_{\beta\alpha}(\theta ,\varphi)
\int^{ }_{ }{ d^3 {\bf q} \over( 2\pi)^ 3} 
\sum^{ }_{\gamma,\gamma^\prime}  \sum^{ }_{d,d'} 
G^{\alpha \gamma}_{dd'}( {\bf q}) G^{\gamma^\prime \beta}_{dd'}( - {\bf q})
\Lambda^{ \gamma \gamma^\prime}_{dd'}( {\bf q}, \omega_\mu). 
\label{appen2_01}
\end{eqnarray}
The sum over {\bf q} in \eref{appen111} has been replaced by an integral over
the first Brillouin zone following the usual rule $\sum ^{} _{\bf q}$
$\rightarrow$ $\int V/(2\pi)^3\, d^3{\bf q}$ where $V$ is the volume of the 
sample. We define 
$ {\cal D}= \left(\mu_ 0/4\pi \right)^2\gamma^ 2_\mu \left(g_L\mu_ B\right)^2$.
$ L(\theta ,\varphi )$ is a symmetric matrix which accounts for the fact that
the symmetry axes of the magnet do not necessarily coincide with the
laboratory frame axes. $\theta$ and $\varphi$ have been defined in
\ref{appendix1_1}. We have
\begin{eqnarray}
\fl
L_{xx}& = & \cos^2\varphi \cos^2\theta +\sin^2\varphi ,
L_{yy}=\sin^2\varphi \cos^2\theta +\cos^2\varphi ,
L_{zz}=\sin^2\theta ,
\label{appen2_02}
\end{eqnarray}
\begin{eqnarray}
\fl
L_{xy}=-\cos\varphi \sin\varphi \sin^2\theta ,
L_{xz}=-\cos\varphi \cos\theta \sin\theta ,
L_{yz}=-\sin\varphi \cos\theta \sin\theta.
\label{appen2_3}
\end{eqnarray}

It is often useful to consider $\tilde  \Lambda ( {\bf q} )$ =
$\sum^{ }_{ d,d'} \tilde  \Lambda _{d,d'} ( {\bf q} )$ and the average coupling
tensor ${\rm G}({\bf q}) = {1/n_d}\sum _{d} {\rm G}_{d}({\bf q})$ since for 
some non
Bravais crystal structures such as the hexagonal closed compact structure
(Dalmas de R\'eotier and Yaouanc 1994) 
\eref{appen2_01} can be written in terms of $\tilde  \Lambda ( {\bf q} )$
and ${\rm G}({\bf q})$, i.e. the sum over $d$ and $d^\prime$ disappears in the 
expression of $\lambda_Z$.\par

The symmetry properties at the muon site and of the magnet itself leads to 
considerable simplification. Examples are
found in Yaouanc \etal 1993a and 1993b, 
Dalmas de R\'eotier and Yaouanc 1994, Bonville \etal 1996, 
Dalmas de R\'eotier \etal 1996, Yaouanc \etal 1996a and in 
\sref{dipolar}.\par

As an example of the drastic simplifications which can occur, we suppose that
only one spin-spin correlation matters. This is the case for an Ising system or 
a magnet with its magnetic ions in a cubic environment. Then the following 
formula can be derived:
\begin{eqnarray}
\fl
\lambda_ Z  = v \int^{ }_{ }{ d^3 {\bf q} \over( 2\pi)^ 3} \Delta^2 ({\bf q})
{\int^{\infty}_{-\infty} d\tau \cos(\omega_\mu \tau)
{\left\langle J^z ({\bf q}, \tau) J^z ({\bf -q}) \right\rangle}
\over J(J+1)/3}.
\label{extra8}
\end{eqnarray}  
$\lambda_Z$ depends on the function $\Delta^2 ({\bf q})$ which accounts for 
the coupling between the muon and the lattice spins. In zero-field, when the 
intersite
correlations are neglected ($\Delta^2 ({\bf q})$ = $\Delta^2 \delta ({\bf q})$)
and the spin-spin-correlation is an exponential function characterized by a 
fluctuation rate $\nu$, we recover the motional narrowing result: 
$\lambda_Z$ = $2 \Delta^2 /\nu$, see \eref{explong}.\par

\section{The $\mu$SR response function for hard superconductors}
\label{appendix3}

In this appendix we calculate the magnetic field inside the mixed state of a 
hard superconductor. For such a superconductor the Ginzburg-Landau parameter 
$\kappa$ is much larger than 1 ($\kappa$ = $\lambda/\xi$ where $\lambda$ 
is the penetration 
length and $\xi$ the coherence length).\par

It is well known that a vortex lattice is formed in the mixed state. We suppose 
that this lattice is ideal. This means we disregard pinning and vortex 
``phases'' such as the glassy or liquid states  
(Blatter \etal 1994, Brandt 1995). Pinning can strongly influence the results
of a $\mu$SR experiment as shown in details by Wu \etal 1993.
We consider the case where the field 
${\bf B}_{\rm ext}$ is parallel to one of the three main axes  $\bf a$, $\bf b$ 
and $\bf c$ of the penetration-length tensor. With these simple conditions, the 
vortices are straight field tubes parallel to ${\bf B}_{\rm ext}$ and form a 
regular lattice. We define an orthonormal reference frame ($\bf x,y,z$) such 
that the vortex tubes are also along $z$. We denote $\lambda_a$, $\lambda_b$ 
and $\lambda_c$, the penetration lengths for currents flowing along the $a$, 
$b$ and $c$ axes, respectively. \par

As shown by Kogan (Kogan 1981), for a conventional vortex lattice and 
$B_{\rm ext}\gg B_{c1}$ where $B_{c1}$ is the lower critical field, the angle 
characterizing the lattice depends only on the penetration-length ratio :
\begin{eqnarray} 
\tan \alpha_v & = & {\sqrt 3} (\lambda_x / \lambda_y).
\label{appen3_01}
\end{eqnarray}
As expected, if the penetration length is isotropic, $\alpha_v$ = 60$^\circ$.
Therefore, although $\alpha_v$ is most naturally measured by small angle 
neutron scattering (SANS), one can determine this angle by $\mu$SR. An example 
is given in \sref{upt3}.\par

It is convenient to introduce the Fourier components 
\begin{eqnarray} 
{\bf B(G)} =  \int {\bf B(r)} \exp(-i{\bf G \cdot r}) d^2 {\bf r}/S
\label{appen3_02}
\end{eqnarray}
of the periodic magnetic field  
\begin{eqnarray}
{\bf B(r)} = \sum_{\bf G} {\bf B(G)} \exp(i{\bf G \cdot r}),
\label{appen3_03}
\end{eqnarray}
where $\bf G$ are the vectors of the vortex reciprocal lattice 
and $S$ the surface of 
the vortex lattice unit cell. Since $B_x( {\bf G})$ = $B_y( {\bf G})$ = 0, 
there is no transverse field component. Interestingly, the only non-zero 
component, $B_z({\bf G})$, is the form factor measured by SANS. \par

Using Kogan's formula, the form factor factorizes (Yaouanc \etal 1997a), 
$B_z({G_{pq}})$ = $B_0 \cdot c_{pq}(b)$, where
\begin{eqnarray}  
B_0 & = & {1 \over \pi^2} \left ( { 3 \over 64} \right)^{1/2}
{\Phi_0 \over \lambda_x \lambda_y}.
\label{appen3_04}
\end{eqnarray}
$c_{pq}(b)$ are universal functions of $b = B/B_{c2}$ where $B$ is the mean 
induction, which for $2b\kappa^2 > 1$ may be equated to $B_{\rm ext}$. 
$\{p,q\}$ are the two indices denoting a Bragg peak. 
Recently the $c_{pq}(b)$ functions have been computed numerically in the 
framework of the conventional Ginzburg-Landau theory (Brandt 1997). In terms of
the functions $b_{pq}(b)$ defined by Brandt, we have $c_{pq}(b)$ = 
$b_{pq}(b)/(p^2 -pq +q^2)$.\par

\begin{figure}
\centerline{\epsfbox{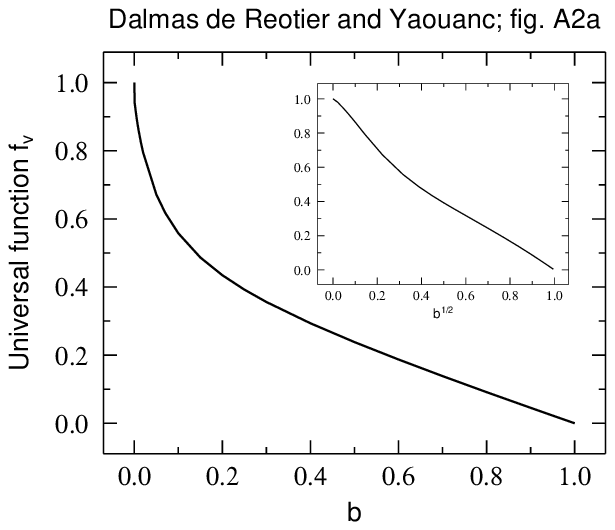}\hfill
\epsfbox{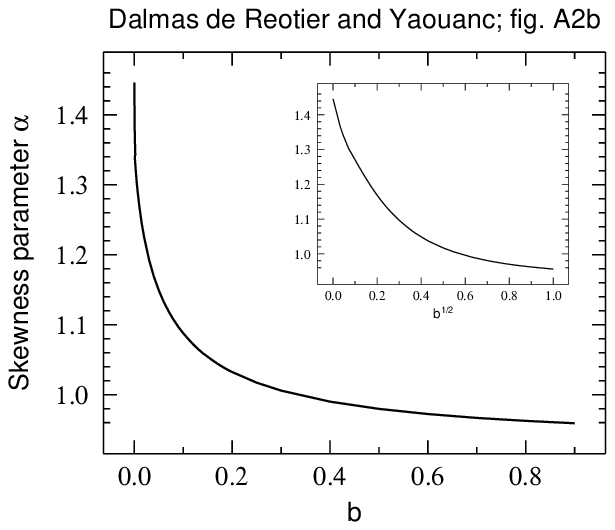}}
\caption{Field dependence of the functions $f_v$ and $\alpha$. 
$b\equiv B/B_{c2}$ where $B$ is the mean induction, which for 
$2b\kappa^2 > 1$ may be equated to $B_{\rm ext}$. In the inserts we present the
functions versus $b^{1/2}$ to extend the low $b$ part.
(adapted from Yaouanc \etal 1997a and Brandt \etal 1997).}
\label{brandt}
\end{figure}

Since a typical penetration length is much larger than a crystal lattice
parameter which is the characteristic distance between adjacent muon stopping
sites, the muon is a good probe for the vortex field distribution. Then 
it is straightforward to compute the distribution using 
\eref{appen117}. This has been done by many authors (for example, 
Sonier \etal 1994 and Greer and Kossler 1995). A characteristic of the flux
line lattice observed in many experiments (for example, 
see \fref{sonier_DISTRIBUTION} and \fref{AEGERTER_champ}) is the
pronounced tail towards high fields due to the vortex cores. A useful 
method to characterize a
distribution is to consider its moments or, since $\langle B_z \rangle$ $\neq$
0, $\Delta_v^n$ = $\langle (B_z - \langle B_z \rangle)^n \rangle$.
The variance which is a measure of the width of the distribution separates into 
two factors, $\sqrt{ \Delta_v^2}$ = $\Delta_0 \cdot f_v (b)$ where
 \begin{eqnarray}  
\Delta_0 & = &0.06092 {\Phi_0 \over \lambda_x \lambda_y}
 \label{appen3_06}
 \end{eqnarray}
is the London limit ($\xi_x,\, \xi_y \to 0$) (Barford and Gunn 1988)
and $f_v(b)$ a universal function which accounts for the core size.
$\Phi_0$ = 2.07 $\times$ 10$^{-15}$ Tm$^2$ is the quantum of flux and the
prefactor 0.06092 is a pure number. 
$f_v(b)$ computed using the data of Brandt 1997 is plotted in
\fref{brandt}. It is strongly field dependent even at low reduced fields 
$b$. The shape of a distribution is characterized by its skewness 
parameter : $\alpha$ = $(\Delta_v^3)^{1/3}/(\Delta_v^2)^{1/2}$. For a symmetric 
distribution $\alpha$ = 0. $\alpha (b)$ is also 
presented in \fref{brandt}. Note the strong field dependence of $\alpha$ near
$b = 0$. We have $\alpha(b = 0)$ = 1.446.\par

The results presented in \fref{brandt} have been deduced from the numerical
solution of the conventional Ginzburg-Landau expansion (Brandt 1997). Although
this expansion has its own limit (basically it is  valid close to the 
superconducting phase transition in the temperature-field phase diagram), 
a comparison between the results derived from the Ginzburg-Landau model and the 
commonly used Gaussian cutoff London model shows that the latter model strongly 
underestimates the effect of the field on the variance (Yaouanc \etal 1997a). 
In fact, the Gaussian model does not have any theoretical support.

The motion of the vortices on a time scale shorter than a few microsecond may
significantly effect the field distribution as probed by muons. 
Since the amplitude of such
fluctuations is field dependent, the variance can exhibits a strong field
dependence as shown by Song 1995. \par

The Ginzburg-Landau model does not include the possibility of a Knight shift.
Because this shift is expected to be different in the superconducting and in 
the normal regions, and since normal (vortex core) and superconducting regions 
coexist in the mixed state, the measured field distribution is a convolution 
of the distributions due to the Knight shift and the flux line 
lattice. This point was first raised by Feyerherm \etal 1994.\par

Since a SANS experiment measures $|B_z({\bf G})|$, one may estimate that SANS 
and $\mu$SR experiments measure the same physical quantity. This idea seems to 
be supported by the fact that $c_{10}$ and $f_v$ have approximately the same 
field dependence (Yaouanc \etal 1997a). But there is at least one important
difference. In order to observe a Bragg peak by SANS, the correlation length of
the vortex lattice must be sufficient. 
This is not required to observe a vortex
field distribution by $\mu$SR which is a local technique. \par

A ``pure'' field distribution is not observed because of broadening effects
due to the finite lifetime of the muon, the defects of the vortex lattice, the 
demagnetization field and the nuclear dipolar broadening. We refer the reader 
to Riseman \etal 1995 and Schneider \etal 1995 for more information.\par

Up to now, we have only considered a three dimensional vortex lattice. But when
the anisotropy of the penetration length is strong, it is more appropriate to
describe the vortices as two dimensional ``pancakes'' vortices in individual
coupled layers. Experimentally, Bi$_{2 +x}$Sr$_{2 - x}$CaCu$_2$O$_{8+ \delta}$
corresponds to this case (see \sref{fusion}). As first noticed by 
Brandt 1991, the misalignment between the layers leads to a variance 
smaller than for a standard three dimensional flux line lattice. A comparison
between simulated field distributions for three and two dimensional vortex 
lattices has been made by Schneider \etal 1995.\par

One of the most interesting parameters deduced from the investigation of a 3D
field distribution is the penetration length $\lambda$; indeed its 
determination is one way to probe the nature of the low energy excitations and 
therefore the pairing state of a superconductor. The relation between 
$\lambda$ and these excitations is reviewed by Gross \etal 1986 and 
Gross-Alltag \etal 1991.\par

\ack
We would like first to take the opportunity to thank the accelerator crew 
and the instrument scientists at the ISIS and PSI facilities for 
enabling us to perform $\mu$SR experiments.
It is a pleasure to acknowledge helpful discussions with C M Aegerter, 
C Bernhard and Ch Niedermayer. Many thanks go to P Bonville, J P Boucher, 
I A Campbell, S F J Cox, P C M Gubbens, J A Hodges, M Lavagna and S W Lovesey 
who carefully read the whole or part of the manuscript. 
We are indebted to the following colleagues for providing us with data used to 
draw figures presented in this review : C M Aegerter, A Amato, S J Blundell, 
A Keren, K Kojima, G Solt and J E Sonier.\par 
\clearpage
\section*{References}
\begin{harvard}
\item[] Aegerter C M, Lee S L, Keller H, Forgan E M and Lloyd S H 1996 \PR B
{\bf 54} R15661--4
\item[] Aegerter C M 1997, private communication
\item[] Aeppli G, Yoshizawa H, Endoh Y, Bucher E, Hufnagl J, Onuki Y and
Komatsubara T 1986 \PRL {\bf 57} 122--5 
\item[] Aeppli G, Bucher E, Broholm C, Kjems J K, Baumann J and Hufnagl J 1988 
\PRL {\bf 60} 615--8
\item[] Amato A, Baines C, Feyerherm R, Flouquet J, Gygax F N, Lejay P, 
Schenck A and Zimmermann U 1993 {\it Physica} B {\bf  186$\&$188} 276--8
\item[] Amato A, Feyerherm R, Gygax F N, Schenck A, Flouquet J and Lejay P
1994 \PR B {\bf 50} 619--22
\item[] Amato A, Feyerherm R, Gygax F N, Schenck A, L\"ohneysen H v and
Schlager H G 1995 \PR B {\bf 52} 54--6
\item[] Amato A, Feyerherm R, Gygax F N and Schenck A, 1997, {\it Hyperfine 
Interact.} {\bf 104} 165--70
\item[] Amato A, 1997 private communication
\item[] Anderson P W 1954 \JPSJ {\bf 9} 316--39
\item[] Aristov D N and Maleyev S V 1994 \ZP B {\bf 93} 181--7
\item[] Attenborough M, Hall I, Nikolov O, Brown S R and Cox S F J 1996
\PR B {\bf 54} 6448--56
\item[] Barford W and Gunn J M F 1988 {\it Physica} C {\bf 156} 515--22
\item[] Barsov S G, Gasnikova G P, Getalov A L, Koptev V P, Kotov S A, Kuz'min
L A, Men'shikov A Z, Mikirtych'yants S M and Shcherbakov G V
1993 JETP lett. {\bf 57} 672--5
\item[] Barsov S G, Getalov A L, Koptev V P, Kotov S A, Kuz'min L A, 
Mikirtych'yants S M and Shcherbakov G V
1994 JETP lett. {\bf 60} 796--9
\item[] Basov D N, Liang R, Bonn D A, Hardy W N, Dabrowski B, Quijada M, Tanner
D B, Rice J P, Ginsberg D M and Timusk T 1995 \PRL {\bf 74} 598--601
\item[] Bernal O O, MacLaughlin D E, Amato A, Feyerherm R, Gygax FN, Schenck A,
Heffner R H, Le L P, Nieuwenhuys G J, Andraka B, L\"ohneysen H v, Stockert O
and Ott H R 1996 \PR B {\bf 54} 13000--8 
\item[] Bernhard C, Wenger C, Niedermayer Ch, Pooke D M, Tallon J, 
Kotaka Y, Shimoyama J, Kishio K, Noakes D R, Stronach C E, Sembiring T and 
Ansaldo E J 1995a \PR B {\bf 52} R7050--3 
\item[] Bernhard C, Niedermayer Ch, Binninger U, Hofer A, Wenger Ch, Tallon J
L, Williams G V M, Ansaldo E J, Budnick J I, Stronach C E, Noakes D R and
Blankson-Mills M A 1995b \PR B {\bf 52} 10488--98
\item[] Bernhard C, Tallon J L, Bucci C, De Renzi R, Guidi G, Williams G V M
and Niedermayer Ch 1996 \PRL {\bf 77} 2304--7
\item[] Berzin A A, Morozov A I and Sigov A S 1993 {\it Phys. Solid state} 
{\bf 35} 1463--5
\item[] Blatter G, Feigel'man M V, Geshkenbein V B, Larkin A I and Vinokur V M 
1994 \RMP {\bf 66} 1125--388
\item[] Blundell S J, Pattenden P A, Valladares R M, Pratt F L,  
Sugano T and Hayes W 1994 {\it Solid State Com.} {\bf 92} 569--72
\item[] Blundell S J, Pattenden P A, Pratt F L, Valladares R M, Sugano T and
Hayes W 1995 {\it Europhys. Lett.} {\bf 31} 573--8
\item[] Blundell S J, Sugano T, Pattenden P A, Pratt F L, Valladares R M, 
Chow K H, Uekusa H, Ohashi Y and Hayes W 1996 \JPCM {\bf 8}
L1--6
\item[] Blundell S J, Pratt F L, Pattenden P A, Kurmoo M, Chow K H, Takagi S,
Jest\"adt Th and Hayes W 1997a \JPCM {\bf 9} L119--24
\item[] Blundell S J, Pattenden P A, Pratt F L, Chow K H, Hayes W and Sugano T,
1997b, {\it Hyperfine Interact.} {\bf 104} 251--6 
\item[] Birrer P, Gygax F N, Hitti B, Lippelt E, Schenck A, Weber M, Cattani D,
Cors J, Decroux M and Fischer \O\ 1993 \PR B {\bf 48} 16589--99
\item[] Birrer P, Torikai E and Nishiyama K 1995 \JMMM {\bf 140-144} 
2177--8
\item[] Bishop D 1996 {\it Nature} {\bf 382} 760--1 and {\it Science} {\bf 273} 
1811--1
\item[] Bonville P, Dalmas de R\'eotier P, Yaouanc A, Polatsek G, Gubbens P C
M and Mulders A M 1996 \JPCM {\bf 8} 7755-70
\item[] Bonville P, LeBras G, Dalmas de R\'eotier P, Yaouanc A, Calemczuk R,
Paulsen C, Kasaya M and Aliev F G 1997 {\it Physica} B {\bf  230$\&$232} 266--8
\item[] Borgs P, Kehr K W and Heitjans P 1995 \PR B {\bf 52} 6668--83 
\item[] Borsa F, Carretta P, Cho J H, Chou F C, Hu Q, Johnston D C, Lascialfari
A, Torgeson D R, Gooding R J, Salem N M and Vos K J E 1995 \PR B {\bf 52}
7334--45
\item[] Brandt E H 1991 \PRL {\bf 66} 3213--6
\item[] \dash 1995 {\it Rep. Prog. Phys.} {\bf 58} 1465--594
\item[] \dash 1997 \PRL {\bf 78} 2208--11
\item[] Brandt E H, Dalmas de R\'eotier P and Yaouanc A 1997, private
communication
\item[] Brown S R, Attenborough M, Hall I, Nikolov O and Cox S F J 1996
{\it Phil. Mag. Lett.}  {\bf 73} 195--9
\item[] Cameron L M and Sholl C A 1994 \JPCM {\bf 6} 3261--72
\item[] Campbell I A, Amato A, Gygax F N, Herlach D, Schenck A, Cywinski R and
Kilcoyne S H 1994 \PRL {\bf 72} 1291--4
\item[] Caspary R, Hellmann, Keller M, Sparn G, Wassilew C, K\"ohler R, Geibel
C, Schank C, Steglich F and Philips N E 1993 \PRL {\bf 71} 2146--9
\item[] Celio M 1986 \PRL {\bf 56} 2720--3
\item[] Chappert J and Grynszpan R I (ed) 1984
{\sl Muon and Pions in Materials Research\/} (Amsterdam: North-Holland).
\item[] Chappert J and Yaouanc A 1986 in {\sl Topics in Current Physics}
vol 40 ed Gonser U 297--316 (Berlin: Springer-Verlag)
\item[] Chattopadhyay T, Scott C A and L\"ohneysen H v 1995 \JMMM {\bf 140-144} 
1259--60
\item[] Chow K H, Pattenden P A, Blundell S J, Hayes W, Pratt F L, Jest\"adt Th,
Green M A, Millburn J E, Rosseinsky M J, Hitti B, Dunsiger S R, Kiefl R F, 
Chen C and Chowdhury A J S 1996 \PR B {\bf 53} R14725--8
\item[] Cloots R, Ansaldo E J and Ausloos M 1994 {\it Physica} C {\bf 221}
104--8
\item[] Condon J H 1966 \PR {\bf 145} 526--35
\item[] Condon J H and Walstedt R E 1968 \PRL {\bf 21} 612--4
\item[] Cooke D W, Smith J L, Blundell S J, Chow K H, Pattenden P A, Pratt F
L, Cox S F J, Brown S R, Morrobel-Sosa A, Lichti R L, Gupta L C, Nagarajan R,
Hossain Z, Mazumdar C and Godart C 1995 \PR B {\bf 52} R3864--7
\item[] Cooke D W, Bennett B L, Lawson A C, Huber J G, Ootens J, Boekema C,
Flint J A and Lichti R L 1996 {\it Phil. Mag.} B {\bf 74} 259--67
\item[] Corti M, Rigamonti A, Tabak F, Carretta P, Licci F and Raffo L 1995
\PR B {\bf 52} 4226--36
\item[] Cox S F J 1987 \JPC {\bf 20} 3187--319
\item[] Crabtree G W, and Nelson D R April 1997 {\it Physics Today} 38--44
\item[] Cristofolini L, Lappas A, Vavekis K, Prassides K, DeRenzi R, Ricco M,
Schenck A, Amato A, Gygax F N, Kosaka M and Tanigaki K 1995
\JPCM {\bf 7} L567--73
\item[] Crook M R and Cywinski R 1997 \JPCM {\bf 9} 1149--58
\item[] Cubitt R, Forgan E M, Yang G, Lee S L, Paul D McK, Mook H A, Yethiraj
M, Kes P H, Li T W, Menovsky A A, Tarnawski Z and Mortensen K 1993 {\it Nature} 
{\bf 365} 407--11
\item[] Cywinski R, Han Z P, Bewley R, Cubitt R, Wylie M T, Forgan E M, Lee S
L, Warden M and Kilcoyne S H 1994 {\it Physica} C {\bf 233} 273--80 
\item[] Cywinski R, Coles B R, Kilcoyne S H and J-G Park 1995 
{\it Physica} B {\bf  206$\&$207} 412--4
\item[] Dalmas de R\'eotier P, Yaouanc A, Gubbens P C M and
L'H\'eritier P 1990, private communication
\item[] Dalmas de R\'eotier 1990 PhD thesis, INPG, Grenoble, unpublished 
\item[] Dalmas de R\'eotier P and Yaouanc A 1992 \JPCM {\bf 4} 4533--56
\item[] \dash 1994 \PRL {\bf 72} 290--3
\item[] \dash 1995 \PR B {\bf 52} 9155--8
\item[] Dalmas de R\'eotier P, Yaouanc A and Meshkov S V 1992 \PL A
{\bf 162} 206--12
\item[] Dalmas de R\'eotier P, Yaouanc A and Frey E 1994 \PR B {\bf 50} 3033--6
\item[] Dalmas de R\'eotier P, Huxley A, Yaouanc A, Flouquet J, Bonville P,
Imbert P, Pari P, Gubbens P~C~M and Mulders A M 1995 \PL {\bf 205A} 239--43
\item[] Dalmas de R\'eotier P, Yaouanc A and Bonville P 1996 \JPCM {\bf 8} 
5113--23
\item[] Dalmas de R\'eotier P, Yaouanc A, Calemczuk R, Huxley A.D.,
Marcenat C, Bonville P, Lejay P, Gubbens P C M and Mulders A M 1997 \PR B
{\bf 55} 2737--40
\item[] des Cloizeaux J and Pearson J J 1962 \PR {\bf 128} 2131--5
\item[] de Visser A, Keizer R J, van Harrevelt R, Menovsky A A, Franse J J M,
Amato A, Gygax F N, Pinkpank M and Schenck A 1997 {\it Physica} B {\bf 230-232}
53--5 
\item[] Dunsiger S R, Kiefl R F, Chow K H, Gaulin B D, Gingras M J P, Greedan J
E, Keren A, Kojima K, Luke G M, MacFarlane W A, Raju N P, Sonier J E, Uemura Y
J and Wu W D 1996 \PR B {\bf 54} 9019--22
\item[]  Duginov V N, Grebinnik V G, Hory\'n R, Kirillov B F, Klamut J,
Krivosheev I A, Mamedov T N, Olshevsky V G, Pirogov A V, Pomjakushin V Yu,
Ponomarev A N, Zaleski A J and Zhukov V A 
1995 \JMMM {\bf 140-144} 1577--8
\item[]  Duginov V N, Grebinnik V G, Gritsaj K I, Mamedov T N, 
Olshevsky V G, Pomjakushin V Yu, Zhukov V A, Krivosheev I A, Ponomarev A N, 
Nikiforov V N, Seropegin Yu D, Baran M and Szymczak H 1997 \PR B {\bf 55}
12343--6
\item[] Eccleston R S, Brown S R and S B Palmer 1995 \JMMM {\bf 140-144} 
745--6
\item[] Feyerherm R, Amato A, Gygax F N, Schenck A, Geibel C, Steglich F, Sato
N and Komatsubara T 1994 \PRL{\bf 73} 1849--52
\item[] Feyerherm R, Amato A, Geibel C, Gygax F N, Hellmann P, Heffner R H,
MacLaughlin D E, M\"uller-Reisener R, NieuwenHuys G J, Schenck A and Steglich F
1995a {\it Physica} B {\bf  206$\&$207} 596--9
\item[] Feyerherm R, Amato A, Grayevsky A, Gygax F N, Kaplan N and Schenck A
1995b \ZP B {\bf 99} 3--13 
\item[] Feyerherm R, Amato A, Gygax F N, Schenck A, Onuki Y and Sato N 
1995c \JMMM {\bf 140-144} 1175--6
\item[] Fisher K H and Hertz J A 1991 {\sl Spin Glasses} 
(Cambridge University press)
\item[] Flaschin S J, Kratzer A, Burghart F J, Kalvius G M, W\"appling R,
Noakes D R, Kadono R, Watanabe I, Takabatake T, Kobayashi K, Nakamoto G and
Fujii H 1996 \JPCM {\bf 8} 6967-83 
\item[] Frey E and Schwabl F 1988 \ZP B {\bf 71} 355--68
\item[] \dash 1989 \ZP B {\bf 76} 139--9
\item[] \dash 1994 {\it Adv. Phys.} {\bf 43} 577--683
\item[] Garc\'ia-Mu\~noz J L, Lacorre P and Cywinski R 1995a \PR B {\bf 51}
15197--202
\item[] Garc\'ia-Mu\~noz J L, Suaaidi M and Mart\'inez 1995b \PR B {\bf 52}
4288--93
\item[] Georges A and Kotliar G 1992 \PR B {\bf 45} 6479--83
\item[] Gingras M J P and Huse D A 1996 \PR B {\bf 53} 15193--200
\item[] Glazman L I and Koshelev A E 1991 \PR B {\bf 43} 2835--43 
\item[] Gl\"uckler H, Niedermayer Ch, Bernhard C, Binninger U, Recknagel E,
Tallon J L and Budnick J L 1995 {\it Physica} C {\bf 242} 39--45
\item[] Grebinnik V G, Gritsai K I, Duginov V N, Zhukov V A, Kirillov B F,
Koksharov Yu A, Krivosheev I A, Mamedov T N, Nikiforov V N, Nikolsky B A,
Olshevsky V G, Pirogov A V, Pomyakushin V Yu, Ponomarev A N and Suetin V A 1996 
{\it Phys. Atom. Nuclei} {\bf 59} 195--8
\item[] Greer A J and Kossler W J 1995 {\it Low Magnetic Fields in Anisotropic
Superconductors\/} Lecture Notes in Physics m30 (Berlin: Springer-Verlag)
\item[] Gubbens P C M, Moolenaar A A, Dalmas de R\'eotier P, Yaouanc A,
Menovsky A A, Prokes K and Snel C E 1995 \JMMM {\bf 140-144} 1993--4
\item[] Gubbens P C M, Mulders A M, Dalmas de R\'eotier P, Yaouanc A, 
Chevalier B and Menovsky AA 1996, private communication
\item[] Goss Levi B October 1996 {\it Physics Today} 17--20
\item[] Gross F, Chandrasekhar B S, Einzel D, Andres K, Hirschfeld P J, Ott H
R, Beuers J, Fisk Z and Smith J L 1986 \ZP B {\bf 64} 175--88
\item[] Gross-Alltag F, Chandrasekhar B S, Einzel D, Hirschfeld P J and 
Andres K 1991 \ZP B {\bf 82} 243--255
\item[] Haldane F D M 1983a \PL A {\bf 93} 464--8
\item[] Haldane F D M 1983b \PRL {\bf 50} 1153--6
\item[] Halperin B I and Hohenberg P C 1967 \PRL {\bf 19} 700--3
\item[] Hardy W N, Bonn D A, Morgan D C, Liang R X and Zhang K 1993
\PRL {\bf 70} 3999--4002
\item[] Harshman D R, Kleiman R N, Inui M, Espinosa G P, Mitzi D B, Kapitulnik
A, Pfiz T and Williams D Ll 1991 \PRL {\bf 67} 3152--5
\item[] Harshman D R, Brandt E H, Fiory A T, Inui M, Mitzi D B, Schneemeyer
L F and Waszczak J V 1993 \PR B {\bf 47} 2905--8
\item[] Harshman D R and Fiory A T 1994 \PRL {\bf 72} 2501--1
\item[] Harshman D R, Fiory A T, Haddon R C, Kaplan M L, Pfiz T, Koster E,
Shinkoda I and Williams D Ll 1994 \PR B {\bf 49} 12990--7 
\item[] Hayano R S, Uemura Y J, Imazato J, Nishida N, Yamazaki T and Kubo R
1979 \PR B {\bf 20} 850--9
\item[] Heffner R H, Cooke D W, Giordi A L, Hutson R L, Schillaci M E, Rempp H
D, Smith J L, Willis J O, MacLaughlin D E, Boekema C, Lichti R L, Oostens J and
Denison A B 1989 \PR B {\bf 39} 11345--57
\item[] Heffner R H, Le L P, Hundley M F, Neumeier J J, Luke G M, Kojima K,
Nachumi B, Uemura Y J, MacLaughlin D E and Cheong S-W 1996 \PRL{\bf 77} 1869--72
\item[] Heffner R H and Norman M R 1996 {\it Comm. Condens. Matter Phys.}
{\bf 17} 361--408 
\item[] Heffner R H, Le L P, Nieuwenhuys G J, MacLaughlin D E, Amato A, Gygax F
N, Schenck A, Kin J S, Stewart G and Ott H R 1997 {\it Physica} B, in press
\item[] Hillberg M, de Melo M A C, Klauss H H, Wagener W, Litterst F J,
Adelmann P and Czjzek G, 1997 {\it Hyperfine Interact.} {\bf 104} 221--6
\item[] Huxley A D, Dalmas de R\'eotier P, Yaouanc A., Caplan D, Couach M,
Lejay P, Gubbens P C M and Mulders A M 1996 \PR B {\bf 54} R9666--9
\item[] Inui M and Harshman D R 1993 \PR B {\bf 47} 12205--13
\item[] Kadono R, Brewer J H, Chow K, Kreitzman S R, Niedermayer Ch, Riseman T
M, Schneider J W and Yamazaki T 1993 \PR B {\bf 48} 16803--6
\item[] Kadono R, Okajima H, Yamashita A, Ishii K, Yokoo T, Akimitsu J,
Kobayashi N, Hiroi Z, Takano M and Nagamine K 1996 \PR B {\bf 54} R9628--30
\item[] Kadono R 1997 \JPSJ {\bf 66} 505--6
\item[] Kalvius G M, Kratzer A, W\"appling R, Takabatake T, Nakamoto G, Fujii
H, Kiefl R F and Kreitzman S R 1995a {\it Physica} B {\bf  206$\&$207} 807--9
\item[] Kalvius G M, Noakes D R, Kratzer A, M\"unch K H, W\"appling R, 
Tanaka H, Takabatake T and Kiefl R F  1995b {\it Physica} 
B {\bf  206$\&$207} 205--8
\item[] Kalvius G M, Kratzer A, Noakes D R, M\"unch K H, W\"appling R,
Tanaka H, Takabatake T and Kiefl R F 1995c {\it Europhys. Lett.} {\bf 29} 
501--6
\item[] Kambara H, Yoshizumi T, Mamiya T, Kimura N, Settai R, Yamamoto E, Haga
Y and \~{O}nuki Y 1996 {\it Europhys. Lett.} {\bf 36} 545--9
\item[] Kambe S, Raymond S, Regnault L P, Flouquet J, Lejay P and Haen P 1996
\JPSJ {\bf 65} 3294--300
\item[] Karlsson E B 1995 {\it Solid State Phenomena as seen by Muons,
Protons and Excited Nuclei\/} (Oxford: Clarendon )
\item[] Kehr K W, Honig G and Richter D 1978 \ZP B {\bf 32} 49--58
\item[] Keren A, Le L P, Luke G M, Sternlieb B J, Wu W D, Uemura Y J,
Tajima S and Uchida S 1993 \PR B {\bf 48} 12926--35
\item[] Keren A 1994 \PR B {\bf 50} 10039--42
\item[] Keren A, Kojima K, Le L P, Luke G M, Wu W D, Uemura Y J,
Tajima S and Uchida S 1995 \JMMM {\bf 140-144} 1641--2
\item[] Keren A, Kojima K, Le L P, Luke G M, Nachumi B, Wu W D, Uemura Y J,
Takano M, Dabkowska H and Gingas M J P 1996a \PR B {\bf 53} 6451--4
\item[] Keren A, Mendels Ph, Campbell I A and Lord J 1996b \PRL {\bf 77} 
1386--9 
\item[] Kilcoyne S H and Telling M T F 1995 \JMMM {\bf 140-144} 871--2
\item[] Klauss H-H, de Melo M A C, Hillberg M, Litterst F J, Asch L, Kratzer A,
Kalvius G M, Mattenberger K and Vogt O 1995 \JMMM {\bf 140-144} 1163--4
\item[] Kleiman R N, Broholm C, Aeppli G, Bucher E, St\"ucheli N, Bishop D J,
Clausen K N, Mortensen K, Pedersen J S and Howard B 1992 \PRL {\bf 69} 3120--3
\item[] Kogan V G 1981 \PL A {\bf 85} 298--300
\item[] Kojima K, Keren A, Luke G M,  Nachumi B, Wu W D, Uemura Y J,
Azuma M and Takano M 1995a \PRL {\bf 74} 2812--5
\item[] Kojima K, Keren A, Le L P, Luke G M, Nachumi B, Wu W D, Uemura Y J,
Kiyono K, Miyasaka S, Takagi H and Uchida S 1995b \PRL{\bf 74} 3471--4
\item[] Kojima K, Fudamoto Y., Larkin M., Luke G M, Merrin J, Nachumi B, 
Uemura Y J, Motoyama N, Eisaki H., Uchida S, Yamada Y., Endoh Y., Hosoya S,
Sternlieb B.J. and Shirane G. 1997 \PRL {\bf 78} 1787--90
\item[] Kondo S, Johnston D C, Swenson C A, Borsa F, Mahajan A V, Miller L L,
Gu T, Goldman A I, Maple M B, Gajewski D A, Freeman N R, Dilley N R, Dickey R
P, Merrin J, Kojima K, Luke G M, Uemura Y J, Chmaissem O and Jorgensen J D
1997 \PRL {\bf 78} 3729--32
\item[] Kornilov E I and Pomjakushin V Yu 1991 \PL A {\bf 153} 364--7
\item[] K\"otzler J 1986 \JMMM {\bf 54-57} 649--54
\item[] Krivosheev I A, Nezhivoi A A, Nikol'skii B A, Ponomarev A N, Duginov V
N, Ol'shevskii V G and Pomjakushin V Yu 1997 {\it JETP Lett.} {\bf 65} 81--5
\item[] Kuramoto Y and Miyake K 1990 \JPSJ {\bf 59} 2831--40
\item[] Kubo R  1981 {\it Hyperfine Interact.} {\bf 8} 731--8
\item[] Kyogaku M. Kitaoka Y, Asayama K, Geibel C, Schank C and Steglich F 1993
\JPSJ {\bf 62} 4016--30
\item[] Lappas A, Prassides K, Amato A, Feyerherm R, Gygax F N and A Schenck
1994 \ZP B {\bf 96} 223--6
\item[] Lappas A, Prassides K, Amato A, Feyerherm R, Gygax F N and A Schenck
1995 \JMMM {\bf 140-144} 1291--2
\item[] Latroche M, Figiel H, Wiesinger G, Kapusta Cz, Mietniowski P,
Paul-Boncour V, Percheron-Guegan A and Cywinski R 1996 \JPCM {\bf 8}
4603--15
\item[] Le L P, Keren A, Luke G M, Sternlieb B J, Wu W D, Uemura Y J, Brewer J
H, Riseman T M, Upasani R V, Chiang L Y, Kang W, Chaikin P M, Csiba T and
Gr\"uner 1993a \PR B {\bf 48} 7284--96
\item[] Le L P, Keren A, Luke G M, Wu W D, Uemura Y J, Tamura M, Ishikawa M and
Kinoshita M 1993b {\it Chem. Phys. Lett} {\bf 206} 405--8
\item[] Le L P, Heffner R.H., Nieuwenhuys G, Canfield P C and Cho B K,
Amato A, Feyerherm R, Gygax F N,  MacLaughlin D E and Schenck A 1995
{\it Physica} B {\bf  206$\&$207} 552--4
\item[] Le L P, Heffner R.H., Thompson J.D., MacLaughlin D E, Nieuwenhuys G
J, Amato A, Feyerherm R, Gygax F N, Schenck A, Canfield P C and Cho B K
1996a \PR B {\bf 53} R510--3
\item[] Le L P, Heffner R H, MacLaughlin D E, Kojima K, Luke G M, Nachumi B,
Uemura Y J, Sarrao J L and Fisk Z 1996b \PR B {\bf 54} 9538--41
\item[] Lee S L, Zimmermann P, Keller H, Warden M, Savi\'c I M, Schauwecker R,
Zech D, Cubitt R, Forgan E M, Kes P H, Li T W, Menovsky A A and Tarnawski Z 
1993 \PRL {\bf 71} 3862--5
\item[] Lee S L, Warden M, Keller H, Schneider J W, Zech D, Zimmermann P, 
Cubitt R, Forgan E M, Wylie M T, Kes P H, Li T W, Menovsky A A and Tarnawski Z
1995 \PRL {\bf 75} 922--5
\item[] Lee S L, Aegerter C M, Keller H, Willemin M, St\"auble-P\"umping B,
Forgan E M, Lioyd S H, Blatter G, Cubitt R, Li T W and Kes P H 1997 \PR B
{\bf 55 } 5666--9
\item[] Lidstr\"om E, W\"appling R, Hartmann O, Ekstr\"om M and Kalvius G M
1996a \JPCM {\bf 8} 6281-96
\item[] Lidstr\"om E, W\"appling R and Hartmann O 1996b {\it Physica Scripta} 
{\bf 54} 210--5
\item[] Luke G M, Le L P, Sternlieb B J, Wu W D, Uemura Y J, Brewer J H, Kadono
R, Kiefl R F, Kreitzman S R, Riseman T M, Dalichaouch Y, Lee B W, Maple M B,
Seaman C L, Armstrong P E, Ellis R W, Fisk Z and Smith J L 1991 \PL A 
{\bf 157} 173--7 
\item[] Luke G M, Keren A, Kojima K, Le L P, Sternlieb B J, Wu W D and Uemura
Y J 1994 \PRL {\bf 73} 1853--6
\item[] Luke G M, Keren A, Kojima K, Le L P, Wu W D and Uemura Y J, Kalvius 
G M, Kratzer A, Nakamoto G, T. Takabatake and M. Ishikawa 1995
{\it Physica} B {\bf  206$\&$207} 222--4
\item[] Lussier B, Taillefer L, Buyers W J L, Mason T E and Petersen 1996
\PR B {\bf 54 } R6873--6
\item[] Lovesey S W, Karlsson E B and Trohidou K N 1992 \JPCM {\bf 4} 2043-60
\item[] Lovesey S W and Engdahl E 1995 \JPCM {\bf 7} 769-76
\item[] Lovesey S W, Balcar E and Cuccoli A 1995 \JPCM {\bf 7} 2615-31
\item[] MacFarlane W A, Kiefl R F, Dunsiger S, J E Sonier and Fischer J E 1995
\PR B {\bf 52} R6995--7
\item[] MacLaughlin D E, Gupta L C, Cooke D W, Heffner R H, Leon M and
Schillaci M E 1983 \PRL {\bf 51} 927--30
\item[] MacLaughlin D E, Bernal O O and Lukefahr H G 1996 \JPCM {\bf 8} 
9855--70 
\item[] Mermin H D and Wagner H 1966 \PRL {\bf 17} 1133--6
\item[] Matsuda M, Katsumata, K, Kojima K M, Larkin M, Luke G M, Merrin J,
Nachumi B, Uemura Y J, Eisaki H, Motoyama N, Uchida S and Shirane G, 1997 
\PR B {\bf 55} 11953--6
\item[] McMullen T and Zaremba E 1978 \PR B {\bf 18} 3026--40
\item[] Mendels P, Alloul H, Brewer J H, Morris G D, Duty T L, Johnston S,
Ansaldo E J, Collin G, Marucco J F, Niedermayer Ch, Noakes D R and Stronach C E
1994 \PR B {\bf 49} R10035--8
\item[] Mekata M, Asano T, Sugino T, Nakamura H, Asai N, Shiga M, Keren A, 
Kojima K, Luke G M,  Wu W D, Uemura Y J, Dunsinger S and Gingas M
1995 \JMMM {\bf 140-144} 2177--8
\item[] Mezei F and Murani A P 1979 \JMMM {\bf 14} 211--4
\item[] Moussa F, Hennion M, Rodriguez-Carvajal J, Moudden H, Pinsard L, and
Revcolevschi A 1996 \PR B {\bf 54} 15149--55
\item[] Murani A P and Eccleston R S 1996 \PR B {\bf 53} 48--51
\item[] Muzikar P 1997 \JPCM {\bf 9} 1159--79
\item[] Nachumi B, Keren A, Kojima K, Larkin M, Luke G M, Merrin J,
Tchernysh\"ov O, Uemura Y J, Ichikawa N, Goto M and Uchida S 1996 
\PRL{\bf 77} 5421--4
\item[] Nelson D R 1997 {\it Nature} {\bf 385} 675--6
\item[] Noakes D R, Ismail A, Ansaldo E J, Brewer J H, Luke G M, Mendels P and
Poon S J 1995 \PL A {\bf 199} 107--112 
\item[] Niedermayer Ch, Bernhard C, Binninger U, Gl\"uckler H, Tallon J L, 
Ansaldo E J and Budnick J I 1993 \PRL {\bf 71} 1764--7
\item[] Nieuwenhuys G J, Mentink S A M, Menovsky A A, Amato A, Feyerherm R,
Gygax F N, Heffner R H, Le L P, MacLaughlin D E and Schenck A 1995 
{\it Physica} B {\bf  206$\&$207} 470--2
\item[] Park J-G, Coles B R, Scott C and Cywinski R 1996 {\it Physica} B 
{\bf  223$\&$224} 189--191
\item[] Pattenden P A, Valladares R M, Pratt F L, Blundell S J, Fisher A J,
Hayes W and Sugano T 1995 {\it Synthetic Metals} {\bf 71} 1823--4
\item[] P\'epin C and Lavagna M 1997, submitted for publication
\item[] Perring T G, Aeppli G, Hayden S M, Carter S A, Remeika J P and Cheong
S-W 1996 \PRL {\bf 77} 711--4
\item[] Pomjakushin V Yu, Zakharov A A, Amato A, Duginov V N, Gygax F N,
Herlach D, Ponomarev A N and Schenck A 1996 {\it Physica} C {\bf 272} 250--6
\item[] Prassides K, Lappas A, Buchgeister M and Verges P 1995
{\it Europhys. Lett.} {\bf 29} 641--6
\item[] Pratt F L, Sasaki T, Toyota N and Nagamine 1995 \PRL {\bf 74} 3892--5
\item[] Prokes K, Svoboda P, Sechovsky V, Br\"uck E, Amato A, Feyerherm R,
Gygax F N, Schenck A, Maletta H and de Boer F R 1995 \JMMM {\bf 140-144} 1381--2
\item[] Rainford B D, Cywinski R and Dakin S J 1995a \JMMM {\bf 140-144}
805--6
\item[] Rainford B D, Adroja D T, W\"appling R, Kalvius G M and Kratzer A
1995b {\it Physica} B {\bf  206$\&$207} 202--4
\item[] Regnault L P, Erkelens W A, Rossat-Mignod J, Lejay P and Flouquet J
1988 \PR B {\bf 38} 4481--7
\item[] Rice T M, Gopalan S and Sigrist M 1993 {\it Europhys. Lett.} {\bf 23}
445--9
\item[] Riseman T M, Brewer J H, Chow K h, Hardy W N, Kiefl R F, Kreitzman S R,
Liang R, MacFarlane W A, Mendels P, Morris G D, Rammer J and Schneider J W
1995 \PR B {\bf 52} 10569--80
\item[] Ryu S, Kapitulnik A and Doniach S 1996 \PRL {\bf 77} 2300--3 
\item[] Ryu S and Stroud D 1996 \PR B {\bf 54} 1320--33
\item[] Sauls J A 1994 {\it Adv. Phys.} {\bf 43} 113--41
\item[] Schatz G and Weidinger A 1995 {\sl Nuclear Condensed Matter Physics}
(Chichester: John Wiley \& Sons)
\item[] Schenck A 1985 {\sl Muon Spin Rotation Spectroscopy\/} (Bristol: Adam
Hilger)
\item[] Schenck A, Birrer P, Gygax F N, Hitti B, Lippelt E, Weber M, B\"oni P,
Fischer P, Ott H R and Fisk Z 1990 \PRL {\bf 65} 2454--7
\item[] Schenck A and Gygax F N 1995 in {\sl Handbook of Magnetic Materials}
vol 9 ed Bushow K H J 57--302 (Amsterdam: Elsevier)
\item[] Schneider J W, Schafroth S and Meier P F 1995 \PR B {\bf 52} 3790--3
\item[] Seeger A 1978 in {\sl Topics in Current Physics}
vol 28 ed  Alefeld G and V\"olkl J 349--97 (Berlin: Springer-Verlag)
\item[] Shibata F and Shimoo Y 1995 {\it Physica} A {\bf 215} 87--103
\item[] Shoenberg D 1984 {\sl Magnetic Oscillations in Metals} 
(Cambridge University press )
\item[] Sohma A, Okajima H, Yokoo T, Yamashita A, Akimitsu J, Nishiyama K and
Nagamine K 1995 \JPSJ {\bf 64} 3060--5
\item[] Song Y -Q 1995 {\it Physica} C {\bf 241} 186--90
\item[] Sonier J E, Kiefl R F, Brewer J H, Bonn D A, Carolan J F, Chow K H,
Dosanjh P, Hardy W N, Ruixing L, MacFarlane W A, Mendels P, 
Morris G D, Riseman T M and Schneider J W 1994 \PRL {\bf 72} 744--7
\item[] Sonier J E, Kiefl R F, Brewer J H, Bonn D A, Dunsiger S R, Hardy W N, 
Ruixing L, MacFarlane and Riseman T M 1997 \PR B {\bf 55} 11789--92 
\item[] Solt G 1994 \PL A {\bf 189} 390--4
\item[] \dash 1995 {\it Hyperfine Interact.} {\bf 96} 167--75
\item[] Solt G, Baines C, Egorov V S, Herlach D, Krasnoperov E and Zimmermann U
1996a  \PRL {\bf 76} 2575--8
\item[] \dash 1996b in {\it From Quantum Mechanics to Technology\/},
Lecture Notes in Physics Vol. 477, eds. Petru Z, Przystawa J, Rapcewicz K
(Berlin: Springer-Verlag)
\item[] Stammler Th, Grund Th, Hampele M, Major J, Notter M, Scheuermann R,
Schimmele L and Seeger A 1995 {\it Phil. Mag.} B {\bf 72} 285--94
\item[] Storchak V, Kirillov B F, Pirogov A V, Duginov V N, Grebinnik V G,
Ol'shevsky V G, Pomyakushin V Yu, Lazarev A B, Shilov S N and Zhulov V A
1994 \PL A{\bf 185} 338--42
\item[] Strong S P February 1997 {\it Physics Today} 13--3
\item[] Sulaiman Shukri B, Srinivas Sudha, Sahoo N, Hagelberg F, Das T P,
Torikai E and Nagamine K 1994 \PR B {\bf 49} 9879--84 
\item[] S\"ullow S, Hendrix R W A, Gortenmulder T J, Nieuwenhuys G J, Menovsky
A A, Schenck A and Mydosh J A 1994 {\it Physica} C {\bf 233} 138--42
\item[] Szeto K Y 1987 \PR B {\bf 35} 5209--18
\item[] Tallon J L, Bernhard C, Binninger U, Hofer A, Williams G V M, 
Ansaldo E J, Budnick J I and Niedermayer ch 1995 \PRL {\bf 74} 1008--11
\item[] Tallon J L, Williams G V M, Bernhard C, Pooke D M, Staines M P, Johnson
J D and Meinhold R H 1996 \PR B {\bf 53} R11972--5
\item[] Tamura M, Nakazawa Y, Shiomi D, Nozawa K, Hosokoshi Y, Ishikawa M,
Takahashi M and Kinoshita M 1991 {\it Chem. Phys. Lett.} {\bf 186} 401--4
\item[] Tchernyshyov O, Blaer A S, Keren A, Kojima K, Luke G M,  Wu W D, 
Uemura Y J, Hase M, Uchinokura K, Ajiro Y, Asano T and Mekata M 1995 
\JMMM {\bf 140-144} 1687--8
\item[] Telling M T F and Cywinski R 1995 \JMMM {\bf 140-144}
45--6
\item[] Tsuei C C, Kirtley J P, Chi C C, Yu-Jahnes Lock See, Gupta A, Shaw T,
Sun J Z, Ketchen M B 1994 \PRL {\bf 73} 593--6
\item[] Uemura Y J, Le L P, Luke G M, Sternlieb B J, Wu W D, Brewer J H,
Riseman T M, Seaman C L, Maple M B, Ishikawa M, Hinks D G, Jorgensen J D, Saito
G and Yamochi H 1991 \PRL {\bf 66} 2665--8
\item[] Uemura Y J, Keren A, Le L P, Luke G M,  Wu W D, Kubo Y, Manako T,
Shimakawa Y, Subramanian M, Cobb J L and Markert J T 1993 {\it Nature} {\bf 364}
605--7
\item[] Uemura Y J, Keren A, Kojima K, Le L P, Luke G M,  Wu W D, Ajiro Y,
Asano T, Kuriyama Y, Mekata M, Kikuchi H and Kakurai K 1994
\PRL {\bf 73} 3306--9
\item[] Uemura Y J, Kojima K, Luke G M,  Wu W D, Oszlanyi G, Chauvet O and
Forro L 1995 \PR B {\bf 52} R6991--3
\item[] Vinokur V M, Kes P H and Koshelev A E 1990 {\it Physica} 
C {\bf 168} 29--39
\item[] Walstedt R E and Walker L P 1974 \PR B {\bf 9} 4857--67
\item[] Weber M, Amato A, Gygax F N, Schenck A, Maletta H, Duginov V N,
Grebinnik V G, Lazarev A B, Olshevsky V G, Pomjakushin V Yu, Shilov S N, Zhukov
V A, Kirillov B F, Pirogov A V, Ponomarev A N, Storchak V G, Kapusta S and Bock
J 1993 \PR B {\bf 48} 13022--36
\item[] Wiesinger G, Bauer E, H\"aufler Th, Amato A, Gygax F N and Schenck A
1995 {\it Physica} B {\bf  206$\&$207} 261--3
\item[] Wu W D, Keren A, Le L P, Sternlieb B J, Luke G M, Uemura Y J, Dosanjh P
and Riseman T M 1993 \PR B {\bf 47} 8172--86 
L
\item[] Wu W D, Keren A, Le L P, Luke G M, Sternlieb B J, Uemura Y J, Seaman C
L, Dalichaouch Y and Maple M B 1994 \PRL {\bf 72} 3722--5
\item[] Yang S-H, Kumigashira H, Yokoya T, Chainani A, Takahashi T, Takeya H
and Kadowaki K 1996 \PR B {\bf 53} R11946--8
\item[] Yaouanc A and Dalmas de R\'eotier P 1991 \JPCM {\bf 3} 6195--201
\item[] Yaouanc A, Dalmas de R\'eotier P and Frey E 1993a {\it Europhys. Lett.}
{\bf 21} 93--8 
\item[] \dash 1993b \PR B {\bf 47} 796--809
\item[] Yaouanc A and Dalmas de R\'eotier P 1994 {\it Hyperfine Interact.}
{\bf 87} 1147--51
\item[] \dash 1995 \PR B {\bf 51} 12011--2
\item[] Yaouanc A, Dalmas de R\'eotier P, Gubbens P C M, Mulders A M, Kayzel F
E and Franse J J M 1996a \PR B {\bf 53} 350--3
\item[] Yaouanc A, Dalmas de R\'eotier P, Gubbens P C M, Mulders A M, Kayzel F
E and Franse J J M 1996b, private communication
\item[] Yaouanc A, Dalmas de R\'eotier P and Brandt E H 1997a \PR B {\bf 55},
11107-10
\item[] Yaouanc A, Dalmas de R\'eotier P Huxley A, Flouquet J,
Bonville P, Gubbens P~C~M and Mulders A M, 1997b private communication
\item[] Yaouanc A, Dalmas de R\'eotier P, Gubbens P~C~M, Kaiser C T, Amato A, 
Baines C, Gygax F N, Schenck A, Huxley A and Flouquet J 1997c, private 
communication
\item[] Zheludev A, Ressouche E, Schweizer J, Turek P, Wan Meixiang and Wang
Hailiang 1994 {\it Sol. Stat. Comm.} {\bf 90} 233-5
\item[] Zimmermann P, Keller H, Lee S L, Savi\'c I M, Warden M, Zech D, Cubitt
R, Forgan E M, Kaldis E, Karpinski J and Kr\"uger C 1995 \PR B {\bf 52} 541--52
\end{harvard}
\end{document}